\documentclass[11pt]{article}
\usepackage{amssymb}
\usepackage{amsfonts}
\usepackage{amsmath}
\usepackage{amssymb}
\usepackage{color}
\usepackage{graphicx,epsfig}
\usepackage{caption}
\usepackage{rotating}

\newtheorem{theorem}{Theorem}

\newtheorem{corollary}[theorem]{Corollary}

\newtheorem{definition}[theorem]{Definition}

\newtheorem{remark}[theorem]{Remark}

\newenvironment{proof}[1][Proof]{\noindent \textbf{#1.} }{\  \rule{0.5em}{0.5em}}
\input{tcilatex}
\begin{document}

\title{\textbf{Robust semiparametric inference for polytomous logistic
regression with complex survey design}}%
\author{Castilla, E.$^{1}$; Ghosh, A$^{2}$; Martin, N.$^{1}$ and Pardo, L.$%
^{1}$ \\
%EndAName
$^{1}${\small Complutense University of Madrid, Madrid, Spain} \\
$^{2}${\small Indian Statistical Institute, Kolkata, India}}
\date{}
\maketitle

\begin{abstract}
Analyzing polytomous response from a complex survey scheme, like stratified or cluster sampling is very crucial in several socio-economics applications. We present a class of  minimum quasi weighted density power divergence estimators for the polytomous logistic regression model with such a complex survey.
This family of semiparametric estimators is a robust generalization of the maximum quasi weighted likelihood estimator exploiting the advantages of the popular density power divergence measure. Accordingly robust estimators for the design effects are also derived. Robust testing of general linear hypotheses on the regression coefficients are proposed using the new estimators. Their asymptotic distributions and  robustness properties are theoretically studied and also empirically validated through a numerical example and an extensive Monte Carlo study.
\end{abstract}

\quad

%\noindent \textbf{MSC}{\small : }


\section{Introduction\label{sec1}}

In many real-life applications, we come across data that have been
collected through a complex survey scheme, like stratified sampling or
cluster sampling, etc., rather than the simple random sampling. Such
situations commonly arise in large scale data collection, for example,
within several states of a country or even among different countries.
Suitable statistical methods are required to analyze these data by taking
care of the stratified structure of the data; this is because there often
exist several inter and intra-class correlations within such stratification
and ignoring them often lead to erroneous inference. Further, in many
such complex surveys, stratified observations are collected on some
categorical responses having two or more mutually exclusive unordered
categories along with some related covariates and inference about their
relationship is of up-most interest for insight generation and policy
making. Polytomous logistic regression (PLR) model is a useful and popular
tool in such situations to model categorical responses with associated
covariates. However, most of classical literature deals with the cases of
simple random sampling scheme (e.g. McCullagh, 1980; Lesaffre and Albert,
1989; Agresti, 2002; Gupta et al., 2006, 2008). The application of PLR model
under complex survey setting can be found, for example, in Binder (1983),
Roberts et al. (1987), Morel (1989), Morel and Neerchal (2012) and Castilla
et al. (2018); most of them, except the last one, are based on the quasi
maximum likelihood approach.

Even though the maximum quasi weighted  likelihood estimator is the 
base of most of the existing literature on logistic models under complex
survey designs, it is known to be non-robust in the presence of possible
outliers in the data. In practice, with such a complex survey design, it is
quite natural to have some outlying observations that make the likelihood
based inference highly unstable. So, we often may need to make additional
efforts to find and discard the outliers from the data before their
analyses. A robust method providing stable solution even in presence of the
outliers will be really helpful and more efficient in practice. The cited
work by Castilla et al (2018) has developed an alternative minimum
divergence estimator based on $\phi $-divergences (Pardo, 2006), but the
important issue of robustness is still ignored there.

In this paper, we develop a robust estimator under the PLR model with a
complex survey based on a minimum quasi weighted  divergence approach. In
particular, we exploit the nice properties of the density power divergence
(DPD) of Basu et al. (1998). This measure become very popular in recent
literature for yielding highly robust and efficient estimators under various
statistical models; see, for example, Ghosh and Basu (2013, 2016, 2018) and, in particular, the recent paper by Castilla et al (2019)
which discussed a PLR model but under the simple random sampling
scheme.

We first start with the mathematical description of the PLR model with complex survey set-up and a brief discussion about the maximum quasi weighted likelihood estimator of the underlying parameters in Section \ref{sec:MLE}. Then we introduce a class of new robust parameter estimates   for the PLR model with complex survey by minimizing a suitably defined DPD measure in Section \ref{sec3}. The asymptotic distribution of the resulting estimator
and the design effect for the PLR model with complex survey are also
described there. In Section \ref{secWald}, a new family of  Wald-type tests is introduced based on our new estimators for testing linear hypotheses about the parameters of the PLR model. In Section \ref{sec4} we theoretically study the robustness
of the proposed estimators and Wald-type tests through the influence function analysis. After presenting an illustrative example in Section \ref{secEx}, an
extensive simulation study is presented in Section \ref{sec5}. The paper
ends with a brief concluding remark in Section \ref{secCon}. For brevity in presentation, proofs of all the results   are given in Appendix \ref{App}.

%\newpage

\section{Maximum Quasi Weighted  Likelihood Estimator \label{sec:MLE}}

Let us assume that the whole population is partitioned into $H$ distinct
strata and the data consist of $n_{h}$ clusters in stratum $h$ for each $%
h=1,\ldots ,H$. Further, for each cluster $i=1,\ldots ,n_{h}$ in the stratum 
$h$, we have observed the values of a categorical response variable ($Y$)
for $m_{hi}$ units. Assuming $Y$ has $(d+1)$ categories, we denote these
observed responses by a $(d+1)$-dimensional classification vector 
\begin{equation}
\boldsymbol{y}_{hij}=\left( y_{hij1},....,y_{hij,d+1}\right) ^{T},\text{ }%
h=1,...,H,\text{ }i=1,...,n_{h},\text{ }j=1,...,m_{hi},  \label{2.1}
\end{equation}%
with $y_{hijr}$ $=1$ and $y_{hijl}$ $=0$ for $l\in \{1,...,d+1\}-\{r\}$ if
the $j$-th unit selected from the $i$-th cluster of the $h$-th stratum falls
in the $r$-th category. We also have data on $(k+1)$ explanatory
variables which are common for all the individuals in the $i$-th cluster of the $h$-th stratum (very common with dummy or qualitative explanatory variables) to be denoted
as $\boldsymbol{x}_{hi}=\left( x_{hi0},x_{hi1},....,x_{hik}\right) ^{T}$; the first one $x_{hi0}=1$ is associated with the intercept.
Let us denote the sampling weight from the $i$-th cluster of the $h$-th
stratum by $w_{hi}$. For each $i$, $h$ and $j$, the expectation of the $r$%
-th element of the random variable $\boldsymbol{Y}%
_{hij}=(Y_{hij1},...,Y_{hij,d+1})^{T}$, corresponding to the realization $%
\boldsymbol{y}_{hij}$, is given by the PLR model 
\begin{equation}
\pi _{hir}\left( \boldsymbol{\beta }\right) =\mathrm{E}\left[ Y_{hijr}|%
\boldsymbol{x}_{hi}\right] =\Pr \left( Y_{hijr}=1|\boldsymbol{x}_{hi}\right)
=\left\{ 
\begin{array}{ll}
\dfrac{\exp \{\boldsymbol{x}_{hi}^{T}\boldsymbol{\beta }_{r}\}}{1+{%
\sum_{l=1}^{d}}\exp \{\boldsymbol{x}_{hi}^{T}\boldsymbol{\beta }_{l}\}}, & 
r=1,...,d, \\ 
\dfrac{1}{1+{\sum_{l=1}^{d}}\exp \{\boldsymbol{x}_{hi}^{T}\boldsymbol{\beta }%
_{l}\}}, & r=d+1,%
\end{array}%
\right. ,  \label{2.1.0}
\end{equation}%
with $\boldsymbol{\beta }_{r}=\left( \beta _{r0},\beta _{r1},...,\beta
_{rk}\right) ^{T}\in \mathbb{R}^{k+1}$, $r=1,...,d$. Note that, the
expectation of $\boldsymbol{Y}_{hij}$ does not depends on the unit number $j$
(homogeneity), which is not a strong assumption as we generally have random
sampling with the clusters in each stratum. Let $\boldsymbol{\pi }%
_{hi}\left( \boldsymbol{\beta }\right) $ denote the $(d+1)$-dimensional
probability vector with the elements given in (\ref{2.1.0}), i.e., 
\begin{equation}
\boldsymbol{\pi }_{hi}\left( \boldsymbol{\beta }\right) =\left( \pi
_{hi1}\left( \boldsymbol{\beta }\right) ,...,\pi _{hi,d+1}\left( \boldsymbol{%
\beta }\right) \right) ^{T}.  \label{2.2}
\end{equation}%
Then, the associated parameter space for (\ref{2.1.0}) is given by 
\begin{equation*}
\Theta =\{\boldsymbol{\beta }=(\boldsymbol{\beta }_{1}^{T},...,\boldsymbol{%
\beta }_{d}^{T})^{T},\text{ }\boldsymbol{\beta }_{r}=\left( \beta
_{r0},...,\beta _{rk}\right) ^{T}\in \mathbb{R}^{k+1},\text{ }r=1,...,d\}=%
\mathbb{R}^{d(k+1)}.
\end{equation*}

For modeling data exhibiting overdispersion, the quasi-likelihood method is an widely used method, originally defined by Wedderburn (1974). The
quasi-loglikelihood function is constructed without complete distributional
knowledge but through the mean and the variance of independent sampling units, the total of clusters in all the strata $n=\sum_{h=1}^{H}n_{h}$ in the PLR
model with complex sampling as given in (\ref{2.1.0}). It is semiparametric
method, since it only specifies the first two multivariate moments of $%
\boldsymbol{Y}_{hij}$. Let $f_{\boldsymbol{\beta }}(\boldsymbol{y}_{hij}|%
\boldsymbol{x}_{hi})$ be the probability mass function of $\boldsymbol{Y}%
_{hij}$ such that $\boldsymbol{\pi }_{hi}\left( \boldsymbol{\beta }\right) $
is modeled by the PLR model with complex sampling, i.e. since the support of 
$\boldsymbol{Y}_{hij}$, say $\mathbb{Y}_{hi}$, is the set of column vectors of
identity matrix $\boldsymbol{I}_{d+1}$ it holds
\begin{align*}
f_{\boldsymbol{\beta }}(\boldsymbol{y}_{hij}|\boldsymbol{x}_{hi})& =%
\boldsymbol{\pi }_{hi}^{T}\left( \boldsymbol{\beta }\right) \boldsymbol{y}%
_{hij}={\sum_{l=1}^{d}}\pi _{hil}\left( \boldsymbol{\beta }\right) y_{hijl},
\\
\log f_{\boldsymbol{\beta }}(\boldsymbol{y}_{hij}|\boldsymbol{x}_{hi})&
=\log \left( \boldsymbol{\pi }_{hi}^{T}\left( \boldsymbol{\beta }\right) 
\boldsymbol{y}_{hij}\right) ={\sum_{l=1}^{d}\log }\pi _{hil}\left( 
\boldsymbol{\beta }\right) y_{hijl}=\log \boldsymbol{\pi }_{hi}^{T}\left( 
\boldsymbol{\beta }\right) \boldsymbol{y}_{hij},
\end{align*}%
where we use the notation $\log \boldsymbol{\pi }_{hi}\left( \boldsymbol{%
\beta }\right) =\left( \log \pi _{hi1}\left( \boldsymbol{\beta }\right)
,...,\log \pi _{hi,d+1}\left( \boldsymbol{\beta }\right) \right) ^{T}$. It
is important to be aware that the correct loglikelihood should be 
$
\mathcal{\ell (}\boldsymbol{\beta })={\sum\limits_{h=1}^{H}}{%
\sum\limits_{i=1}^{n_{h}}}\log f_{\boldsymbol{\beta }}(\underline{%
\boldsymbol{y}}_{hi}|\boldsymbol{x}_{hi}),
$
where $\underline{\boldsymbol{y}}_{hi}=(\boldsymbol{y}_{hi1},...,\boldsymbol{%
y}_{him_{hi}})^{T}$. Under homogeneity assumption within the clusters, since 
$f_{\boldsymbol{\beta }}(\underline{\boldsymbol{y}}_{hi}|\boldsymbol{x}%
_{hi}) $ is unknown, the quasi loglikelihood, $\mathcal{\ell (}\boldsymbol{%
\beta })$, considers the likelihood within each cluster the same as
independent case as an approximation, 
\begin{equation}
\log f_{\boldsymbol{\beta }}(\underline{\boldsymbol{y}}_{hi}|\boldsymbol{x}%
_{hi})\overset{def}{=}{\sum\limits_{j=1}^{m_{hi}}}\log f_{\boldsymbol{\beta }%
}(\boldsymbol{y}_{hij}|\boldsymbol{x}_{hi}),  \label{def0}
\end{equation}%
and thus%
\begin{equation*}
\mathcal{\ell (}\boldsymbol{\beta })={\sum\limits_{h=1}^{H}}{%
\sum\limits_{i=1}^{n_{h}}}{\sum\limits_{j=1}^{m_{hi}}}\log \boldsymbol{\pi }%
_{hi}^{T}\left( \boldsymbol{\beta }\right) \boldsymbol{y}_{hij}={%
\sum\limits_{h=1}^{H}}{\sum\limits_{i=1}^{n_{h}}}\log \boldsymbol{\pi }%
_{hi}^{T}\left( \boldsymbol{\beta }\right) \widehat{\boldsymbol{y}}_{hi},
\end{equation*}%
where%
\begin{align}
\widehat{\boldsymbol{y}}_{hi}& =\boldsymbol{1}_{m_{hi}}^{T}\underline{%
\boldsymbol{y}}_{hi}=\sum\limits_{j=1}^{m_{hi}}\boldsymbol{y}_{hij},
\label{2.201} \quad
\widehat{\boldsymbol{y}}_{hi} =(\widehat{y}_{hi1},...,\widehat{y}%
_{hi,d+1})^{T}=\left(
\sum\limits_{j=1}^{m_{hi}}y_{hij1},...,\sum\limits_{j=1}^{m_{hi}}y_{hij,d+1}%
\right) ^{T},  
\end{align}%
denotes the realization value of counts in the $i$-th cluster of the $h$-th
stratum.

An important feature of the quasi loglikelihood is that the marginal
distributions of $\boldsymbol{Y}_{hij}$ are completely known but the
components of $\underline{\boldsymbol{Y}}_{hi}$, jointly, might be
correlated. This means that distribution of their total, $\widehat{%
\boldsymbol{Y}}_{hi}$, might be also unknown, but the expectation is
obtained as the total of the expectation of $\boldsymbol{Y}_{hij}$, $%
j=1,...,m_{hi}$. The most common assumption is to consider that $\widehat{%
\boldsymbol{Y}}_{hi}$ has a multinomial sampling scheme, which means that $%
\boldsymbol{Y}_{hij}$, $j=1,...,m_{hi}$ are independent random variables and%
\begin{equation}
\boldsymbol{\Sigma }_{hi}=\boldsymbol{\Sigma }_{hi}\left( \boldsymbol{\beta }%
\right) =m_{hi}\boldsymbol{\Delta }(\boldsymbol{\pi }_{hi}\left( \boldsymbol{%
\beta }\right) ),  \label{var1}
\end{equation}%
where $\boldsymbol{\Delta }(\boldsymbol{\pi }_{hi}\left( \boldsymbol{\beta }\right)) =\mathrm{diag}(\boldsymbol{\pi }_{hi}\left( \boldsymbol{\beta }\right) )-\boldsymbol{\pi }_{hi}\left( \boldsymbol{\beta }\right) \boldsymbol{\pi }_{hi}^{T}\left( \boldsymbol{\beta }\right)$. Since (\ref{def0}) is not an approximation, the term \textquotedblleft
quasi\textquotedblright\ should be dropped. A weaker assumption is to
consider that $\widehat{\boldsymbol{Y}}_{hi}$ has a multinomial sampling
scheme with a overdispersion parameter $\nu _{hi}=1+\rho _{hi}^{2}(m_{hi}-1)$, which means that $\boldsymbol{Y}_{hij}$, $j=1,...,m_{hi}$ are correlated
random variables  ($Cor[\boldsymbol{Y}_{hia},\boldsymbol{Y}%
_{hib}]=\rho _{hi}^{2}$, $a\neq b$, $a,b\in \{1,...,m_{hi}\}$) and%
\begin{equation}
\boldsymbol{\Sigma }_{hi}=\boldsymbol{\Sigma }_{hi}(\nu _{hi},\boldsymbol{%
\beta })=\nu _{hi}m_{hi}\boldsymbol{\Delta }(\boldsymbol{\pi }_{hi}\left( 
\boldsymbol{\beta }\right) ),  \label{var2}
\end{equation}%
but the distribution of $\widehat{\boldsymbol{Y}}_{hi}$ is not in principle
used for the estimators. Distributions such as Dirichlet Multinomial, Random
Clumped and $m$-inflated belong this family (see Alonso et al. (2017), Morel
and Neerchal (2012) and Raim et al. (2015) for details). The weakest
assumption is to consider that $\widehat{\boldsymbol{Y}}_{hi}$ has an
unknown distribution, with $\boldsymbol{Y}_{hij}$, $j=1,...,m_{hi}$ being
possibly correlated but with no specific pattern. It is worth of
mentioning that $\nu _{hi}$ plays here a role of nuisance parameter and it
is possible to consider a model with additional nuisance parameters, which
are more complex than (\ref{var2}) but simpler than the option of completely
unknown distribution for $\widehat{\boldsymbol{Y}}_{hi}$ (see Morel and
Koehler (1995) for details).

Taking into account weights for each cluster, the quasi weighted
loglikelihood is defined as 
\begin{equation}
\mathcal{\ell (}\boldsymbol{\beta },w)={\sum\limits_{h=1}^{H}}{%
\sum\limits_{i=1}^{n_{h}}}w_{hi}\log \boldsymbol{\pi }_{hi}^{T}\left( 
\boldsymbol{\beta }\right) \widehat{\boldsymbol{Y}}_{hi}.  \label{llike}
\end{equation}%
Then, the maximum quasi weighted likelihood estimator of $\boldsymbol{\beta }
$, say $\widehat{\boldsymbol{\beta }}_{P}$, is obtained by maximizing the quasi weighted loglikelihood, $\mathcal{\ell (}\boldsymbol{\beta },w)$, with
respect to $\boldsymbol{\beta }$. The corresponding estimating equation is
then given by 
\begin{equation}
{\sum\limits_{h=1}^{H}}{\sum\limits_{i=1}^{n_{h}}}w_{hi}\frac{\partial 
\boldsymbol{\pi }_{hi}^{T}\left( \boldsymbol{\beta }\right) }{\partial 
\boldsymbol{\beta }}\mathrm{diag}^{-1}(\boldsymbol{\pi }_{hi}\left( 
\boldsymbol{\beta }\right) )\left[ \widehat{\boldsymbol{Y}}_{hi}-m_{hi}%
\boldsymbol{\pi }_{hi}\left( \boldsymbol{\beta }\right) \right] =\boldsymbol{%
0}_{d(k+1)},  \label{2.4}
\end{equation}%
with 
\begin{align*}
\frac{\partial \boldsymbol{\pi }_{hi}^{T}\left( \boldsymbol{\beta }\right) }{%
\partial \boldsymbol{\beta }}& =\mathcal{\boldsymbol{\Delta }}^{\ast }(%
\boldsymbol{\pi }_{hi}\left( \boldsymbol{\beta }\right) )\otimes \boldsymbol{%
x}_{hi}, \quad \boldsymbol{\Delta }^{\ast }(\boldsymbol{\pi }_{hi}\left( \boldsymbol{\beta }
\right) ) =\left( \boldsymbol{I}_{d},\boldsymbol{0}_{d}\right) \boldsymbol{%
\Delta }(\boldsymbol{\pi }_{hi}\left( \boldsymbol{\beta }\right) ) .
\end{align*}%
 The system of equations (\ref{2.4}) can be written as $%
\boldsymbol{u}\left( \boldsymbol{\beta }\right) =\boldsymbol{0}_{d(k+1)}$,
where 
\begin{align}
\boldsymbol{u}\left( \boldsymbol{\beta }\right) &
=\sum\limits_{h=1}^{H}\sum\limits_{i=1}^{n_{h}}\boldsymbol{u}_{hi}\left( 
\boldsymbol{\beta },\boldsymbol{x}_{hi}\right) ,  \label{Un} \quad
\boldsymbol{u}\left( \boldsymbol{\beta },\boldsymbol{x}_{hi}\right)  =w_{hi}%
\left[ \widehat{\boldsymbol{Y}}_{hi}^{\ast }-m_{hi}\boldsymbol{\pi }%
_{hi}^{\ast }\left( \boldsymbol{\beta }\right) \right] \otimes \boldsymbol{x}%
_{hi},  
\end{align}%
with superscript $^{\ast }$ here, the vector (matrix) obtained by deleting\
the last row from the initial vector (matrix) is denoted; thus $\boldsymbol{%
\pi }_{hi}^{\ast }\left( \boldsymbol{\beta }\right) =\left( \pi _{hi1}\left( 
\boldsymbol{\beta }\right) ,...,\pi _{hid}\left( \boldsymbol{\beta }\right)
\right) ^{T}$ and $\widehat{\boldsymbol{Y}}_{hi}^{\ast }=\left( \widehat{Y}_{hi1}^{\ast
},...,\widehat{Y}_{hid}^{\ast }\right) ^{T}$. The derivation from (\ref{2.4}%
) to (\ref{Un}) is in Appendix and some additional details can be
found in Morel (1989).

\section{The Minimum Quasi Weighted Density Power Divergence estimators \label{sec3}}

Let $f_{\boldsymbol{\beta }}(\boldsymbol{y}_{hij}|\boldsymbol{x}_{hi})$\ be
the probability mass function of $\boldsymbol{Y}_{hij}|\boldsymbol{x}_{hi}$
as defined in the previous section, $g(\boldsymbol{y}_{hij}|\boldsymbol{x}%
_{hi})$ an unknown and true probability mass function of $\boldsymbol{Y}%
_{hij}|\boldsymbol{x}_{hi}$ and $\mathbb{Y}_{hi}$ the support. The DPD based
on the probability mass functions of a single observation of the sample, between  $f_{%
\boldsymbol{\beta }}(\boldsymbol{y}_{hij}|\boldsymbol{x}_{hi})$ and $g(%
\boldsymbol{y}_{hij}|\boldsymbol{x}_{hi})$, is given  by, for $\lambda >0$,
\begin{align*}
d_{\lambda }\left( g(\boldsymbol{y}_{hij}|\boldsymbol{x}_{hi}),f_{%
\boldsymbol{\beta }}(\boldsymbol{y}_{hij}|\boldsymbol{x}_{hi})\right) &
=\int\nolimits_{\mathbb{Y}_{hi}}\left( f_{\boldsymbol{\beta }}^{\lambda +1}(%
\boldsymbol{y}|\boldsymbol{x}_{hi})-\frac{\lambda +1}{\lambda }f_{%
\boldsymbol{\beta }}^{\lambda }(\boldsymbol{y}|\boldsymbol{x}_{hi})g(%
\boldsymbol{y}|\boldsymbol{x}_{hi})+\frac{1}{\lambda }g^{\lambda +1}(%
\boldsymbol{y}|\boldsymbol{x}_{hi})\right) d\boldsymbol{y} \\
& =\int\nolimits_{\mathbb{Y}_{hi}}f_{\boldsymbol{\beta }}^{\lambda }(%
\boldsymbol{y}|\boldsymbol{x}_{hi})dF(\boldsymbol{y}|\boldsymbol{x}_{hi})-%
\frac{\lambda +1}{\lambda }\int\nolimits_{\mathbb{Y}_{hi}}f_{\boldsymbol{%
\beta }}^{\lambda }(\boldsymbol{y}|\boldsymbol{x}_{hi})dG(\boldsymbol{y}|%
\boldsymbol{x}_{hi})+K \\
& =d_{\lambda }^{\ast }\left( g(\boldsymbol{y}_{hij}|\boldsymbol{x}_{hi}),f_{%
\boldsymbol{\beta }}(\boldsymbol{y}_{hij}|\boldsymbol{x}_{hi})\right) +K,
\end{align*}%
where $K$ is a constant not depending on $\boldsymbol{\beta }$, $F(%
\boldsymbol{y}|\boldsymbol{x}_{hi})$ and $G(\boldsymbol{y}|\boldsymbol{x}_{hi})$
are the distribution functions corresponding to the densities $f(\boldsymbol{y}|%
\boldsymbol{x}_{hi})$ and $g(\boldsymbol{y}|\boldsymbol{x}_{hi})$, respectively,
and%
\begin{equation*}
d_{\lambda }^{\ast }\left( g(\boldsymbol{y}_{hij}|\boldsymbol{x}_{hi}),f_{%
\boldsymbol{\beta }}(\boldsymbol{y}_{hij}|\boldsymbol{x}_{hi})\right) =E[f_{%
\boldsymbol{\beta }}^{\lambda }(\boldsymbol{Y}_{hij}|\boldsymbol{x}_{hi})]-%
\frac{\lambda +1}{\lambda }\int\nolimits_{\mathbb{Y}_{hi}}f_{\boldsymbol{%
\beta }}^{\lambda }(\boldsymbol{y}|\boldsymbol{x}_{hi})dG(\boldsymbol{y}|%
\boldsymbol{x}_{hi})
\end{equation*}%
is the kernel of $d_{\lambda }\left( g(\boldsymbol{y}_{hij}|\boldsymbol{x}%
_{hi})\text{, }f_{\boldsymbol{\beta }}(\boldsymbol{y}_{hij}|\boldsymbol{x}%
_{hi})\right) $. In practice, since $G$ is unknown, it must be estimated
from the sample, which is in this case a single individual, so%
\begin{equation*}
d_{\lambda }^{\ast }\left( \widehat{g}(\boldsymbol{y}_{hij}|\boldsymbol{x}%
_{hi}),f_{\boldsymbol{\beta }}(\boldsymbol{y}_{hij}|\boldsymbol{x}%
_{hi})\right) =E[f_{\boldsymbol{\beta }}^{\lambda }(\boldsymbol{Y}_{hij}|%
\boldsymbol{x}_{hi})]-\frac{\lambda +1}{\lambda }f_{\boldsymbol{\beta }%
}^{\lambda }(\boldsymbol{y}_{hij}|\boldsymbol{x}_{hi}).
\end{equation*}%
Based on Ghosh and Basu (2013), the \textquotedblleft kernel of the ordinary
DPD\textquotedblright\ between the probability mass functions $\widehat{g}(%
\boldsymbol{y}_{hij}|\boldsymbol{x}_{hi})$ and $f_{\boldsymbol{\beta }}(%
\boldsymbol{y}_{hij}|\boldsymbol{x}_{hi})$ for the whole sample is defined
as a total discrepancy given by
\begin{equation*}
d_{\lambda }^{\ast }(\widehat{g},f_{\boldsymbol{\beta }})=\sum_{h=1}^{H}%
\sum_{i=1}^{n_{h}}\sum_{j=1}^{m_{hi}}\left( E\left[ f_{\boldsymbol{\beta }%
}^{\lambda }(\boldsymbol{Y}_{hij}|\boldsymbol{x}_{hi})\right] -\frac{\lambda
+1}{\lambda }f_{\boldsymbol{\beta }}^{\lambda }(\boldsymbol{y}_{hij}|%
\boldsymbol{x}_{hi})\right) ,\quad \text{for }\lambda >0.
\end{equation*}%
Since the support of $\boldsymbol{Y}_{hij}|\boldsymbol{x}_{hi}$, is the set
of column vectors of identity matrix $\boldsymbol{I}_{d+1}$, it holds 
\begin{equation*}
E\left[ f_{\boldsymbol{\beta }}^{\lambda }(\boldsymbol{Y}_{hij}|\boldsymbol{x%
}_{hi})\right] =E\left[ \boldsymbol{\pi }_{hi}^{\lambda ,T}\left( 
\boldsymbol{\beta }\right) \boldsymbol{Y}_{hij}\right] =\sum_{l=1}^{d+1}\pi
_{hil}^{\lambda +1}\left( \boldsymbol{\beta }\right) .
\end{equation*}

In the current paper it is defined for the first time the \textquotedblleft
kernel of the quasi weighted DPD\textquotedblright\ between the probability
mass functions $\widehat{g}(\boldsymbol{y}_{hij}|\boldsymbol{x}_{hi})$ and $%
f_{\boldsymbol{\beta }}(\boldsymbol{y}_{hij}|\boldsymbol{x}_{hi})$ in the
whole sample as a weighted sum%
\begin{equation*}
d_{\lambda }^{\ast }(\widehat{g},f_{\boldsymbol{\beta }},w)=\sum_{h=1}^{H}%
\sum_{i=1}^{n_{h}}\sum_{j=1}^{m_{hi}}w_{hi}\left( E\left[ f_{\boldsymbol{%
\beta }}^{\lambda }(\boldsymbol{Y}_{hij}|\boldsymbol{x}_{hi})\right] -\frac{%
\lambda +1}{\lambda }f_{\boldsymbol{\beta }}^{\lambda }(\boldsymbol{y}_{hij}|%
\boldsymbol{x}_{hi})\right) ,\quad \text{for }\lambda >0,
\end{equation*}%
whose expression for the PLR model with complex sampling is%
\begin{equation}
d_{\lambda }^{\ast }(\widehat{g},f_{\boldsymbol{\beta }},w)=\sum_{h=1}^{H}%
\sum_{i=1}^{n_{h}}w_{hi}\boldsymbol{\pi }_{hi}^{\lambda ,T}\left( 
\boldsymbol{\beta }\right) \left( m_{hi}\boldsymbol{\pi }_{hi}\left( 
\boldsymbol{\beta }\right) -\frac{\lambda +1}{\lambda }\widehat{\boldsymbol{y%
}}_{hi}\right) ,\quad \text{for }\lambda >0.  \label{def}
\end{equation}

Based on $d_{\lambda }^{\ast }(\widehat{g},f_{\boldsymbol{\beta }},w)$ the
minimum quasi weighted DPD estimator is formally defined as follows.

\begin{definition}
The minimum quasi weighted DPD estimator, say $\widehat{\boldsymbol{\beta }}%
_{\lambda ,Q}$, of $\boldsymbol{\beta }$ is defined as%
\begin{equation*}
\widehat{\boldsymbol{\beta }}_{\lambda ,Q}=\arg \min_{\boldsymbol{\beta }\in 
%TCIMACRO{\U{211d} }%
%BeginExpansion
\mathbb{R}
%EndExpansion
^{d(k+1)}}d_{\lambda }^{\ast }(\widehat{g},f_{\boldsymbol{\beta }},w).
\end{equation*}
\end{definition}

At the particular choice $\lambda\rightarrow 0$, the DPD measure, defined as the limit
of (\ref{def}) coincides (in limit) with the weighted
quasi-loglikelihood, $\mathcal{\ell (}\boldsymbol{\beta },w)$, given in (\ref{llike}); thus the minimum quasi weighted DPD estimator of $\boldsymbol{%
\beta }$ at $\lambda =0$ coincides with the maximum weighted quasi likelihood
estimator. With the same philosophy, the following result generalizes $%
\boldsymbol{u}(\boldsymbol{\beta },\boldsymbol{x}_{hi})$ given in (\ref{Un}%
) which plays an important role for the derivation of the asymptotic
distribution of $\widehat{\boldsymbol{\beta}}_{\lambda, Q}$.

\begin{theorem}
\label{th:u} The minimum quasi weighted DPD estimate of $\boldsymbol{\beta }$%
, say $\widehat{\boldsymbol{\beta }}_{\lambda ,Q}$, can be obtained by
solving the system of equations $\boldsymbol{u}_{\lambda }(\boldsymbol{\beta 
})=\boldsymbol{0}_{d(k+1)}$, where 
\begin{align}
\boldsymbol{u}_{\lambda }(\boldsymbol{\beta })&
=\sum_{h=1}^{H}\sum_{i=1}^{n_{h}}\boldsymbol{u}_{\lambda }(\boldsymbol{\beta 
},\boldsymbol{x}_{hi}),  \label{u0} \\
\boldsymbol{u}_{\lambda }(\boldsymbol{\beta },\boldsymbol{x}_{hi})& =\left[
w_{hi}\boldsymbol{\Delta }^{\ast }(\boldsymbol{\pi }_{hi}\left( \boldsymbol{%
\beta }\right) )\mathrm{diag}^{\lambda -1}\{\boldsymbol{\pi }_{hi}\left( 
\boldsymbol{\beta }\right) \}\{\widehat{\boldsymbol{y}}_{hi}-m_{hi}%
\boldsymbol{\pi }_{hi}\left( \boldsymbol{\beta }\right) \}\right] \otimes 
\boldsymbol{x}_{hi}.  \label{u}
\end{align}
\end{theorem}

Let $\boldsymbol{U}_{\lambda }(\boldsymbol{\beta },\boldsymbol{X})$ be a random
variable generator of (\ref{u}) associated with a generic random explanatory
variable $\boldsymbol{X}$, with no stratum and cluster assignment. In what
is to follow, 
\begin{equation*}
\boldsymbol{U}_{\lambda }(\boldsymbol{\beta },\boldsymbol{x}_{hi})=\left[
w_{hi}\boldsymbol{\Delta }^{\ast }(\boldsymbol{\pi }_{hi}\left( \boldsymbol{%
\beta }\right) )\mathrm{diag}^{\lambda -1}\{\boldsymbol{\pi }_{hi}\left( 
\boldsymbol{\beta }\right) \}\{\widehat{\boldsymbol{Y}}_{hi}-m_{hi}%
\boldsymbol{\pi }_{hi}\left( \boldsymbol{\beta }\right) \}\right] \otimes 
\boldsymbol{x}_{hi}
\end{equation*}%
denotes $\boldsymbol{U}_{\lambda }(\boldsymbol{\beta },\boldsymbol{X})|%
\boldsymbol{X=x}_{hi}$. An important property is that the
estimating equation given in Theorem \ref{th:u} is unbiased. This property is
very important to consider these estimators  similar to the maximum
quasi weighted likelihood ones in the construction of the asymptotic properties, which was not the case for other distance based estimators such as the one proposed in Castilla et al (2018).

\subsection{Asymptotic distribution and estimates of the design effect} 

The following results are generalized for the PLR model with
complex sampling and random explanatory variables. Without any loss of
generality it is assumed that $\widehat{\Pr }(\boldsymbol{X=x}_{hi})=\frac{1%
}{n}$ is estimated from the sample of strata in order to get assymptotic
results for the fixed explanatory variables (as a particular case of random
explanatory variables).

\begin{theorem}
\label{Th1} Let $\widehat{\boldsymbol{\beta }}_{\lambda ,Q}$ the minimum
quasi weighted DPD estimate of $\boldsymbol{\beta }$ in the PLR model (\ref%
{2.1.0}) under a complex survey. Then we have%
\begin{equation*}
\sqrt{n}(\widehat{\boldsymbol{\beta }}_{\lambda ,Q}-\boldsymbol{\beta }_{0})%
\overset{\mathcal{L}}{\underset{n\mathcal{\rightarrow }\infty }{%
\longrightarrow }}\mathcal{N}\left( \boldsymbol{0}_{d(k+1)},\mathbf{\Psi}%
_{\lambda }^{-1}\left( \boldsymbol{\beta }_{0}\right) \mathbf{\Omega}_{\lambda
}\left( \boldsymbol{\beta }_{0}\right) \mathbf{\Psi}_{\lambda }^{-1}\left( 
\boldsymbol{\beta }_{0}\right) \right) ,
\end{equation*}%
where $\boldsymbol{\beta }_{0}$ is the true parameter value and 
\begin{align}
\mathbf{\Omega}_{\lambda }\left( \boldsymbol{\beta }\right) & =\lim_{n\rightarrow
\infty }\mathbf{\Omega}_{n,\lambda }\left( \boldsymbol{\beta }\right)
=\lim_{n\rightarrow \infty }\frac{1}{n}\sum_{h=1}^{H}\sum_{i=1}^{n_{h}}%
\mathbf{\Omega}_{hi,\lambda }\left( \boldsymbol{\beta },\boldsymbol{\Sigma }%
_{hi}\right) ,  \label{hs} \\
\mathbf{\Psi}_{\lambda }\left( \boldsymbol{\beta }\right) & =\lim_{n\rightarrow
\infty }\mathbf{\Psi}_{n,\lambda }\left( \boldsymbol{\beta }\right)
=\lim_{n\rightarrow \infty }\frac{1}{n}\sum_{h=1}^{H}\sum_{i=1}^{n_{h}}%
\mathbf{\Psi}_{hi,\lambda }\left( \boldsymbol{\beta }\right) ,  \label{gs}
\end{align}%
where%
\begin{align}
\mathbf{\Omega}_{hi,\lambda }\left( \boldsymbol{\beta },\boldsymbol{\Sigma }%
_{hi}\right) & =w_{hi}^{2}\boldsymbol{\Delta }^{\ast }(\boldsymbol{\pi }%
_{hi}\left( \boldsymbol{\beta }\right) )\mathrm{diag}^{\lambda -1}\{%
\boldsymbol{\pi }_{hi}\left( \boldsymbol{\beta }\right) \}\boldsymbol{\Sigma 
}_{hi}\mathrm{diag}^{\lambda -1}\{\boldsymbol{\pi }_{hi}\left( \boldsymbol{%
\beta }\right) \}\boldsymbol{\Delta }^{\ast T}(\boldsymbol{\pi }_{hi}\left( 
\boldsymbol{\beta }\right) )\otimes \boldsymbol{x}_{hi}\boldsymbol{x}%
_{hi}^{T},  \label{hhi} \\
\boldsymbol{\Sigma }_{hi}& =\mathrm{Var}[\widehat{\boldsymbol{Y}}_{hi}], 
\notag \\
\mathbf{\Psi}_{hi,\lambda }\left( \boldsymbol{\beta }\right) & =\left\{ 
\begin{array}{lc}
w_{hi}m_{hi}\boldsymbol{\Delta }^{\ast }(\boldsymbol{\pi }_{hi}\left( 
\boldsymbol{\beta }\right) )\mathrm{diag}^{\lambda -1}\{\boldsymbol{\pi }%
_{hi}\left( \boldsymbol{\beta }\right) \}\boldsymbol{\Delta }^{\ast T}(%
\boldsymbol{\pi }_{hi}\left( \boldsymbol{\beta }\right) )\otimes \boldsymbol{%
x}_{hi}\boldsymbol{x}_{hi}^{T}, & \lambda >0 \\ 
w_{hi}m_{hi}\boldsymbol{\Delta }(\boldsymbol{\pi }_{hi}^{\ast }\left( 
\boldsymbol{\beta }\right) )\otimes \boldsymbol{x}_{hi}\boldsymbol{x}%
_{hi}^{T}, & \lambda =0%
\end{array}%
\right. .  \label{ghi}
\end{align}
\end{theorem}

Notice that the expression of $\mathbf{\Psi}_{\lambda =0}\left( \boldsymbol{%
\beta }\right) $ is the same as the so called Fisher information matrix for
multinomial sampling. In addition, if $\widehat{\boldsymbol{Y}}_{hi}$ has a
multinomial sampling scheme $\mathbf{\Omega}_{\lambda =0}\left( \boldsymbol{\beta 
}\right) =\mathbf{\Psi}_{\lambda =0}\left( \boldsymbol{\beta }\right) $ and $%
\mathbf{\Omega}_{\lambda =0}^{-1}\left( \boldsymbol{\beta }\right) \mathbf{\Psi}%
_{\lambda =0}\left( \boldsymbol{\beta }\right) \mathbf{\Omega}_{\lambda
=0}^{-1}\left( \boldsymbol{\beta }\right) $ is the inverse of the Fisher
 information matrix.

Consistency is usually considered as a minimal requirement for an inference
procedure.  The following result is useful as a tool for
estimating $\mathbf{\Omega}_{\lambda }\left( \boldsymbol{\beta }\right) $ and $%
\mathbf{\Psi}_{\lambda }\left( \boldsymbol{\beta }\right) $ consistently
plugging a consistent estimator into $\boldsymbol{\beta }$, and hence also $%
\mathbf{\Omega}_{\lambda }^{-1}\left( \boldsymbol{\beta }\right) \mathbf{\Psi}%
_{\lambda }\left( \boldsymbol{\beta }\right) \mathbf{\Omega}_{\lambda
}^{-1}\left( \boldsymbol{\beta }\right) $ again as a (double) plug-in
estimator.

\begin{corollary}
\label{cor1}The following ones are (weak) consistent estimators as $n$ goes
to infinity:

\begin{description}
\item[a)] $\widehat{\boldsymbol{\beta }}_{\lambda ,Q}$ is a consistent
estimator of the true regression coefficient $\boldsymbol{\beta }_{0}$.

\item[b)] $\mathbf{\Psi}_{n,\lambda }(\widehat{\boldsymbol{\beta }}_{\lambda
,Q})=\frac{1}{n}\sum_{h=1}^{H}\sum_{i=1}^{n_{h}}\mathbf{\Psi}_{hi,\lambda }(%
\widehat{\boldsymbol{\beta }}_{\lambda ,Q})$ is a consistent estimator of $%
\mathbf{\Psi}_{\lambda }\left( \boldsymbol{\beta }_{0}\right) $.

\item[c)] $\mathbf{\Omega}_{n,\lambda }(\widehat{\boldsymbol{\beta }}_{\lambda
,Q},\{\widehat{\boldsymbol{\Sigma }}_{hi}\}_{h=1,...,H;i=1,...,n_{h}})=\frac{%
1}{n}\sum_{h=1}^{H}\sum_{i=1}^{n_{h}}\mathbf{\Omega}_{hi,\lambda }(\widehat{%
\boldsymbol{\beta }}_{\lambda ,Q},\widehat{\boldsymbol{\Sigma }}_{hi})$ with%
\begin{align}
\mathbf{\Omega}_{hi,\lambda }(\widehat{\boldsymbol{\beta }}_{\lambda ,Q},\widehat{%
\boldsymbol{\Sigma }}_{hi})& =w_{hi}^{2}\boldsymbol{\Delta }^{\ast }(%
\boldsymbol{\pi }_{hi}(\widehat{\boldsymbol{\beta }}_{\lambda ,Q}))\mathrm{%
diag}^{\lambda -1}\{\boldsymbol{\pi }_{hi}(\widehat{\boldsymbol{\beta }}%
_{\lambda ,Q})\}\widehat{\boldsymbol{\Sigma }}_{hi}  \notag \\
& \times \mathrm{diag}^{\lambda -1}\{\boldsymbol{\pi }_{hi}(\widehat{%
\boldsymbol{\beta }}_{\lambda ,Q})\}\boldsymbol{\Delta }^{\ast T}(%
\boldsymbol{\pi }_{hi}(\widehat{\boldsymbol{\beta }}_{\lambda ,Q}))\otimes 
\boldsymbol{x}_{hi}\boldsymbol{x}_{hi}^{T},  \label{Hhi}
\end{align}%
is a consistent estimator of $\mathbf{\Omega}_{\lambda }\left( \boldsymbol{\beta }%
_{0}\right) $, whenever $\Sigma _{hi}=\mathrm{Var}[\widehat{\boldsymbol{Y}}%
_{hi}]$ is consistently estimated through $\widehat{\boldsymbol{\Sigma }}%
_{hi}$ for all $(h,i)\in \{1,...,H\}\times \{1,...,m_{hi}\}$.
\end{description}
\end{corollary}

The following two cases of $\mathbf{\Omega}_{n,\lambda }(\widehat{\boldsymbol{%
\beta }}_{\lambda ,Q},\{\widehat{\boldsymbol{\Sigma }}_{hi}%
\}_{h=1,...,H;i=1,...,n_{h}})$ are taken into account for two cases of $%
\boldsymbol{\Sigma }_{hi}$ for which there exists a consistent estimator:

\begin{itemize}
\item if $\widehat{\boldsymbol{Y}}_{hi}$ has (an ordinary) multinomial
sampling scheme then%
\begin{equation*}
\mathbf{\Omega}_{n,\lambda }(\widehat{\boldsymbol{\beta }}_{\lambda ,Q})=\frac{1}{%
n}\sum_{h=1}^{H}\sum_{i=1}^{n_{h}}\mathbf{\Omega}_{hi,\lambda }(\widehat{%
\boldsymbol{\beta }}_{\lambda ,Q})
\end{equation*}%
where $\mathbf{\Omega}_{hi,\lambda }(\widehat{\boldsymbol{\beta }}_{\lambda ,Q})$
is $\mathbf{\Omega}_{hi,\lambda }(\widehat{\boldsymbol{\beta }}_{\lambda ,Q},%
\widehat{\boldsymbol{\Sigma }}_{hi})$ given in (\ref{Hhi}) with 
\begin{equation*}
\widehat{\boldsymbol{\Sigma }}_{hi}=\boldsymbol{\Sigma }_{hi}(\widehat{%
\boldsymbol{\beta }}_{\lambda ,Q})=m_{hi}\boldsymbol{\Delta }(\boldsymbol{%
\pi }_{hi}(\widehat{\boldsymbol{\beta }}_{\lambda ,Q}));
\end{equation*}

\item if $\widehat{\boldsymbol{Y}}_{hi}$ has an overdispersed multinomial
sampling scheme then%
\begin{equation*}
\mathbf{\Omega}_{n,\lambda }(\widehat{\boldsymbol{\beta }}_{\lambda ,Q},\{%
\widetilde{\nu }_{hi}\}_{h=1,...,H;i=1,...,n_{h}})=\frac{1}{n}%
\sum_{h=1}^{H}\sum_{i=1}^{n_{h}}\mathbf{\Omega}_{hi,\lambda }(\widehat{%
\boldsymbol{\beta }}_{\lambda ,Q},\widetilde{\nu }_{hi}),
\end{equation*}%
where $\widetilde{\nu }_{hi}$ is a consistent estimator of the
overdispersion parameter $\nu _{hi}$ which is established later in Corollary \ref{cor2}, and $\mathbf{\Omega}_{n,\lambda }(\widetilde{\nu }_{hi},\widehat{%
\boldsymbol{\beta }}_{\lambda ,Q})$ is $\mathbf{\Omega}_{hi,\lambda }(\widehat{%
\boldsymbol{\beta }}_{\lambda ,Q},\widehat{\boldsymbol{\Sigma }}_{hi})$
given in (\ref{Hhi}) with 
\begin{equation*}
\widehat{\boldsymbol{\Sigma }}_{hi}=\boldsymbol{\Sigma }_{hi}(\widetilde{\nu 
}_{hi},\widehat{\boldsymbol{\beta }}_{\lambda ,Q})=\widetilde{\nu }%
_{hi}m_{hi}\boldsymbol{\Delta }(\boldsymbol{\pi }_{hi}(\widehat{\boldsymbol{%
\beta }}_{\lambda ,Q}));
\end{equation*}
\end{itemize}

\begin{remark}
If $\widehat{\boldsymbol{Y}}_{hi}$ has a multinomial sampling scheme,
Theorem \ref{Th1} can be similarly formulated taking the assumption that $%
\widehat{\Pr }(\boldsymbol{X=x}_{hi})=\frac{1}{m}$, where $%
m=\sum_{h=1}^{H}\sum_{i=1}^{n_{h}}m_{hi}$, for each individual of all the
clusters in the sample, and taking for the estimation equation $m$ summands
rather than $n$, i.e. plugging ${\sum\limits_{j=1}^{m_{hi}}}\boldsymbol{y}%
_{hij}=\widehat{\boldsymbol{y}}_{hi}$ into (\ref{u0}) and considering the
system of equations $\boldsymbol{u}_{\lambda }(\boldsymbol{\beta })=%
\boldsymbol{0}_{d(k+1)}$, where%
\begin{align*}
\boldsymbol{u}_{\lambda }(\boldsymbol{\beta })&
=\sum_{h=1}^{H}\sum_{i=1}^{n_{h}}{\sum\limits_{j=1}^{m_{hi}}}\boldsymbol{v}%
_{j,\lambda }(\boldsymbol{\beta },\boldsymbol{x}_{hi}), \\
\boldsymbol{v}_{j,\lambda }(\boldsymbol{\beta },\boldsymbol{x}_{hi})& =\left[
w_{hi}\boldsymbol{\Delta }^{\ast }(\boldsymbol{\pi }_{hi}\left( \boldsymbol{%
\beta }\right) )\mathrm{diag}^{\lambda -1}\{\boldsymbol{\pi }_{hi}\left( 
\boldsymbol{\beta }\right) \}\{\boldsymbol{y}_{hij}-\boldsymbol{\pi }%
_{hi}\left( \boldsymbol{\beta }\right) \}\right] \otimes \boldsymbol{x}_{hi}.
\end{align*}%
Hence, it holds%
\begin{equation*}
\sqrt{m}(\widehat{\boldsymbol{\beta }}_{\lambda ,Q}-\boldsymbol{\beta }_{0})%
\overset{\mathcal{L}}{\underset{m\mathcal{\rightarrow }\infty }{%
\longrightarrow }}\mathcal{N}\left( \boldsymbol{0}_{d(k+1)},\mathbf{\Psi}%
_{\lambda }^{-1}\left( \boldsymbol{\beta }_{0}\right) \mathbf{\Omega}_{\lambda
}\left( \boldsymbol{\beta }_{0}\right) \mathbf{\Psi}_{\lambda }^{-1}\left( 
\boldsymbol{\beta }_{0}\right) \right) ,
\end{equation*}%
with%
\begin{align*}
\mathbf{\Omega}_{\lambda }\left( \boldsymbol{\beta }\right) & =\lim_{m\rightarrow
\infty }\left\{ 
\begin{array}{lc}
\begin{array}{l}
\frac{1}{m}\sum_{h=1}^{H}\sum_{i=1}^{n_{h}}{m}_{hi}w_{hi}^{2}\boldsymbol{%
\Delta }^{\ast }(\boldsymbol{\pi }_{hi}\left( \boldsymbol{\beta }\right) )%
\mathrm{diag}^{\lambda -1}\{\boldsymbol{\pi }_{hi}\left( \boldsymbol{\beta }%
\right) \}\boldsymbol{\Delta }(\boldsymbol{\pi }_{hi}\left( \boldsymbol{%
\beta }\right) ) \\ 
\times \mathrm{diag}^{\lambda -1}\{\boldsymbol{\pi }_{hi}\left( \boldsymbol{%
\beta }\right) \}\boldsymbol{\Delta }^{\ast T}(\boldsymbol{\pi }_{hi}\left( 
\boldsymbol{\beta }\right) )\otimes \boldsymbol{x}_{hi}\boldsymbol{x}%
_{hi}^{T},%
\end{array}
& \lambda >0 \\ 
\frac{1}{m}\sum_{h=1}^{H}\sum_{i=1}^{n_{h}}w_{hi}^{2}m_{hi}\boldsymbol{%
\Delta }(\boldsymbol{\pi }_{hi}^{\ast }\left( \boldsymbol{\beta }\right)
)\otimes \boldsymbol{x}_{hi}\boldsymbol{x}_{hi}^{T}, & \lambda =0%
\end{array}%
\right.  \\
\mathbf{\Psi}_{\lambda }\left( \boldsymbol{\beta }\right) & =\lim_{m\rightarrow
\infty }\left\{ 
\begin{array}{lc}
\frac{1}{m}\sum_{h=1}^{H}\sum_{i=1}^{n_{h}}w_{hi}m_{hi}\boldsymbol{\Delta }%
^{\ast }(\boldsymbol{\pi }_{hi}\left( \boldsymbol{\beta }\right) )\mathrm{%
diag}^{\lambda -1}\{\boldsymbol{\pi }_{hi}\left( \boldsymbol{\beta }\right)
\}\boldsymbol{\Delta }^{\ast T}(\boldsymbol{\pi }_{hi}\left( \boldsymbol{%
\beta }\right) )\otimes \boldsymbol{x}_{hi}\boldsymbol{x}_{hi}^{T}, & 
\lambda >0 \\ 
\frac{1}{m}\sum_{h=1}^{H}\sum_{i=1}^{n_{h}}w_{hi}m_{hi}\boldsymbol{\Delta }(%
\boldsymbol{\pi }_{hi}^{\ast }\left( \boldsymbol{\beta }\right) )\otimes 
\boldsymbol{x}_{hi}\boldsymbol{x}_{hi}^{T}, & \lambda =0%
\end{array}%
\right. .
\end{align*}%
The formal proof is omitted, but the derivation of the expressions is almost
the same considering $\boldsymbol{U}_{\lambda }(\boldsymbol{\beta },%
\boldsymbol{x}_{hi})={\sum\limits_{j=1}^{m_{hi}}}\boldsymbol{V}_{j,\lambda }(%
\boldsymbol{\beta },\boldsymbol{x}_{hi})$, with $\boldsymbol{V}_{j,\lambda }(%
\boldsymbol{\beta },\boldsymbol{x}_{hi})$ i.i.d. random variables $%
j=1,...,m_{hi}$. This idea matches the philosophy of the asymptotic result
developed in Castilla et al. (2019), where for $H=1$ and $w_{1i}=1$, $%
i=1,...,m_{1i}$.
\end{remark}

The following result is useful for any sample of polytomous logistic
regression with complex sample design, more general in comparison with
Corollary \ref{cor1}, since it is not necessary to get any consistent
estimators for $\boldsymbol{\Sigma }_{hi}$.

\begin{theorem}
\label{Th1b}The estimator $\widehat{\mathbf{\Omega}}_{n,\lambda }(\widehat{\boldsymbol{\beta }}_{\lambda
,Q}) =\frac{1}{n}\sum_{h=1}^{H}\sum_{i=1}^{n_{h}}\widehat{\mathbf{\Omega}}%
_{hi,\lambda }(\widehat{\boldsymbol{\beta }}_{\lambda ,Q})$, with
\begin{align*}
\widehat{\mathbf{\Omega}}_{hi,\lambda }(\widehat{\boldsymbol{\beta }}_{\lambda
,Q})& =\boldsymbol{U}_{\lambda }(\widehat{\boldsymbol{\beta }}_{\lambda ,Q},%
\boldsymbol{x}_{hi})\boldsymbol{U}_{\lambda }^{T}(\widehat{\boldsymbol{\beta 
}}_{\lambda ,Q},\boldsymbol{x}_{hi}) \\
& =\left[ w_{hi}^{2}\boldsymbol{\Delta }^{\ast }(\boldsymbol{\pi }_{hi}(%
\widehat{\boldsymbol{\beta }}_{\lambda ,Q}))\mathrm{diag}^{\lambda -1}\{%
\boldsymbol{\pi }_{hi}(\widehat{\boldsymbol{\beta }}_{\lambda ,Q})\}\{%
\widehat{\boldsymbol{Y}}_{hi}-m_{hi}\boldsymbol{\pi }_{hi}(\widehat{%
\boldsymbol{\beta }}_{\lambda ,Q})\}\right. \\
& \left. \times \{\widehat{\boldsymbol{Y}}_{hi}-m_{hi}\boldsymbol{\pi }_{hi}(%
\widehat{\boldsymbol{\beta }}_{\lambda ,Q})\}^{T}\mathrm{diag}^{\lambda -1}\{%
\boldsymbol{\pi }_{hi}(\widehat{\boldsymbol{\beta }}_{\lambda ,Q})\}%
\boldsymbol{\Delta }^{\ast T}(\boldsymbol{\pi }_{hi}(\widehat{\boldsymbol{%
\beta }}_{\lambda ,Q}))\right] \otimes \boldsymbol{x}_{hi}\boldsymbol{x}%
_{hi}^{T},
\end{align*}%
is consistent for $\mathbf{\Omega}_{\lambda }\left( \boldsymbol{\beta }\right)$ as $n$ goes to infinity.
\end{theorem}

\begin{corollary}
\label{cor2}Let $\widehat{\boldsymbol{Y}}_{hi}$ be a random variable with
overdisepersed multinomial sampling scheme with a common overdispersion parameter $\nu $ and $m_{hi}=\overline{m}$,%
\begin{align*}
\boldsymbol{\Sigma }_{hi}& =\boldsymbol{\Sigma }_{hi}\left( \nu ,\boldsymbol{%
\beta }\right) =\nu \overline{m}\boldsymbol{\Delta }(\boldsymbol{\pi }%
_{hi}\left( \boldsymbol{\beta }\right) ), \\
\nu & =1+\rho ^{2}(\overline{m}-1),
\end{align*}%
then, for $\nu $ and $\rho ^{2}$:

\begin{description}
\item[a)] \textquotedblleft robust and consistent estimators based on the
estimating equation\textquotedblright\ are given respectively by%
\begin{align}
\widetilde{\nu }_{n,\lambda }^{E}=\widetilde{\nu }_{n,\lambda }^{E}(\widehat{%
\boldsymbol{\beta }}_{\lambda ,Q})& =\frac{1}{d(k+1)}\mathrm{trace}\left( 
\mathbf{\Omega}_{n,\lambda }^{-1}(\widehat{\boldsymbol{\beta }}_{\lambda ,Q})%
\widehat{\mathbf{\Omega}}_{n,\lambda }(\widehat{\boldsymbol{\beta }}_{\lambda
,Q})\right) ,  \label{vE} \\
\widetilde{\rho }_{n,\lambda }^{2,E}& =\frac{\widetilde{\nu }_{n,\lambda
}^{E}(\widehat{\boldsymbol{\beta }}_{\lambda ,Q})-1}{\overline{m}-1},  \notag
\end{align}%
where the matrices of interest are as follows, the one associated with
multinomial sampling%
\begin{align*}
\mathbf{\Omega}_{n,\lambda }(\widehat{\boldsymbol{\beta }}_{\lambda ,Q})& =\frac{1%
}{n}\sum_{h=1}^{H}\sum_{i=1}^{n_{h}}\mathbf{\Omega}_{hi,\lambda }(\widehat{%
\boldsymbol{\beta }}_{\lambda ,Q}), \\
\mathbf{\Omega}_{hi,\lambda }(\widehat{\boldsymbol{\beta }}_{\lambda ,Q})& =%
\overline{m}w_{hi}^{2}\boldsymbol{\Delta }^{\ast }(\boldsymbol{\pi }%
_{hi}\left( \boldsymbol{\beta }\right) )\mathrm{diag}^{\lambda -1}\{%
\boldsymbol{\pi }_{hi}\left( \boldsymbol{\beta }\right) \}\boldsymbol{\Delta 
}(\boldsymbol{\pi }_{hi}\left( \boldsymbol{\beta }\right) ) \\
& \times \mathrm{diag}^{\lambda -1}\{\boldsymbol{\pi }_{hi}\left( 
\boldsymbol{\beta }\right) \}\boldsymbol{\Delta }^{\ast T}(\boldsymbol{\pi }%
_{hi}\left( \boldsymbol{\beta }\right) )\otimes \boldsymbol{x}_{hi}%
\boldsymbol{x}_{hi}^{T},
\end{align*}%
and Theorem \ref{Th1b} for overdispersed multinomial sampling.

\item[b)] \textquotedblleft robust and consistent estimators based on the
method of moments\textquotedblright\ are given by%
\begin{align}
\widetilde{\nu }_{n,\lambda }^{M}& =\widetilde{\nu }_{n,\lambda }^{M}(%
\widehat{\boldsymbol{\beta }}_{\lambda ,Q})=\frac{1}{nd}\sum_{h=1}^{H}%
\sum_{i=1}^{n_{h}}\sum_{j=1}^{d+1}\frac{(\widehat{Y}_{hij}-\overline{m}\pi
_{hij}(\widehat{\boldsymbol{\beta }}_{\lambda ,Q}))^{2}}{\overline{m}\pi
_{hij}(\widehat{\boldsymbol{\beta }}_{\lambda ,Q})},  \label{vM} \\
\widetilde{\rho }_{n,\lambda }^{2,M}& =\frac{\widetilde{\nu }_{n,\lambda
}^{M}(\widehat{\boldsymbol{\beta }}_{\lambda ,Q})-1}{\overline{m}-1}.  \notag
\end{align}
\end{description}
\end{corollary}

The preceding results were established for $n=\sum\nolimits_{h=1}^{H}n_{h}$
tending to infinity which implies in practice that $H$ is fixed and there
exists $\eta _{h}=\lim_{n\rightarrow \infty }\frac{n_{h}}{n}\in (0,1)$, $%
\sum\nolimits_{h=1}^{H}\eta _{h}=1$. This means that it is approriate to
consider in the place of Theorem \ref{Th1}, the following one, assuming that 
$\widehat{\Pr }(\boldsymbol{X}_{h}\boldsymbol{=x}_{hi})=\frac{1}{n_{h}}$.

\begin{theorem}
\label{th2} Let $\widehat{\boldsymbol{\beta }}_{\lambda ,Q}$ the quasi MDPDE
of $\boldsymbol{\beta }$ in the PLR model (\ref{2.1.0}) under a complex
survey. Then we have%
\begin{equation*}
\sqrt{n}(\widehat{\boldsymbol{\beta }}_{\lambda ,Q}-\boldsymbol{\beta }_{0})%
\overset{\mathcal{L}}{\underset{n\mathcal{\rightarrow }\infty }{%
\longrightarrow }}\mathcal{N}\left( \boldsymbol{0}_{d(k+1)},\mathbf{\Psi}%
_{\lambda }^{-1}\left( \boldsymbol{\beta }_{0}\right) \mathbf{\Omega}_{\lambda
}\left( \boldsymbol{\beta }_{0}\right) \mathbf{\Psi}_{\lambda }^{-1}\left( 
\boldsymbol{\beta }_{0}\right) \right) ,
\end{equation*}%
where $\boldsymbol{\beta }_{0}$ is the true parameter value and%
\begin{align*}
\mathbf{\Omega}_{\lambda }\left( \boldsymbol{\beta }\right) & =\sum_{h=1}^{H}\eta
_{h}\mathbf{\Omega}_{\lambda }^{(h)}\left( \boldsymbol{\beta }\right)
=\sum_{h=1}^{H}\eta _{h}\lim_{n_{h}\rightarrow \infty }\frac{1}{n_{h}}%
\sum_{i=1}^{n_{h}}\mathbf{\Omega}_{i,\lambda }^{(h)}\left( \boldsymbol{\beta },%
\boldsymbol{\Sigma }_{hi}\right) , \\
\mathbf{\Psi}_{\lambda }\left( \boldsymbol{\beta }\right) & =\sum_{h=1}^{H}\eta
_{h}\mathbf{\Psi}_{\lambda }^{(h)}\left( \boldsymbol{\beta }\right)
=\sum_{h=1}^{H}\eta _{h}\lim_{n_{h}\rightarrow \infty }\frac{1}{n_{h}}%
\sum_{i=1}^{n_{h}}\mathbf{\Psi}_{i,\lambda }^{(h)}\left( \boldsymbol{\beta }%
\right) ,
\end{align*}%
where $\mathbf{\Omega}_{i,\lambda }^{(h)}\left( \boldsymbol{\beta },\boldsymbol{%
\Sigma }_{hi}\right) $ and $\mathbf{\Psi}_{i,\lambda }^{(h)}\left( \boldsymbol{%
\beta }\right) $\ have the same expressions (not meaning) as $\mathbf{\Omega}%
_{hi,\lambda }\left( \boldsymbol{\beta },\boldsymbol{\Sigma }_{hi}\right) $\
and $\mathbf{\Psi}_{hi,\lambda }\left( \boldsymbol{\beta }\right) $\
respectively, given in (\ref{hhi}), (\ref{ghi}).
\end{theorem}

The idea of the previous theorem matches the philosophy of the asymptotic
result developed in Castilla et al. (2018), where other family of
divergences was considered and the derivations of the results were quite
different. It is also appropriate to consider the following results in the
place of Corollary \ref{cor1}, b and c.

\begin{corollary}
\label{cor1b}The following ones are (weak) consistent estimators as $n_{h}$
goes to infinity, for every $h=1,...,H$:

\begin{description}
\item[a)] $\mathbf{\Psi}_{\lambda }^{(h)}(\widehat{\boldsymbol{\beta }}%
_{\lambda ,Q})=\frac{1}{n_{h}}\sum_{i=1}^{n_{h}}\mathbf{\Psi}_{i,\lambda
}^{(h)}(\widehat{\boldsymbol{\beta }}_{\lambda ,Q})$ is a consistent
estimator of $\mathbf{\Psi}_{\lambda }^{(h)}\left( \boldsymbol{\beta }%
_{0}\right) $, for every $h=1,...,H$.

\item[b)] $\mathbf{\Omega}_{\lambda }^{(h)}(\widehat{\boldsymbol{\beta }}%
_{\lambda ,Q},\{\widehat{\boldsymbol{\Sigma }}_{hi}%
\}_{h=1,...,H:i=1,...,n_{h}})=\frac{1}{n_{h}}\sum_{i=1}^{n_{h}}\mathbf{\Omega}%
_{i,\lambda }^{(h)}(\widehat{\boldsymbol{\beta }}_{\lambda ,Q},\widehat{%
\boldsymbol{\Sigma }}_{hi})$ is a consistent estimator of $\mathbf{\Omega}%
_{\lambda }^{(h)}\left( \boldsymbol{\beta }_{0}\right) $, for every $%
h=1,...,H$, whenever $\Sigma _{hi}=\mathrm{Var}[\widehat{\boldsymbol{Y}}%
_{hi}]$ is consistently estimated through $\widehat{\boldsymbol{\Sigma }}%
_{hi}$ for all $(h,i)\in \{1,...,H\}\times \{1,...,m_{hi}\}$.
\end{description}
\end{corollary}

The following result is useful for any sample of polytomous logistic
regression with complex sample design, more general in comparison with
Corollary \ref{cor1b}, since it is not neccessary to get any consistent
estimators for $\boldsymbol{\Sigma }_{hi}$.

\begin{theorem}
\label{Th2b}The estimator%
\begin{align*}
\widehat{\mathbf{\Omega}}_{\lambda }^{(h)}(\widehat{\boldsymbol{\beta }}_{\lambda
,Q})& =\frac{1}{n_{h}}\sum_{i=1}^{n_{h}}\widehat{\mathbf{\Omega}}_{i,\lambda
}^{(h)}(\widehat{\boldsymbol{\beta }}_{\lambda ,Q}), \\
\widehat{\mathbf{\Omega}}_{i,\lambda }^{(h)}(\widehat{\boldsymbol{\beta }}%
_{\lambda ,Q})& =\boldsymbol{U}_{\lambda }(\widehat{\boldsymbol{\beta }}%
_{\lambda ,Q},\boldsymbol{x}_{hi})\boldsymbol{U}_{\lambda }^{T}(\widehat{%
\boldsymbol{\beta }}_{\lambda ,Q},\boldsymbol{x}_{hi}) \\
& =\left[ w_{hi}^{2}\boldsymbol{\Delta }^{\ast }(\boldsymbol{\pi }_{hi}(%
\widehat{\boldsymbol{\beta }}_{\lambda ,Q}))\mathrm{diag}^{\lambda -1}\{%
\boldsymbol{\pi }_{hi}(\widehat{\boldsymbol{\beta }}_{\lambda ,Q})\}\{%
\widehat{\boldsymbol{Y}}_{hi}-m_{hi}\boldsymbol{\pi }_{hi}(\widehat{%
\boldsymbol{\beta }}_{\lambda ,Q})\}\right. \\
& \left. \{\widehat{\boldsymbol{Y}}_{hi}-m_{hi}\boldsymbol{\pi }_{hi}(%
\widehat{\boldsymbol{\beta }}_{\lambda ,Q})\}^{T}\mathrm{diag}^{\lambda -1}\{%
\boldsymbol{\pi }_{hi}(\widehat{\boldsymbol{\beta }}_{\lambda ,Q})\}%
\boldsymbol{\Delta }^{\ast T}(\boldsymbol{\pi }_{hi}(\widehat{\boldsymbol{%
\beta }}_{\lambda ,Q}))\right] \otimes \boldsymbol{x}_{hi}\boldsymbol{x}%
_{hi}^{T},
\end{align*}%
is consistent of $\mathbf{\Omega}_{\lambda }^{(h)}\left( \boldsymbol{\beta }%
\right) $, as $n_{h}$ goes to infinity, for every $h=1,...,H$.
\end{theorem}

\begin{corollary}
\label{cor3}Let $\widehat{\boldsymbol{Y}}_{hi}$ be a random variable with
overdisepersed multinomial sampling scheme with an overdispersion parameter $%
\nu _{h}$, specific for each stratum,\ and $m_{hi}=\overline{m}_{h}$,%
\begin{align*}
\boldsymbol{\Sigma }_{hi}\left( \boldsymbol{\beta }\right) & =\nu _{h}%
\overline{m}_{h}\boldsymbol{\Delta }(\boldsymbol{\pi }_{hi}\left( 
\boldsymbol{\beta }\right) ), \\
\nu _{h}& =1+\rho _{h}^{2}(\overline{m}_{h}-1),
\end{align*}%
then, for $\nu _{h}$ and $\rho _{h}^{2}$:

\begin{description}
\item[a)] \textquotedblleft robust and consistent estimators based on the
estimating equation\textquotedblright\ are given respectively by%
\begin{align*}
\widetilde{\nu }_{h,\lambda }^{E}(\widehat{\boldsymbol{\beta }}_{\lambda
,Q})& =\frac{1}{d(k+1)}\mathrm{trace}\left( \mathbf{\Omega}_{\lambda }^{(h),-1}(%
\widehat{\boldsymbol{\beta }}_{\lambda ,Q})\widehat{\mathbf{\Omega}}_{\lambda
}^{(h)}(\widehat{\boldsymbol{\beta }}_{\lambda ,Q})\right) , \\
\widetilde{\rho }_{h,\lambda }^{2,E}& =\frac{\widetilde{\nu }_{h,\lambda
}^{E}(\widehat{\boldsymbol{\beta }}_{\lambda ,Q})-1}{\overline{m}_{h}-1},
\end{align*}%
where the matrices of interest are the one associated with multinomial
sampling for the $h$-th stratum, 
\begin{align*}
\mathbf{\Omega}_{\lambda }^{(h)}(\widehat{\boldsymbol{\beta }}_{\lambda ,Q})& =%
\frac{1}{n_{h}}\sum_{i=1}^{n_{h}}\mathbf{\Omega}_{i,\lambda }^{(h)}(\widehat{%
\boldsymbol{\beta }}_{\lambda ,Q}), \\
\mathbf{\Omega}_{i,\lambda }^{(h)}\left( \boldsymbol{\beta },\boldsymbol{\Sigma }%
_{hi}\right) & =\overline{m}_{h}w_{hi}^{2}\boldsymbol{\Delta }^{\ast }(%
\boldsymbol{\pi }_{hi}(\widehat{\boldsymbol{\beta }}_{\lambda ,Q}))\mathrm{%
diag}^{\lambda -1}\{\boldsymbol{\pi }_{hi}(\widehat{\boldsymbol{\beta }}%
_{\lambda ,Q})\}\boldsymbol{\Delta }(\boldsymbol{\pi }_{hi}(\widehat{%
\boldsymbol{\beta }}_{\lambda ,Q})) \\
& \times \mathrm{diag}^{\lambda -1}\{\boldsymbol{\pi }_{hi}(\widehat{%
\boldsymbol{\beta }}_{\lambda ,Q})\}\boldsymbol{\Delta }^{\ast T}(%
\boldsymbol{\pi }_{hi}\widehat{\boldsymbol{\beta }}_{\lambda ,Q}))\otimes 
\boldsymbol{x}_{hi}\boldsymbol{x}_{hi}^{T}
\end{align*}%
and Theorem \ref{Th2b} for overdispersed multinomial sampling.

\item[b)] \textquotedblleft robust and consistent estimators based on the
method of moments\textquotedblright\ are given by%
\begin{align*}
\widetilde{\nu }_{h,\lambda }^{M}(\widehat{\boldsymbol{\beta }}_{\lambda
,Q})& =\frac{1}{n_{h}d}\sum_{i=1}^{n_{h}}\sum_{j=1}^{d+1}\frac{(\widehat{Y}%
_{hij}-\overline{m}_{h}\pi _{hij}(\widehat{\boldsymbol{\beta }}_{\lambda
,Q}))^{2}}{\overline{m}_{h}\pi _{hij}(\widehat{\boldsymbol{\beta }}_{\lambda
,Q})}, \\
\widetilde{\rho }_{h,\lambda }^{2,M}& =\frac{\widetilde{\nu }_{h,\lambda
}^{M}(\widehat{\boldsymbol{\beta }}_{\lambda ,Q})-1}{\overline{m}_{h}-1}.
\end{align*}
\end{description}
\end{corollary}

\section{Testing linear hypotheses for the PLR coefficients with complex survey design \label{secWald}}

Based on the asymptotic distribution of the minimum quasi weighted DPD estimator, $\widehat{%
\boldsymbol{\beta }}_{\lambda ,Q},$ presented in Theorem \ref{th2}, we now introduce and  study  a family of Wald-type test
statistics for testing
\begin{equation}
H_{0}:\boldsymbol{M}^{T}\boldsymbol{\beta =l}\text{ against }H_{0}:%
\boldsymbol{M}^{T}\boldsymbol{\beta \neq l}\text{ }  \label{A}
\end{equation}%
where $\boldsymbol{M}$ is a $d(k+1)\times r$ full row-rank matrix with $%
r\leq d(k+1)$ and the right-hand-side $r-$vector consists of constants that
in many situations are $\boldsymbol{l=0}_{r}.$ For solving the problem of
testing given in (\ref{A}) we define a family of Wald-type
test statistics based on quasi MDPDE as follows.

\begin{definition}
Let $\widehat{\boldsymbol{\beta }}_{\lambda ,Q}$ the minimum quasi weighted DPD estimator
of $\boldsymbol{\beta }$ in the PLR model (\ref{2.1.0}) under a complex
survey and we denote 
\begin{align*}
\widehat{\mathbf{Q}}_{n,\lambda }(\widehat{\boldsymbol{\beta }}_{\lambda
,Q})& =\frac{1}{n}\mathbf{\Psi}_{n,\lambda }^{-1}(\widehat{\boldsymbol{\beta }}%
_{\lambda ,Q})\widehat{\mathbf{\Omega}}_{n,\lambda }(\widehat{\boldsymbol{\beta }}%
_{\lambda ,Q})\mathbf{\Psi}_{n,\lambda }^{-1}(\widehat{\boldsymbol{\beta }}%
_{\lambda ,Q}) \\
& =\left( n\mathbf{\Psi}_{n,\lambda }(\widehat{\boldsymbol{\beta }}_{\lambda
,Q})\right) ^{-1}n\widehat{\mathbf{\Omega}}_{n,\lambda }(\widehat{\boldsymbol{%
\beta }}_{\lambda ,Q})\left( n\mathbf{\Psi}_{n,\lambda }(\widehat{\boldsymbol{%
\beta }}_{\lambda ,Q})\right) ^{-1},
\end{align*}%
where%
\begin{align*}
n\widehat{\mathbf{\Omega}}_{n,\lambda }(\widehat{\boldsymbol{\beta }}_{\lambda
,Q})& =\sum_{h=1}^{H}\sum_{i=1}^{n_{h}}\left[ w_{hi}^{2}\boldsymbol{\Delta }%
^{\ast }(\boldsymbol{\pi }_{hi}(\widehat{\boldsymbol{\beta }}_{\lambda ,Q}))%
\mathrm{diag}^{\lambda -1}\{\boldsymbol{\pi }_{hi}(\widehat{\boldsymbol{%
\beta }}_{\lambda ,Q})\}\{\widehat{\boldsymbol{Y}}_{hi}-m_{hi}\boldsymbol{%
\pi }_{hi}(\widehat{\boldsymbol{\beta }}_{\lambda ,Q})\}\right.  \\
& \left. \times \{\widehat{\boldsymbol{Y}}_{hi}-m_{hi}\boldsymbol{\pi }_{hi}(%
\widehat{\boldsymbol{\beta }}_{\lambda ,Q})\}^{T}\mathrm{diag}^{\lambda -1}\{%
\boldsymbol{\pi }_{hi}(\widehat{\boldsymbol{\beta }}_{\lambda ,Q})\}%
\boldsymbol{\Delta }^{\ast T}(\boldsymbol{\pi }_{hi}(\widehat{\boldsymbol{%
\beta }}_{\lambda ,Q}))\right. \otimes \boldsymbol{x}_{hi}\boldsymbol{x}%
_{hi}^{T}, \\
n\mathbf{\Psi}_{n,\lambda }(\widehat{\boldsymbol{\beta }}_{\lambda ,Q})&
=\left\{ 
\begin{array}{lc}
\sum_{h=1}^{H}\sum_{i=1}^{n_{h}}w_{hi}m_{hi}\boldsymbol{\Delta }^{\ast }(%
\boldsymbol{\pi }_{hi}\left( \boldsymbol{\beta }\right) )\mathrm{diag}%
^{\lambda -1}\{\boldsymbol{\pi }_{hi}\left( \boldsymbol{\beta }\right) \}%
\boldsymbol{\Delta }^{\ast T}(\boldsymbol{\pi }_{hi}\left( \boldsymbol{\beta 
}\right) )\otimes \boldsymbol{x}_{hi}\boldsymbol{x}_{hi}^{T}, & \lambda >0
\\ 
\sum_{h=1}^{H}\sum_{i=1}^{n_{h}}w_{hi}m_{hi}\boldsymbol{\Delta }(\boldsymbol{%
\pi }_{hi}^{\ast }\left( \boldsymbol{\beta }\right) )\otimes \boldsymbol{x}%
_{hi}\boldsymbol{x}_{hi}^{T}, & \lambda =0%
\end{array}%
\right. .
\end{align*}%
The family of Wald-type test statistics for testing the null hypothesis
given in (\ref{A}) is given by 
\begin{equation}
W_{n}(\widehat{\boldsymbol{\beta }}_{\lambda ,Q})=(\boldsymbol{M}^{T}%
\widehat{\boldsymbol{\beta }}_{\lambda ,Q}-\boldsymbol{l})^{T}\left( 
\boldsymbol{M}^{T}\widehat{\mathbf{Q}}_{n,\lambda }(\widehat{\boldsymbol{%
\beta }}_{\lambda ,Q})\boldsymbol{M}\right) ^{-1}(\boldsymbol{M}^{T}\widehat{%
\boldsymbol{\beta }}_{\lambda ,Q}-\boldsymbol{l})  \label{B}
\end{equation}
\end{definition} 

For $\lambda =0,$ $\widehat{\boldsymbol{\beta }}_{\lambda =0,Q},$
is the  maximum quasi weighted likelihood estimator of $\boldsymbol{\beta }$,
with estimating equations given in (\ref{2.4}). It is not difficult to see
that $\mathbf{Q}_{\lambda =0}(\widehat{\boldsymbol{\beta }}_{\lambda =0,Q})$
is the Fisher information matrix and $W_{n}(\widehat{\boldsymbol{\beta }}%
_{\lambda =0,Q})$ will be the classical Wald test statistic.

\begin{theorem}
Under the null hypothesis given in (\ref{A})  the asymptotic distribution
of the Wald-type test statistics $W_{n}(\widehat{\boldsymbol{\beta }}%
_{\lambda ,Q})$, defined in (\ref{B}), is  chi-square  with $r$
degrees of freedom.
\end{theorem}

The proof is immediate using the asymptotic distribution of the minimum quasi weighted DPD estimator, presented in Theorem \ref{th2}, and taking into account the consistence of the matrix $\mathbf{Q}_{\lambda }\left( \widehat{\boldsymbol{\beta }}_{\lambda ,Q}\right) $ presented in Corollary \ref{cor1b}. Based on the previous theorem the null hypothesis given in (\ref{A}) will be rejected if 
\begin{equation}
W_{n}\left( \widehat{\boldsymbol{\beta }}_{\lambda ,Q}\right) >\chi
_{r,\alpha }^{2}.  \label{C}
\end{equation}

Next we present a result in order to give an approximation of
the power function for the test statisitcs given in (\ref{C}). Let $\boldsymbol{\beta }^{\ast }$ $\in \Theta $ be such that $%
\boldsymbol{M}^{T}\boldsymbol{\beta }^{\ast }\neq \boldsymbol{c,}$ i.e., $%
\boldsymbol{\beta }^{\ast }$ does not belong to the null parameter space. Let us denote
\begin{equation*}
l_{\boldsymbol{\beta }_{1}}(\boldsymbol{\beta }_{2})=\left( \boldsymbol{M}%
^{T}\boldsymbol{\beta }_{1}-\boldsymbol{l}\right) ^{T}\left( \boldsymbol{M}%
^{T}\widehat{\mathbf{Q}}_{n,\lambda }\left( \boldsymbol{\beta }_{2}\right) 
\boldsymbol{M}\right) ^{-1}\left( \boldsymbol{M}^{T}\boldsymbol{\beta }_{1}-%
\boldsymbol{l}\right) .
\end{equation*}%
Then we have the following result.

\begin{theorem}\label{th:c}
Let $\boldsymbol{\beta }^{\ast }\in \Theta $ with $\boldsymbol{M}^{T}%
\boldsymbol{\beta }^{\ast }\neq \boldsymbol{l}$ the true value of the
parameter so that $\widehat{\boldsymbol{\beta }}_{\lambda ,Q}\underset{%
n\rightarrow \infty }{\overset{P}{\longrightarrow }}\boldsymbol{\beta }%
^{\ast }$. The power function of the test statistic given in (\ref{C}), at $%
\boldsymbol{\beta }^{\ast }$, is given by 
\begin{equation}
Po_{W_{n}\left( \widehat{\boldsymbol{\beta }}_{\lambda ,Q}\right) }\left( 
\boldsymbol{\beta }^{\ast }\right) =1-\phi _{n}\left( \frac{1}{\sigma \left( 
\boldsymbol{\beta }^{\ast }\right) }\left( \frac{\chi _{r,\alpha }^{2}}{%
\sqrt{n}}-\sqrt{n}l_{\boldsymbol{\beta }^{\ast }}(\boldsymbol{\beta }^{\ast
})\right) \right)  \label{D}
\end{equation}%
where $\phi _{n}\left( x\right) $ tends uniformly to the standard normal
distribution $\phi \left( x\right) $ and $\sigma \left( \boldsymbol{\beta }%
^{\ast }\right) $ is given by 
\begin{equation*}
\sigma \left( \boldsymbol{\beta }^{\ast }\right) ^{2}=\left. \frac{\partial
l_{\boldsymbol{\beta }}(\boldsymbol{\beta }^{\ast })}{\partial \boldsymbol{%
\beta }^{T}}\right\vert _{\boldsymbol{\beta }=\boldsymbol{\beta }^{\ast }}%
\widehat{\mathbf{Q}}_{n,\lambda }\left( \boldsymbol{\beta }^{\ast }\right)
\left. \frac{\partial l_{\boldsymbol{\beta }}(\boldsymbol{\beta }^{\ast })}{%
\partial \boldsymbol{\beta }}\right\vert _{\boldsymbol{\beta }=\boldsymbol{%
\beta }^{\ast }}.
\end{equation*}
\end{theorem}

It is clear that 
\begin{equation*}
\lim_{n\rightarrow \infty }Po_{W_{n}\left( \widehat{\boldsymbol{\beta }}%
_{\lambda ,Q}\right) }\left( \boldsymbol{\beta }^{\ast }\right) =1
\end{equation*}%
for all $\alpha \in \left( 0,1\right) $. Therefore, the Wald-type test
statistics are consistent in the sense of Fraser.

\begin{remark}
Based on the previous theorem we can obtain the sample size necessary to get
a fix power $Po_{W_{n}\left( \widehat{\boldsymbol{\beta }}_{\lambda
,Q}\right) }\left( \boldsymbol{\beta }^{\ast }\right) =\pi _{0}.$ By (\ref{D}%
) we must solve the equation 
\begin{equation*}
1-\pi _{0}=\phi \left( \frac{1}{\sigma \left( \boldsymbol{\beta }^{\ast
}\right) }\left( \frac{\chi _{r,\alpha }^{2}}{\sqrt{n}}-\sqrt{n}l_{%
\boldsymbol{\beta }^{\ast }}(\boldsymbol{\beta }^{\ast })\right) \right)
\end{equation*}%
and we get that $n=\left[ n^{\ast }\right] +1$ with 
\begin{equation*}
n^{\ast }=\frac{A+B+\sqrt{A(A+2B)}}{2l_{\boldsymbol{\beta }^{\ast }}(%
\boldsymbol{\beta }^{\ast })^{2}}
\end{equation*}
being 
$A=\sigma ^{2}\left( \boldsymbol{\beta }^{\ast }\right) \left( \phi
^{-1}\left( 1-\pi _{0}\right) \right) ^{2}\text{ and }B=\frac{1}{2}\chi
_{r,\alpha }^{2}l_{\boldsymbol{\beta }^{\ast }}(\boldsymbol{\beta }^{\ast }).$
\end{remark}

We may also find an approximation of the power of the Wald-type tests given in (\ref{C}) at an alternative close to the null hypothesis. Let $\boldsymbol{\beta }_{n}$ such that $\boldsymbol{M}^{T}\boldsymbol{\mathbf{\beta }_{n}\neq l}$ be a given alternative and let $\boldsymbol{\mathbf{\beta }}_{0}$ be the element such that $\boldsymbol{M}^{T}\boldsymbol{\mathbf{\beta }_{0}}=\boldsymbol{l}$ closest to $\boldsymbol{\beta }_{n}$ in the Euclidean distance sense. A first possibility to introduce contiguous alternative hypotheses is to consider a fixed $\boldsymbol{d}\in \mathbb{R}^{d(k+1)}$ and to permit $\boldsymbol{\beta }_{n}$ moving towards $\boldsymbol{\beta }_{0}$ as $n$ increases in the following way 
\begin{equation}
H_{1,n}:\boldsymbol{\beta}=\boldsymbol{\beta }_{n}, \quad \text{where} \quad \boldsymbol{\beta }_{n}=\boldsymbol{\beta }_{0}+n^{-1/2}\boldsymbol{d}.  \label{E}
\end{equation}%
A second approach is to relax the condition $\boldsymbol{M}^{T}\boldsymbol{\mathbf{\beta }_{0}}=\boldsymbol{l}$ defining the null hypothesis$.$ Let $\boldsymbol{\delta }\in \mathbb{R}^{r}$ and consider the following sequence, 
$\boldsymbol{\beta }_{n},$ of parameters moving towards $\boldsymbol{\beta }_{0}$ according to 
\begin{equation}
H_{1,n}^{\ast }:\boldsymbol{\beta}=\boldsymbol{\beta }_{n}, \quad \text{where} \quad \boldsymbol{M}^{T}\boldsymbol{\mathbf{\beta }_{n}}-\boldsymbol{l}=n^{-1/2}\boldsymbol{\delta }.  \label{F}
\end{equation}%
Note that 
\begin{equation}
\boldsymbol{M}^{T}\boldsymbol{\mathbf{\beta }_{n}}-l\boldsymbol{=M^{T}\boldsymbol{\beta }_{0}+M^{T}}n^{-1/2}\boldsymbol{\boldsymbol{d}}-\boldsymbol{l}=M\boldsymbol{^{T}}n^{-1/2}\boldsymbol{\boldsymbol{d}}.\label{G}
\end{equation}%
Then the equivalence between the two hypotheses is given by $M\boldsymbol{^{T}\boldsymbol{d}}=\boldsymbol{\delta }.$

If we denote by $\chi_{r}^{2}(\Delta)$ the non central chi-square distribution with $r$ degrees of freedom and noncentrality parameter $\Delta, $ we can state the following theorem.

\begin{theorem}\label{th:wald}
We have,

\begin{enumerate}
\item[i)] $W_{n}\left( \widehat{\boldsymbol{\beta }}_{\lambda ,Q}\right) \underset{n\longrightarrow \infty }{\overset{\mathcal{L}}{\longrightarrow }}\chi _{r}^{2}\left( \Delta _{1}\right) $under $H_{1,n}$ given in (\ref{E}),
with 
\begin{equation*}
\Delta _{1}=\boldsymbol{d}^{T}\boldsymbol{M}\left[ M\boldsymbol{^{T}\widehat{%
\mathbf{Q}}_{n,\lambda }\left( \boldsymbol{\beta }_{0}\right) M}\right]
^{-1}M\boldsymbol{^{T}d.}
\end{equation*}

\item[ii)] $W_{n}\left( \widehat{\boldsymbol{\beta }}_{\lambda ,Q}\right)\underset{n\longrightarrow \infty }{\overset{\mathcal{L}}{\longrightarrow }}\chi _{r}^{2}\left( \Delta _{2}\right) $ under $H_{1,n}^{\ast }$ given in (\ref{F}), with 
\begin{equation*}
\Delta _{2}=\boldsymbol{\delta }^{T}\left( M\boldsymbol{^{T}\widehat{\mathbf{Q}}_{n,\lambda }\left( \boldsymbol{\beta }_{0}\right) M}\right) ^{-1}\boldsymbol{\delta .}
\end{equation*}
\end{enumerate}
\end{theorem}

\section{Influence Function\label{sec4}}

Influence function is a classical tool to measure robustness of an estimator (Hampel et al., 1968). However, the present set-up of complex survey is not as simple as the iid set-up; in fact the observations within a cluster of a stratum are iid but  the observations in different cluster and stratum are independent non-homogeneous.  So we need to modify the definition of the influence function accordingly.  Recently, Ghosh and Basu (2013, 2016, 2018) have discussed the extended definition of the influence function for  the independent but non-homogeneous observations;  we will extend their approach to define the influence function in the present case of PLR model under complex design.

\subsection{Influence Function of the minimum quasi weighted DPD estimator}

We first need to define the statistical functional corresponding to the minimum quasi weighted DPD estimator
as the minimizer of the DPD between the true and model densities. 
Assume the set-up and notations of Section 3 with the DPD kernel being given by 
$d_{\lambda }^{\ast }\left( g(\boldsymbol{y}_{hij}|\boldsymbol{x}_{hi}),f_{\boldsymbol{\beta }}(\boldsymbol{y}_{hij}|\boldsymbol{x}_{hi})\right)$
for individual densities  $g(\boldsymbol{y}_{hij}|\boldsymbol{x}_{hi})$ and $f_{\boldsymbol{\beta }}(\boldsymbol{y}_{hij})$.
Then, following Ghosh and Basu (2013), the minimum quasi weighted DPD estimator functional is to be defined by the minimizer of  the total weighted DPD measure given by 
\begin{eqnarray}
d_\lambda(g, f_{\boldsymbol{\beta}}, w) &=& \sum_{h=1}^H\sum_{i=1}^{n_h} 
d_{\lambda }^{\ast }\left( g(\boldsymbol{y}_{hij}|\boldsymbol{x}_{hi}),f_{\boldsymbol{\beta }}(\boldsymbol{y}_{hij}|\boldsymbol{x}_{hi})\right)
\nonumber\\
&=& \sum_{h=1}^{H}\sum_{i=1}^{n_{h}}w_{hi} \left( 
m_{hi}\boldsymbol{\pi }_{hi}^{\lambda ,T}\left(\boldsymbol{\beta }\right)\boldsymbol{\pi }_{hi}\left(\boldsymbol{\beta }\right) 
-\frac{\lambda +1}{\lambda }\int \boldsymbol{\pi }_{hi}^{\lambda ,T}\left(\boldsymbol{\beta }\right){\boldsymbol{y}}_{hi}
dG({\boldsymbol{y}}_{hi}|{\boldsymbol{x}}_{hi})\right),
\quad \text{for }\lambda >0.  \nonumber\\
\label{def_func}
\end{eqnarray}

\begin{definition}
We consider the PLR model with complex survey defined in (\ref{2.1.0}). 
The minimum quasi weighted DPD estimator functional, ${\boldsymbol{T}}_{\lambda,Q}(\boldsymbol{G})$, of $\boldsymbol{\beta}$ 
at $\boldsymbol{G}=(G(\boldsymbol{y}_{hij}|\boldsymbol{x}_{hi}), ~ h=1, \ldots, H, i=1, \ldots, n_h)$ is defined as%
\begin{equation*}
{\boldsymbol{T}}_{\lambda,Q}(\boldsymbol{G}) = \arg \min_{\boldsymbol{\beta}\in \Theta }d_\lambda(g, f_{\boldsymbol{\beta}}, w),
\end{equation*}
where $d_\lambda(g, f_{\boldsymbol{\beta}}, w)$ is as defined above  in (\ref{def_func}).
\end{definition}

Note that, by the property of the DPD measure, it is immediate that the minimum quasi weighted DPD estimator functional 
${\boldsymbol{T}}_{\lambda, Q}(\boldsymbol{G})$ is Fisher consistent at the assumed PLR model (\ref{2.1.0}),
i.e., ${\boldsymbol{T}}_{\lambda, Q}(\boldsymbol{\pi}(\boldsymbol{\beta})) =\boldsymbol{\beta}$ 
for all parameter values $\boldsymbol{\beta}$. 
Also, following the calculations for Theorem \ref{th:u}, one can see that the minimum quasi weighted DPD estimator functional 
${\boldsymbol{T}}_{\lambda, Q}(\boldsymbol{G})$ can also be derived as a solution to the estimating equations
%$\boldsymbol{u}_{\lambda}^\ast\left(\boldsymbol{g}, \boldsymbol{\beta}\right) =\boldsymbol{0}_{d(k+1)}$, where
\begin{align}
\boldsymbol{u}_{\lambda}^*\left(\boldsymbol{g}, \boldsymbol{\beta}\right) 
& =\sum\limits_{h=1}^{H}\sum \limits_{i=1}^{n_{h}}
\boldsymbol{u}_{\lambda,hi}^*\left({g}_{hi}, \boldsymbol{\beta}\right)=\boldsymbol{0}_{d(k+1)},  \label{4.09} 
\end{align}
where ${g}_{hi}=g(\boldsymbol{y}_{hij}|\boldsymbol{x}_{hi})$ and 
\begin{align}
\boldsymbol{u}_{\lambda,hi}^*\left(g_{hi}, \boldsymbol{\beta}\right) & 
=\left[w_{hi}m_{hi}\boldsymbol{\Delta }^{\ast }(\boldsymbol{\pi }_{hi}\left( \boldsymbol{\beta }\right) )
\mathrm{diag}^{\lambda -1}\{\boldsymbol{\pi }_{hi}\left( \boldsymbol{\beta }\right) \}
\{g_{hi}-\boldsymbol{\pi }_{hi}\left( \boldsymbol{\beta }\right) \}\right] \otimes 
\boldsymbol{x}_{hi}.  \label{4.010}
\end{align}
Note that, these estimating equations are unbiased at the model probability $\boldsymbol{g}=\boldsymbol{\pi}(\boldsymbol{\beta})$.

Now, in order to define the influence function of the minimum DPD functional ${\boldsymbol{T}}_{\lambda, Q}(\boldsymbol{g})$,
we note that the functional itself depends on the sample sizes and cluster weights
and so it IF will have the same dependence in analogue to the non-homogeneous case of Ghosh and Basu (2013, 2016, 2018).
Also note that, the contamination can be in any one particular cluster within one stratum
or simultaneously in many of them (or all). For simplicity, let us first assume that the  
contamination is only in one cluster probability ${g}_{h_0i_0}$ for some fix $h_0$ and $i_0$.
Consider the contaminated probability vector 
${g}_{h_0i_0, \epsilon} = (1-\epsilon) {g}_{h_0i_0} + \epsilon \delta_{\boldsymbol{t}}$,
where $\epsilon$ is the contamination proportion and $ \delta_{\boldsymbol{t}}$ is the degenerate probability at
the outlier point $\boldsymbol{t} =(t_1, \ldots, t_{d+1})^T \in \{0, 1\}^{d+1}$ with $\sum_{s=1}^{d+1}t_s=1$. 
Denote the corresponding contaminated full probability vector as $\boldsymbol{g}_{\epsilon}$
which is same as $\boldsymbol{g}$ except ${g}_{h_0i_0}$ being replaced by ${g}_{h_0i_0,\epsilon}$
and let the corresponding contaminated distribution vector be $\boldsymbol{G}_\epsilon$.
Then, the corresponding influence function is defined as
$$
\mathcal{IF}(\boldsymbol{t}, {\boldsymbol{T}}_{\lambda, Q}, \boldsymbol{g})
= \lim_{\epsilon\rightarrow 0}
\frac{{\boldsymbol{T}}_{\lambda, Q}(\boldsymbol{G}_{\epsilon}) -{\boldsymbol{T}}_{\lambda, Q}(\boldsymbol{G})}{\epsilon}
= \frac{\partial}{\partial\epsilon}{\boldsymbol{T}}_{\lambda, Q}(\boldsymbol{G}_{\epsilon})\bigg|_{\epsilon=0}.
$$

In order to calculate this influence function, we start with the estimating equation for ${\boldsymbol{T}}_{\lambda, Q}$.
Note that, ${\boldsymbol{T}}_{\lambda, Q}(\boldsymbol{G}_{\epsilon})$ satisfies the equations
$$
\boldsymbol{u}_{\lambda}^*\left(\boldsymbol{g}_{\epsilon}, {\boldsymbol{T}}_{\lambda, Q}(\boldsymbol{G}_{\epsilon})\right) 
=\boldsymbol{0}_{d(k+1)}.
$$
Now, differentiating it with respect to $\epsilon$ art $\epsilon=0$ and simplifying, we get the
required influence function as given by 
\begin{eqnarray}
\mathcal{IF}(\boldsymbol{t}, {\boldsymbol{T}}_{\lambda, Q}, \boldsymbol{g})=
%\boldsymbol{\Psi}_{n,\lambda}^*(\boldsymbol{g}, \boldsymbol{\beta})^{-1}\frac{1}{n}
%\left[\left(\frac {w_{h_0i_0}m_{h_0i_0}}{\tau}\right)^{\lambda+1}\boldsymbol{\Delta}(\boldsymbol{\pi}_{h_0i_0}^{\ast}
%\left(\boldsymbol{\beta}\right))\boldsymbol{f}^{*}_{\lambda}(\delta_{\boldsymbol{t}}, \boldsymbol{\pi}_{h_0i_0}^{\ast}
%\left( \boldsymbol{\beta}\right)) \otimes \boldsymbol{x}_{h_0i_0} 
%- \boldsymbol{u}_{\lambda,h_0i_0}^*\left(\boldsymbol{g}_{h_0i_0}, \boldsymbol{\beta}\right)\right]
\boldsymbol{\Psi}_{n,\lambda}^*(\boldsymbol{g}, \boldsymbol{\beta})^{-1}\frac{1}{n}
\left[\boldsymbol{u}_{\lambda,h_0i_0}^*\left(\delta_{\boldsymbol{t}}, \boldsymbol{\beta}\right)
- \boldsymbol{u}_{\lambda,h_0i_0}^*\left({g}_{h_0i_0}, \boldsymbol{\beta}\right)\right],
\end{eqnarray}
where 
\begin{eqnarray}
\boldsymbol{\Psi}_{n,\lambda}^*(\boldsymbol{g}, \boldsymbol{\beta})= - \frac{1}{n} \sum\limits_{h=1}^{H}\sum \limits_{i=1}^{n_{h}}
\frac{\partial}{\partial\boldsymbol{\beta}}\boldsymbol{u}_{\lambda,hi}^*\left({g}_{hi}, \boldsymbol{\beta}\right).
\end{eqnarray}
Note that, at the model probability $\boldsymbol{g}=\boldsymbol{\pi}(\boldsymbol{\beta})$, 
we have $\boldsymbol{u}_{\lambda,h_0i_0}^*\left(\boldsymbol{\pi}_{h_0i_0}(\boldsymbol{\beta}), \boldsymbol{\beta}\right) 
= \boldsymbol{0}$ and 
$\boldsymbol{\Psi}_{n,\lambda}^*(\boldsymbol{\pi}(\boldsymbol{\beta}), \boldsymbol{\beta})=\boldsymbol{\Psi}_{n,\lambda}(\boldsymbol{\beta})$,
as defined in Theorem \ref{Th1}. Hence, the influence function of the proposed minimum quasi weighted DPD estimator
at the model probability simplifies to 
\begin{eqnarray}
\mathcal{IF}(\boldsymbol{t}, {\boldsymbol{T}}_{\lambda, Q}, \boldsymbol{\pi}(\boldsymbol{\beta}))
=\boldsymbol{\Psi}_{n,\lambda}(\boldsymbol{\beta})^{-1}\frac{1}{n}
\boldsymbol{u}_{\lambda,h_0i_0}^*\left(\delta_{\boldsymbol{t}}, \boldsymbol{\beta}\right).
\end{eqnarray}
Clearly, by the assumed form of the model probability $\boldsymbol{\pi}(\boldsymbol{\beta})$
this influence function is bounded for all $\lambda>0$ but unbounded at $\lambda=0$.
So, the proposed minimum quasi weighted DPD estimators are expected to be robust for any $\lambda>0$ but 
the corresponding maximum quasi weighted likelihood estimator (at $\lambda=0$) is non-robust against data contamination.

Similarly, one can show that, if there is contamination in some of the clusters within some stratum
indexed by ${h,i}\in \Gamma \subseteq \{h=1, \ldots, H; i=1, \ldots, n_h\}$ at the contamination points $\boldsymbol{t}_{hi}$, 
the corresponding influence function of the proposed minimum quasi weighted DPD estimator
at $\boldsymbol{g}$ and the model $\pi(\boldsymbol{\beta})$ are respectively given by 
\begin{eqnarray}
\mathcal{IF}((\boldsymbol{t}: \{h, i\}\in \Gamma), {\boldsymbol{T}}_{\lambda, Q}, \boldsymbol{g})=
\boldsymbol{\Psi}_{n,\lambda}^*(\boldsymbol{g}, \boldsymbol{\beta})^{-1}\frac{1}{n}
\sum_{\{h,i\}\in \Gamma}\left[\boldsymbol{u}_{\lambda,hi}^*\left(\delta_{\boldsymbol{t}_{hi}}, \boldsymbol{\beta}\right)
- \boldsymbol{u}_{\lambda,hi}^*\left({g}_{hi}, \boldsymbol{\beta}\right)\right],
\end{eqnarray}
and 
\begin{eqnarray}
\mathcal{IF}((\boldsymbol{t}: \{h, i\}\in \Gamma), {\boldsymbol{T}}_{\lambda, Q}, \boldsymbol{\pi}(\boldsymbol{\beta}))
=\boldsymbol{\Psi}_{n,\lambda}(\boldsymbol{\beta})^{-1}\frac{1}{n}
\sum_{\{h,i\}\in \Gamma}\boldsymbol{u}_{\lambda,hi}^*\left(\delta_{\boldsymbol{t}_{hi}}, \boldsymbol{\beta}\right).
\end{eqnarray}
The boundedness and robustness implications for these influence functions are exactly the same as before.

%However, due to the complicated form of these influence functions, 
%the implications about the robustness of the proposed MDPDE not directly visible.
%We will justify this robustness more clearly through empirical illustrations in the next section.

\subsection{Influence Function of the Wald-type Tests in PLRM}

Let us now study the robustness of proposed Wald-type test through the influence function of the corresponding test statistics in (\ref{B}).
Considering the notation and set-up of Section 2 and 3, and define 
$$\mathbf{Q}_{\lambda }(\boldsymbol{\beta}) = \frac{1}{n}\mathbf{\Psi}_{\lambda}^{-1}({\boldsymbol{\beta}})
	{\mathbf{\Omega}}_{\lambda }({\boldsymbol{\beta }})\mathbf{\Psi}_{\lambda }^{-1}({\boldsymbol{\beta }}).$$ 
Then the Wald-type test functional $W_\lambda(\boldsymbol{G})$ corresponding to (\ref{B}) for testing the null hypothesis given in (\ref{A}) 
at the true distribution vector $\boldsymbol{G}$ is defined as  
\begin{equation}
W_{\lambda}(\boldsymbol{G})=(\boldsymbol{M}^{T}{\boldsymbol{T}}_{\lambda, Q}(\boldsymbol{G})-\boldsymbol{l})^{T}
\left(\boldsymbol{M}^{T}{\mathbf{Q}}_{\lambda }({\boldsymbol{T}}_{\lambda, Q}(\boldsymbol{G}))\boldsymbol{M}\right) ^{-1}
(\boldsymbol{M}^{T}{\boldsymbol{T}}_{\lambda, Q}(\boldsymbol{G})-\boldsymbol{l}),  \label{B1}
\end{equation}
where ${\boldsymbol{T}}_{\lambda, Q}$ is the minimum quasi weighted DPD estimator functional as defined in the previous subsection.
Also, let $\boldsymbol{\beta }_{0}$ denote the true null parameter value for the hypothesis in  (\ref{A}).

Let us now derive the influence function of the test functional $W_\lambda$. 
As before, here also the contamination can be any particular cluster and strata (a given $h_0, i_0$) combination 
or in many (or all) of them. The influence function of general Wald-type tests under such non-homogeneous set-up 
has been extensively studied in Basu et al.~(2018).
Here, we follow the general theory of Basu et al.~(2018) to conclude that the first order influence functions of $W_\lambda$,
defined as the first order derivative of its value at the contaminated distribution with respect to $\epsilon$ at $\epsilon=0$,
in either case of contamination become identically zero at the null distribution $\boldsymbol{g}=\boldsymbol{\pi}(\boldsymbol{\beta}_0)$.
Therefore, the first order influence function is not informative in this case of Wald-type tests, 
and we need to investigate the second order influence function of $W_\lambda$. 

The second order influence function $IF^{(2)}$, which measures the second order approximation to the asymptotic bias due to infinitesimal contamination, 
is defined  as the second order derivative of the value of $W_\lambda$ at the contaminated distribution with respect to 
the contamination proportion $\epsilon$ at $\epsilon=0$. Again, following Basu et al.~(2018), 
we can derive these second order influence functions of the Wald-type tests in either case of contaminations; 
at the null distribution $\boldsymbol{g}=\boldsymbol{\pi}(\boldsymbol{\beta}_0)$, they simplifies to  
\begin{align*} 
&\mathcal{IF}^{(2)}(\boldsymbol{t}, W_{\lambda}, \boldsymbol{\pi}(\boldsymbol{\beta}_0))
 =2\mathcal{IF}(\boldsymbol{t}, {\boldsymbol{T}}_{\lambda, Q}, \boldsymbol{\pi}(\boldsymbol{\beta}_0))^{T}
\boldsymbol{M}\left(\boldsymbol{M}^{T}{\mathbf{Q}}_{\lambda }(\boldsymbol{\beta}_0)\boldsymbol{M}\right) ^{-1}\boldsymbol{M}^{T}
\mathcal{IF}(\boldsymbol{t}, {\boldsymbol{T}}_{\lambda, Q}, \boldsymbol{\pi}(\boldsymbol{\beta}_0)), 
\nonumber\\
&\mathcal{IF}^{(2)}((\boldsymbol{t}: \{h, i\}\in \Gamma), W_{\lambda}, \boldsymbol{\pi}(\boldsymbol{\beta}_0))\nonumber
\\
& =2\mathcal{IF}((\boldsymbol{t}: \{h, i\}\in \Gamma), {\boldsymbol{T}}_{\lambda, Q}, \boldsymbol{\pi}(\boldsymbol{\beta}_0))^{T}
\boldsymbol{M}\left(\boldsymbol{M}^{T}{\mathbf{Q}}_{\lambda }(\boldsymbol{\beta}_0)\boldsymbol{M}\right) ^{-1}\boldsymbol{M}^{T}
\mathcal{IF}((\boldsymbol{t}: \{h, i\}\in \Gamma), {\boldsymbol{T}}_{\lambda, Q}, \boldsymbol{\pi}(\boldsymbol{\beta}_0)), 
\nonumber
\end{align*}%
for the two types of contamination as before in the case of estimator. 
Note that, the second order influence functions of the proposed Wald-type tests are a quadratic function of 
the corresponding influence functions of the minimum quasi weighted DPD estimator for any type of contamination. 
Therefore, the boundedness of the influence functions of minimum quasi weighted DPD estimator at $\lambda>0$
also indicates the boundedness of the influence functions of the Wald-type test functional $W_\lambda$ 
indicating their robustness against contamination in any cluster or stratum of the sample data.

\section{Illustrative Example: BMI data set \label{secEx}}
Let us consider an illustrative real-life dataset on BMI which was previously studied in Castilla et al. (2018). This data set, obtained from CANSIM Canada's database and presented in Table \ref{table:CANSIM_data},  shows the body mass indexes of population in Canada in the year 1994.  Each person in the sample is divided by their body mass index category under the international standard: acceptable weight, overweight or obese. The data set consists on a stratified  sample design  with clusters nested on them, with the strata being three different age groups (20--34 years, 35--44 years and 45--64 years) and  the genders (male or female) as the clusters.  The qualitative explanatory variables are valid to distinguish the clusters within the strata. They are given by $\boldsymbol{x}_{h1}^T=(1,0)$, and $\boldsymbol{x}_{h2}^T=(0,1)$, $h=1,\dots,5$ for men and women, respectively.   

\renewcommand{\arraystretch}{1.05}
\begin{table}[t!ht]
\tabcolsep2.8pt \centering
\caption{Body mass index (BMI) data set}
$%
\begin{tabular}{llccc}
\hline
&  & \multicolumn{2}{c}{Body mass index (BMI)} &  \\ \cline{3-5}
Age group & Sex & Acceptable weight (18.5-24.9) &
Overweight (25.0-29.9) & Obese (30 or higher) \\ \hline
20-34 years & Men & \multicolumn{1}{c}{5438} & \multicolumn{1}{c}{4790} &
\multicolumn{1}{c}{1470} \\
& Women & \multicolumn{1}{c}{4910} & \multicolumn{1}{c}{2878} &
\multicolumn{1}{c}{802} \\
35-44 years & Men & \multicolumn{1}{c}{2458} & \multicolumn{1}{c}{3437} &
\multicolumn{1}{c}{1319} \\
& Women & \multicolumn{1}{c}{3100} & \multicolumn{1}{c}{1494} &
\multicolumn{1}{c}{1313} \\
45-64 years & Men$^*$ & \multicolumn{1}{c}{1968} & \multicolumn{1}{c}{3290} &
\multicolumn{1}{c}{1412} \\
& Women$^*$ & \multicolumn{1}{c}{1710} & \multicolumn{1}{c}{1481} &
\multicolumn{1}{c}{1078} \\ \hline
\end{tabular} $
\label{table:CANSIM_data}
\end{table}

To illustrate the robustness of  minimum quasi weighted DPD estimators, we contaminate the BMI data by permuting the categories  overweight and obese in the Men with age range 45--64 years. After obtaining the correpsonding minimum quasi weighted DPD estimators estimates, we compute the  mean absolute standardized deviations ($mssd$) between the estimated parameters and corresponding  estimated probabilities obtained for the modified and original data, as given by $$masd(\widehat{\boldsymbol{\beta}}_{\lambda}^*,\widehat{\boldsymbol{\beta}}_{\lambda})=\frac{1}{4}\sum_{r=1}^{2}\sum_{s=1}^{2}\left|\frac{\widehat{\beta}^*_{\lambda,rs}-\widehat{\beta}_{\lambda,rs}}{\widehat{\beta}_{\lambda,rs}} \right| \quad \text{ and } \quad masd(\widehat{\boldsymbol{\pi}}^*_{\lambda},\widehat{\boldsymbol{\pi}}_{\lambda})=\frac{1}{6}\sum_{r=1}^{3}\sum_{s=1}^{2}\left|\frac{\pi_{rs}(\widehat{\boldsymbol{\beta}}_{\lambda}^*)-\pi_{rs}(\widehat{\boldsymbol{\beta}}_{\lambda})}{\pi_{rs}(\widehat{\boldsymbol{\beta}}_{\lambda})} \right|,$$
where, with superscript $^*$ we denote the contaminated case. Assuming common overdispersion parameter, absolute standardized deviation ($asd$) of the two versions of the intra-cluster correlation estimator are also computed as in Corollary \ref{cor2}.  Their values, as presented in Table \ref{table:CANSIM_dataMSD}, clearly show that the minimum quasi weighted DPD estimators become more robust as $\lambda$ increases. 
Similar results are obtained when permuting the categories for Women instead of Men, as can also be seen in Table  \ref{table:CANSIM_dataMSD}.

\renewcommand{\arraystretch}{1.25}
\begin{table}[h!!]
\caption{Mean standardized deviations under contamination for BMI data \label{table:CANSIM_dataMSD}}
\centering
\begin{tabular}{l rrr rrr}
\hline
& \multicolumn{2}{r}{$\lambda$} \\
\cline{2-7}
& 0         & 0.2     & 0.4     & 0.6     & 0.8     & 1               \\
\hline

\multicolumn{7}{l}{contamination of  Men (45--64 years)}  \\ \hline
$masd$($\widehat{\boldsymbol{\beta}}^*_{\lambda}$,$\widehat{\boldsymbol{\beta}}_{\lambda}$) & 0.24396 & 0.23057 & 0.21731 & 0.20441 & 0.19187 & 0.17969 \\
$masd$($\widehat{\boldsymbol{\pi}}^*_{\lambda}$,$\widehat{\boldsymbol{\pi}}_{\lambda}$)& 0.10170 & 0.09700 & 0.09220  & 0.08750 & 0.08280 & 0.07810 \\
$asd$($\widehat{\rho}^{2*,E}_{\lambda},\widehat{\rho}^{2,E}_{\lambda}$)& 0.64785 & 0.58974 & 0.52955  & 0.46868 & 0.40858 & 0.35044 \\
$asd$($\widehat{\rho}^{2*,M}_{\lambda},\widehat{\rho}^{2,M}_{\lambda}$)& 1.57103 & 1.41605 & 1.25700  & 1.09724 & 0.94068 & 0.79037\\
 \hline
 \multicolumn{7}{l}{contamination of  Women (45--64 years)}  \\ \hline
$masd$($\widehat{\boldsymbol{\beta}}^{*}_{\lambda}$,$\widehat{\boldsymbol{\beta}}_{\lambda}$) & 0.10516 & 0.09484 & 0.08533 & 0.07665 & 0.0687 & 0.06148 \\
$masd$($\widehat{\boldsymbol{\pi}}^{*}_{\lambda}$,$\widehat{\boldsymbol{\pi}}_{\lambda}$)  & 0.03250 & 0.03030 & 0.02810 & 0.02600 & 0.0240 & 0.02210 \\
$asd$($\widehat{\rho}^{2*,E}_{\lambda},\widehat{\rho}^{2,E}_{\lambda}$)&  0.07038 & 0.05132 & 0.03358 & 0.01744 & 0.00321 & 0.00893 \\
$asd$($\widehat{\rho}^{2*,M}_{\lambda},\widehat{\rho}^{2,M}_{\lambda}$)&  0.03358 & 0.01169 & 0.05244 & 0.08819 & 0.11821 & 0.14206\\
\hline
\end{tabular}
\end{table}

%\clearpage

\section{Monte Carlo Simulation Study\label{sec5}}

\subsection{Simulation Scheme\label{sec:SimA}}

We develop a complex design extension of the simulation scheme previously
studied by Castilla et al. (2019), where a simple sample design was
considered. $H$ strata consisting of $n_h=n$ clusters having $m_{hi}=m$ units each, are
taken. Three overdispersed multinomial distributions: the random-clumped
(RC), the $m$-inflated ($m$-I) and the Dirichlet Multinomial (DM)
distributions (Alonso et al., 2017) having the same parameters $\boldsymbol{%
\pi}_{hi}\left(\boldsymbol{\beta}_{0}\right)$ and $\rho$, are then
considered for $\widehat{\boldsymbol{Y}}_{hi}$; which are characterized by 
\begin{align*}
\boldsymbol{E}[\widehat{\boldsymbol{Y}}_{hi}] & =m\boldsymbol{\pi}%
_{hi}\left( \boldsymbol{\beta}_{0}\right) \quad\text{and}\quad\boldsymbol{V}[%
\widehat{\boldsymbol{Y}}_{hi}]=\nu_{m}m\boldsymbol{\Delta}(\boldsymbol{\pi }%
_{hi}\left( \boldsymbol{\beta}_{0}\right) ), \\
\nu_{m} & =1+\rho^{2}(m-1), ~~~~~~~i=1, \ldots, n; ~h=1,2.
\end{align*}
As in Castilla et al. (2019), we further assume that the outcome nominal
variable $\boldsymbol{Y}$ has $d+1=3$ categories depending on $k=2$
explanatory variables through the PLR model probabilities 
\begin{equation*}
\pi _{hir}\left( \boldsymbol{\beta }\right) =\left \{ 
\begin{array}{ll}
\dfrac{\exp \{ \boldsymbol{x}_{hi}^{T}\boldsymbol{\beta }_{r}\}}{1+{%
\sum_{l=1}^{d}}\exp \{ \boldsymbol{x}_{hi}^{T}\boldsymbol{\beta }_{l}\}}, & 
r=1,2 \\ 
\dfrac{1}{1+{\sum_{l=1}^{d}}\exp \{ \boldsymbol{x}_{hi}^{T}\boldsymbol{\beta 
}_{l}\}}, & r=3%
\end{array}%
\right. ,
\end{equation*}%
with $\boldsymbol{\beta }=(\beta_{01},\beta_{11},\beta_{21},\beta_{02},%
\beta_{12},\beta_{22})^T=(0,-0.9,0.1,0.6,-1.2,0.8)^T$ and $\boldsymbol{x}%
_{hi}\overset{iid}{\sim}\mathcal{N}(\boldsymbol{0},\boldsymbol{I})$ for all $%
i=1, \ldots,n$, $h=1,\dots,H$.

%\quad

Considering different values of the intra-cluster correlation parameter ($%
\rho^2$), the number of clusters in each stratum ($n$) and the clusters
sizes ($m$), we simulate data from different scenarios.

\begin{enumerate}
\item[] Scenario 1: $H=2$, $n\in\{10i\}_{i=4}^{15}$, $m=21$, $\rho^2=0.25$, RC
distribution.

\item[] Scenario 1b: $H=2$, $n\in\{10i\}_{i=4}^{15}$, $m=21$, $\rho^2=0.50$,
RC distribution.

\item[] Scenario 2: $H=2$, $n=60$, $m\in \{10i\}_{i=2}^{12}$, $\rho^2=0.25$,
RC m-I and DM distributions.

\item[] Scenario 3: $H=2$, $n=60$, $m=21$, $\rho^2\in\{0.1i\}_{i=0}^9$, RC,
m-I and DM distributions.
\end{enumerate}

In order to study the robustness issue, these simulations are repeated under
contaminated data having $7\%$ outliers. These outliers are generated by
permuting the elements of the outcome variable, such that categories 1, 2, 3
are classified as categories 3, 1, 2 for the outlying observations.

%\newpage

\subsection{Performance of minimum quasi weighted DPD estimators and $\protect\rho^2$ estimates}

\begin{figure}[tbp]
\center
\includegraphics[scale=1.15]{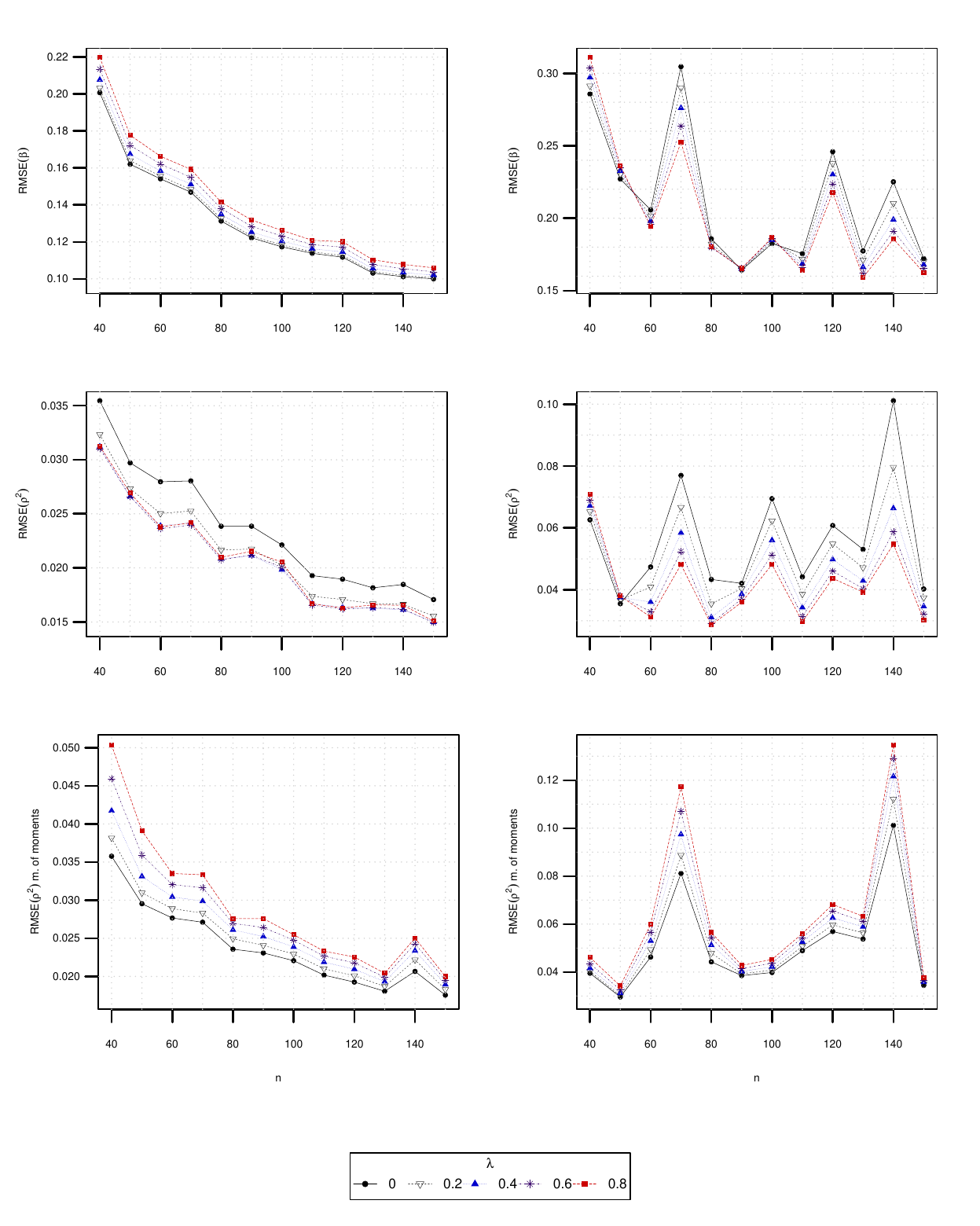}
\caption{Scenario 1: RMSEs of minimum quasi weighted DPD estimators of $\boldsymbol{\protect\beta}$ and $%
\protect\rho^2$ by the equations method and the method of moments. Pure data
(left) and contaminated data (right). RC distribution, $\protect\rho^2=0.25$%
.  }
\label{fig:S1}
\end{figure}

\begin{figure}[tbp]
\center
\includegraphics[scale=1.15]{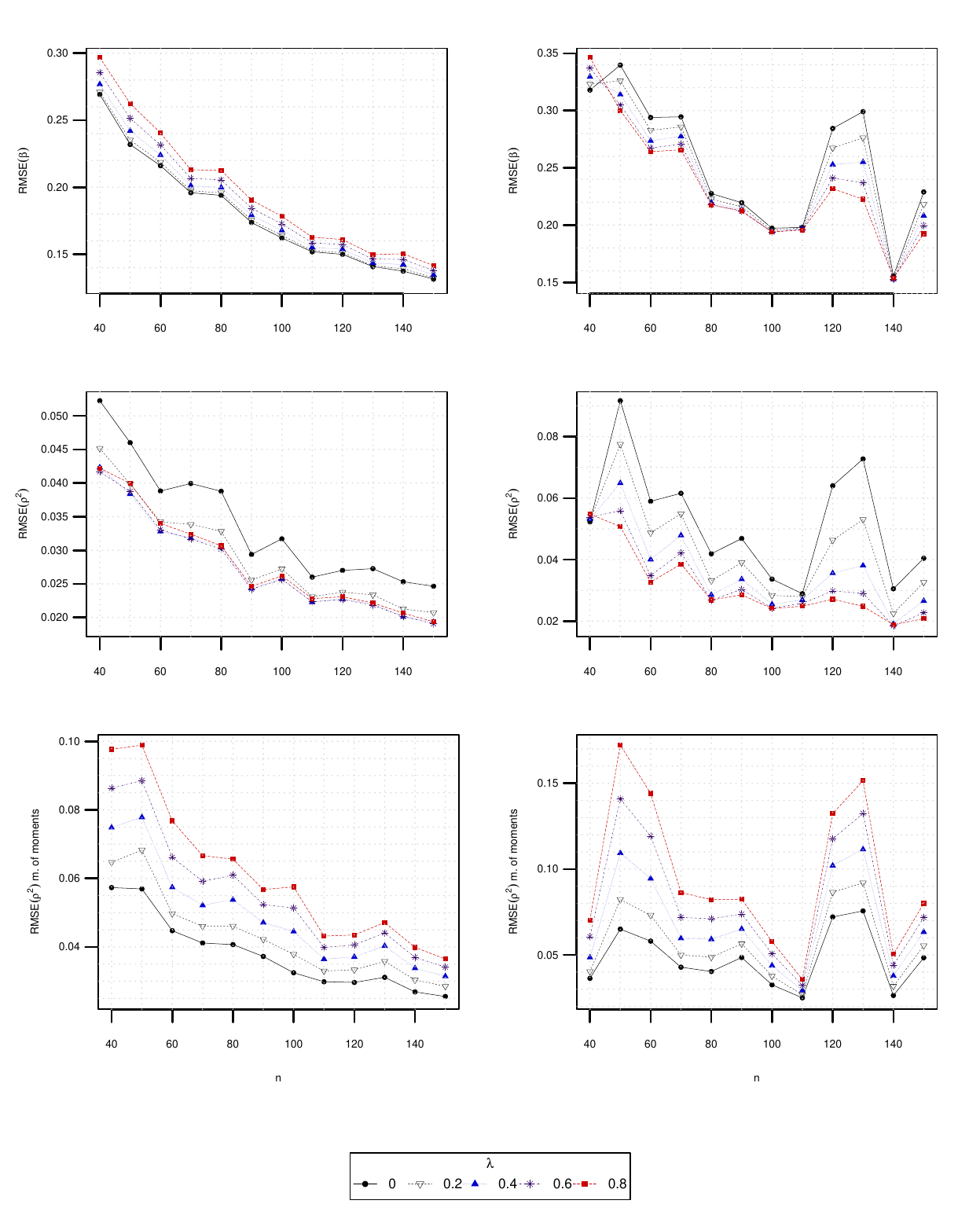}
\caption{Scenario 1b: RMSEs of minimum quasi weighted DPD estimators of $\boldsymbol{\protect\beta}$ and $%
\protect\rho^2$ by the equations method and the method of moments. Pure data
(left) and contaminated data (right). RC distribution, $\protect\rho^2=0.50$%
.}
\label{fig:S1b}
\end{figure}

\begin{figure}[tbp]
\center
\includegraphics[scale=1.15]{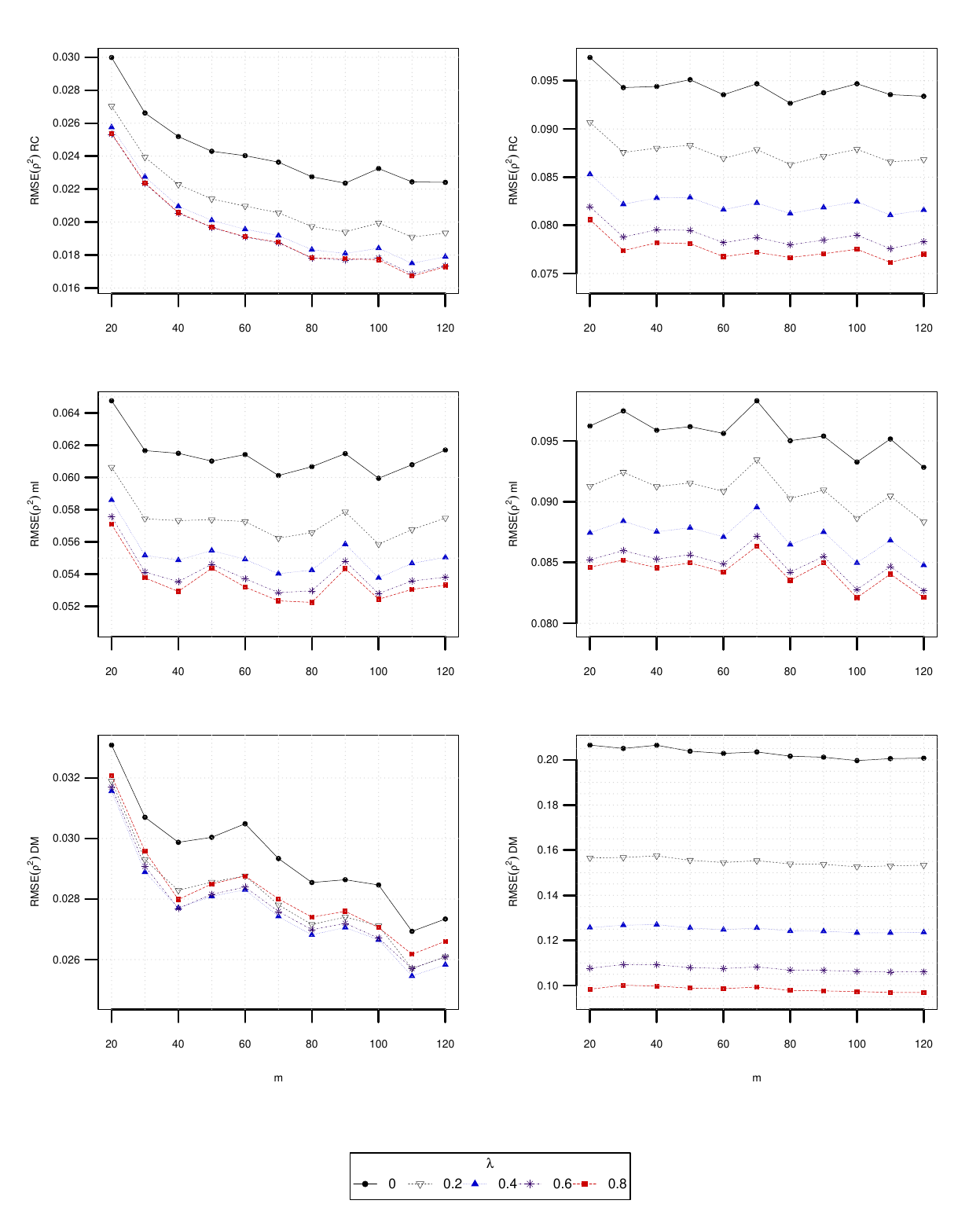}
\caption{Scenario 2: RMSEs of $\protect\rho^2$ by the equations method. Pure
data (left) and contaminated data (right). RC (top), mI (middle) and DM
(bottom) distribution, $\protect\rho^2=0.25, n=60$.}
\label{fig:S2}
\end{figure}

\begin{figure}[tbp]
\center
\includegraphics[scale=1.15]{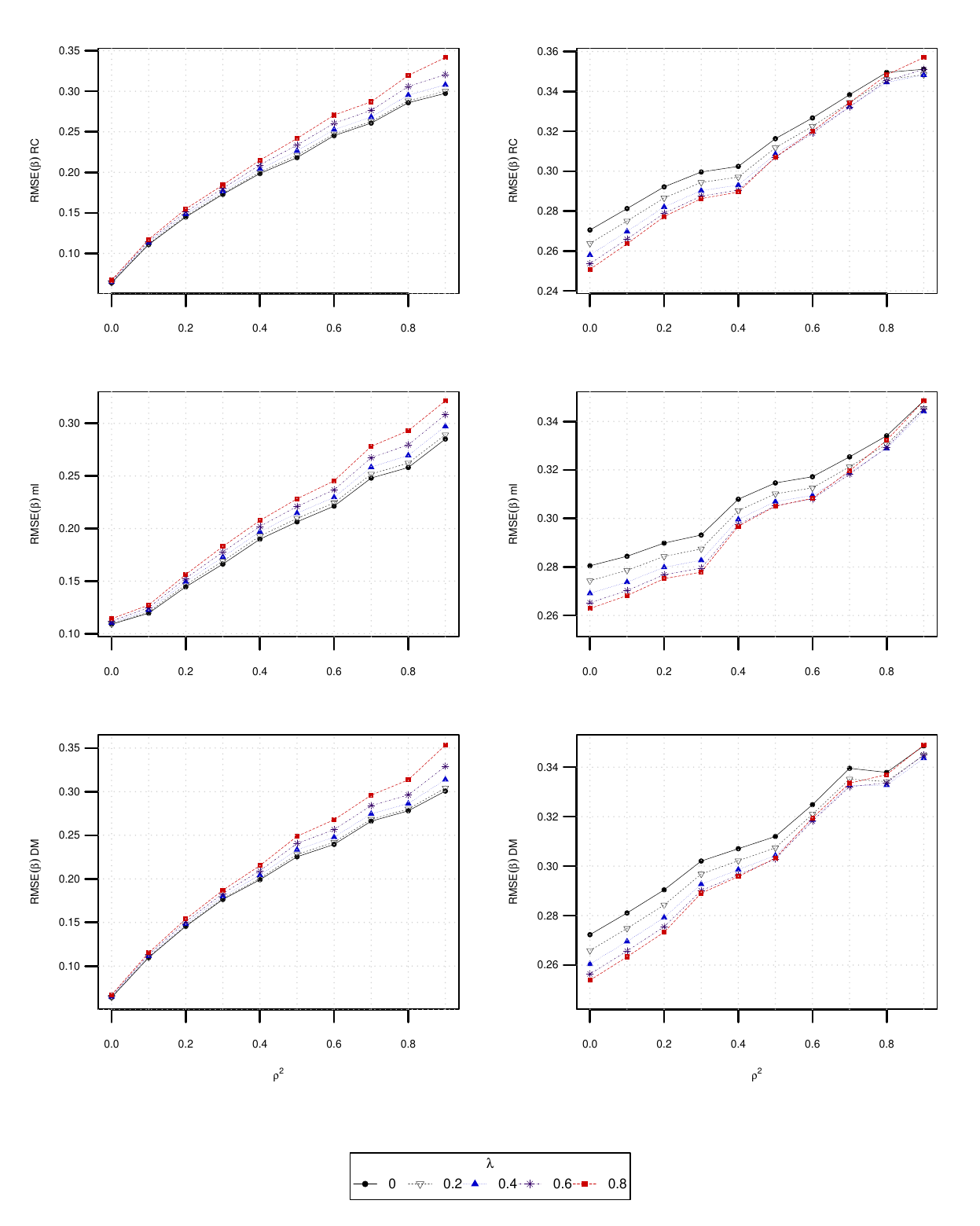}
\caption{Scenario 3: RMSEs of minimum quasi weighted DPD estimators of $\boldsymbol{\protect\beta}$. Pure
data (left) and contaminated data (right). RC (top), mI (middle) and DM
(bottom) distribution, $n=60, m=21$.}
\label{fig:S3}
\end{figure}

For the above scenarios, we compute the minimum quasi weighted DPD estimator of $\boldsymbol{\beta}$%
, for different tuning parameters $\lambda\in\{0,0.2,0.4,0.6,0.8\}$, and the
corresponding estimate of $\rho^2$, both for the methods of moments and the
estimating equations. The root of the mean square error (RMSE) are then
computed based on $1000$ replications (Figures \ref{fig:S1}--\ref{fig:S3}).

%Notice that $\boldsymbol{\beta}$ estimates are obtained from algorithms 6 and 7 with  tolerance parameter $\varepsilon=10^{-4}$ and a maximum number of iterations of $100$.
While classical  maximum quasi weighted likelihood estimator presents the best behaviour under pure data, minimum quasi weighted DPD estimators
with $\lambda>0$ are much more robust. In particular, as $\lambda$
increases, the change on their behaviour is accentuated. This is independent
to the sample size (Figures \ref{fig:S1} and \ref{fig:S1b}), the
intra-cluster correlation and the distribution (Figure \ref{fig:S2})
considered.

\begin{figure}[h]
\center
\includegraphics[scale=0.76]{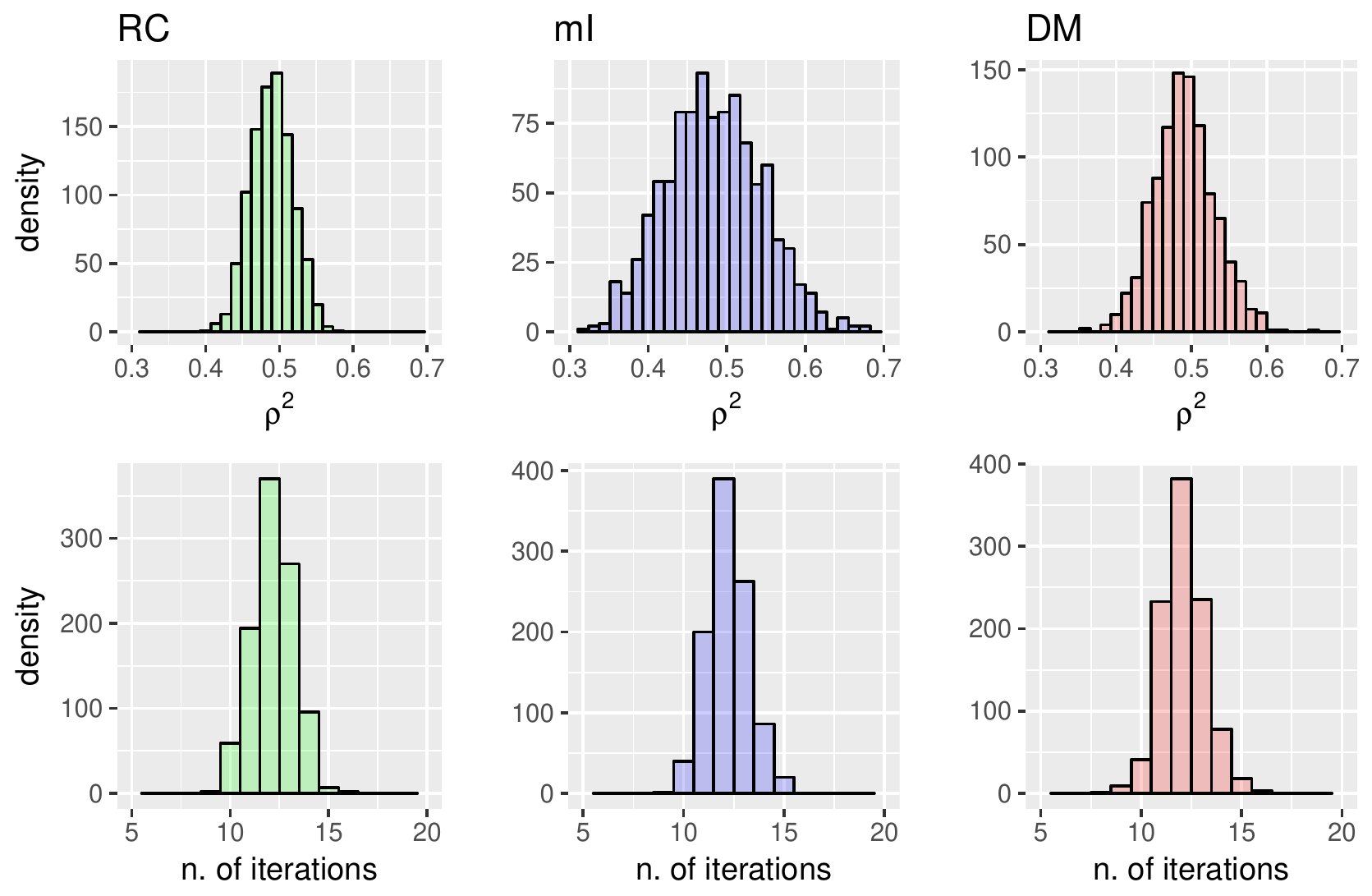}
\caption{ Density plots with estimates obtained from observations of three
distributions, DM, mI, RC, when $\protect\rho^2 = 0.5$ and $\protect\lambda=0.4$. $1,000$ samples.}
\label{fig:hist}
\end{figure}

Estimator of $\rho^2$ by the method of moments seems less precise than the
method of estimating equations, with independence of the tuning parameter
chosen (Figures \ref{fig:S1} and \ref{fig:S1b}). Best estimators of $\rho^2$
by the estimating equations are obtained from minimum quasi weighted DPD estimators with $\lambda>0$, both
for pure and contaminated data and for any of the distributions considered
(Figure \ref{fig:S2}). Error of the estimators of $\rho^2$ tends to be
smaller with the DM and RC distributions in comparison with the mI
distribution, as can be seen in Figure \ref{fig:hist}, where density plots
based on $1,000$ replications are shown for $\rho^2=0.5, n=60, m=21$ and
tuning parameter $\lambda=0.4$. These results on the design effect parameter
are consistent with the previous work of Alonso et al. (2017).

Notice that $\boldsymbol{\beta}$ estimates are obtained through $nlm()$ procedure with tolerance $10^{-6}$ in the software R, used for the whole Monte Carlo study. In Figure \ref{fig:hist} iterations needed to compute the corresponding minimum quasi weighted DPD estimators were also obtained, with not a significant
difference among the different distributions.

%CHECK WHEN TO USE WORD ESTIMATES AND WORD ESTIMATOR AND CHANGE IT WHEN NECESSARY

\begin{figure}[h!!!]
\center
\includegraphics[scale=0.62]{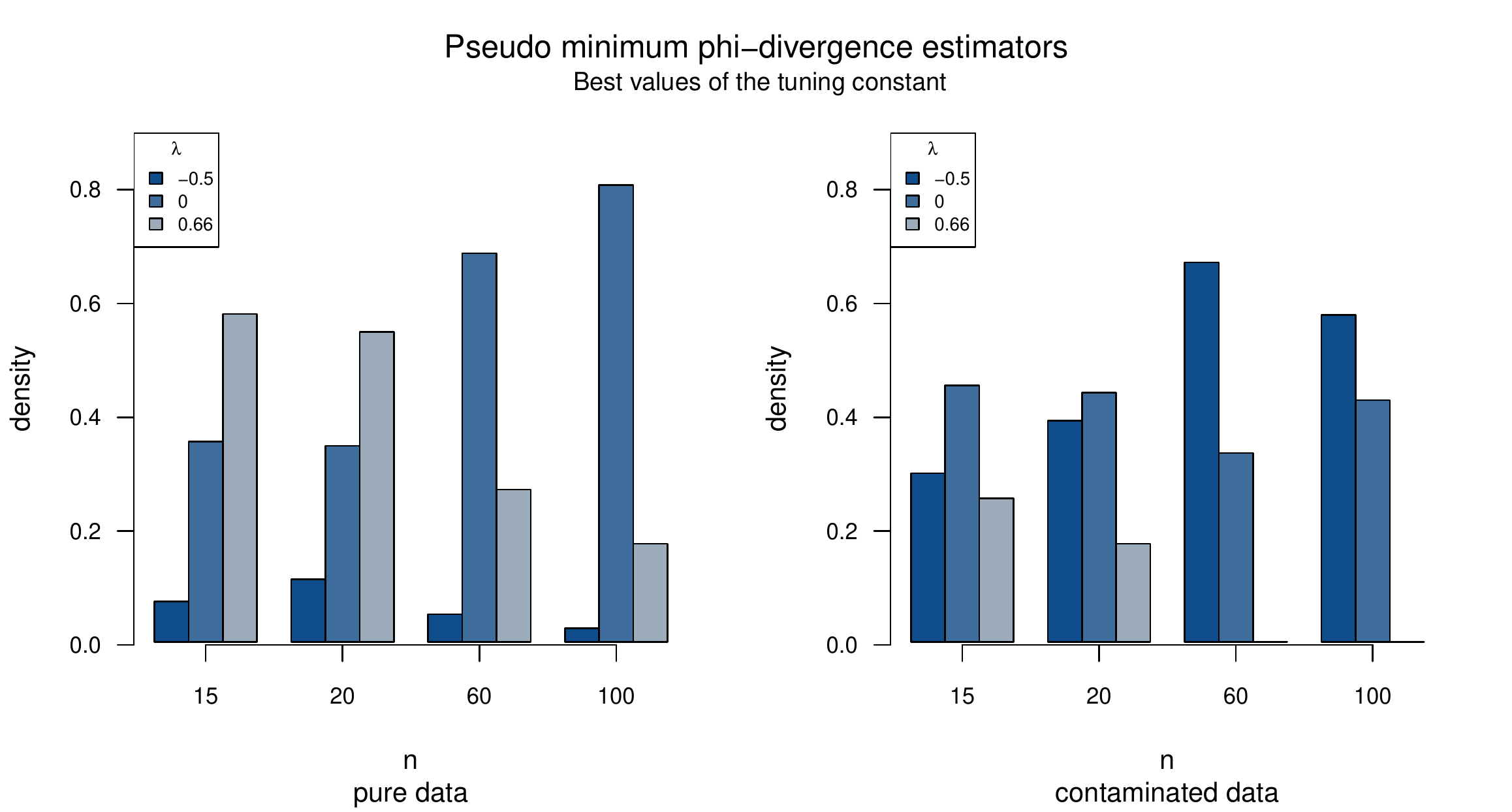} 
\caption{Pseudo minimum phi-divergence estimators: best choice of the tuning parameter, $m=21$ and $\protect\rho^2=0.5.$}
\label{fig:S_bar_phi}
\end{figure}

\begin{figure}[h!!!]
\center
\includegraphics[scale=0.64]{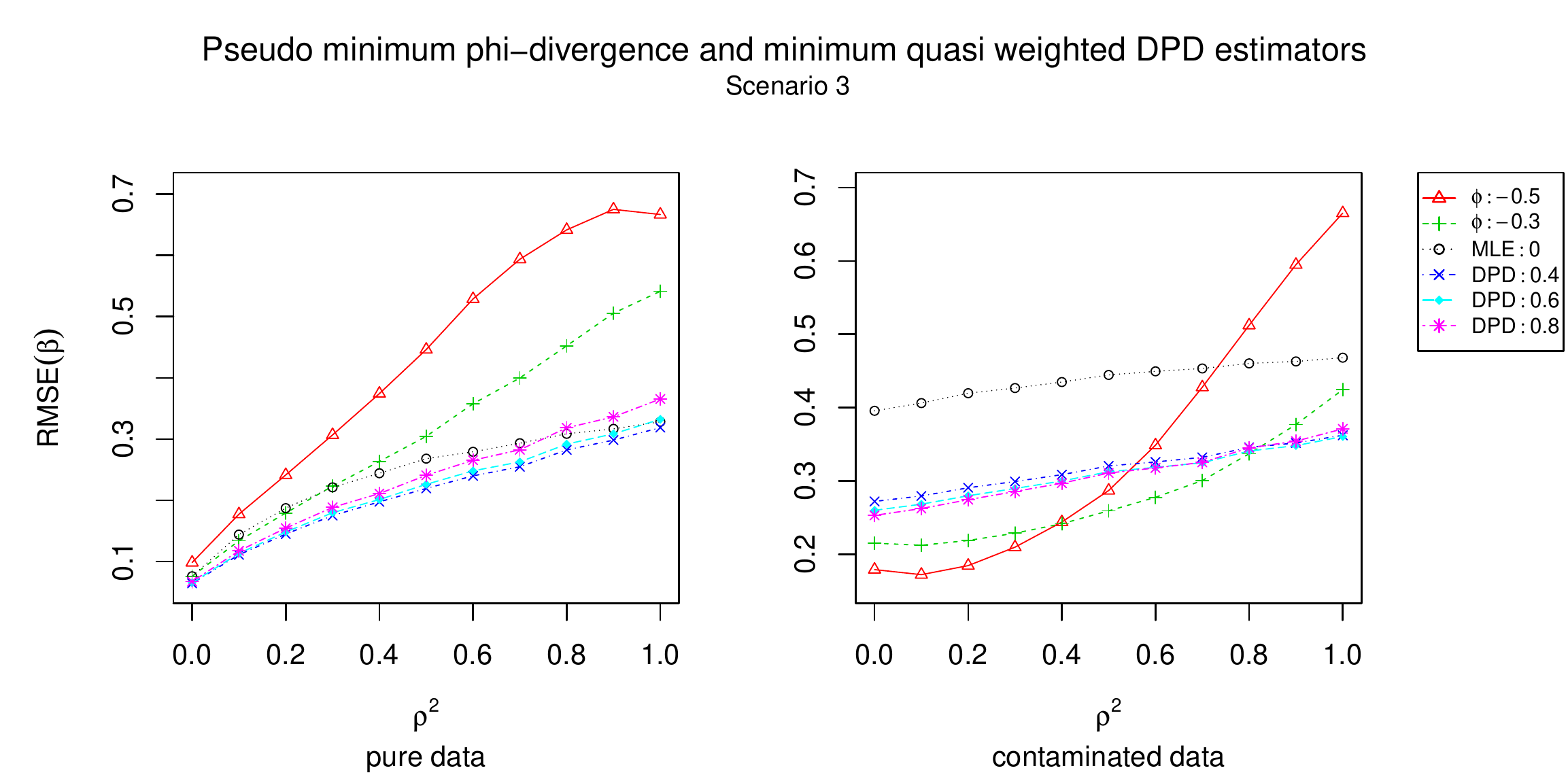} %
\includegraphics[scale=0.63]{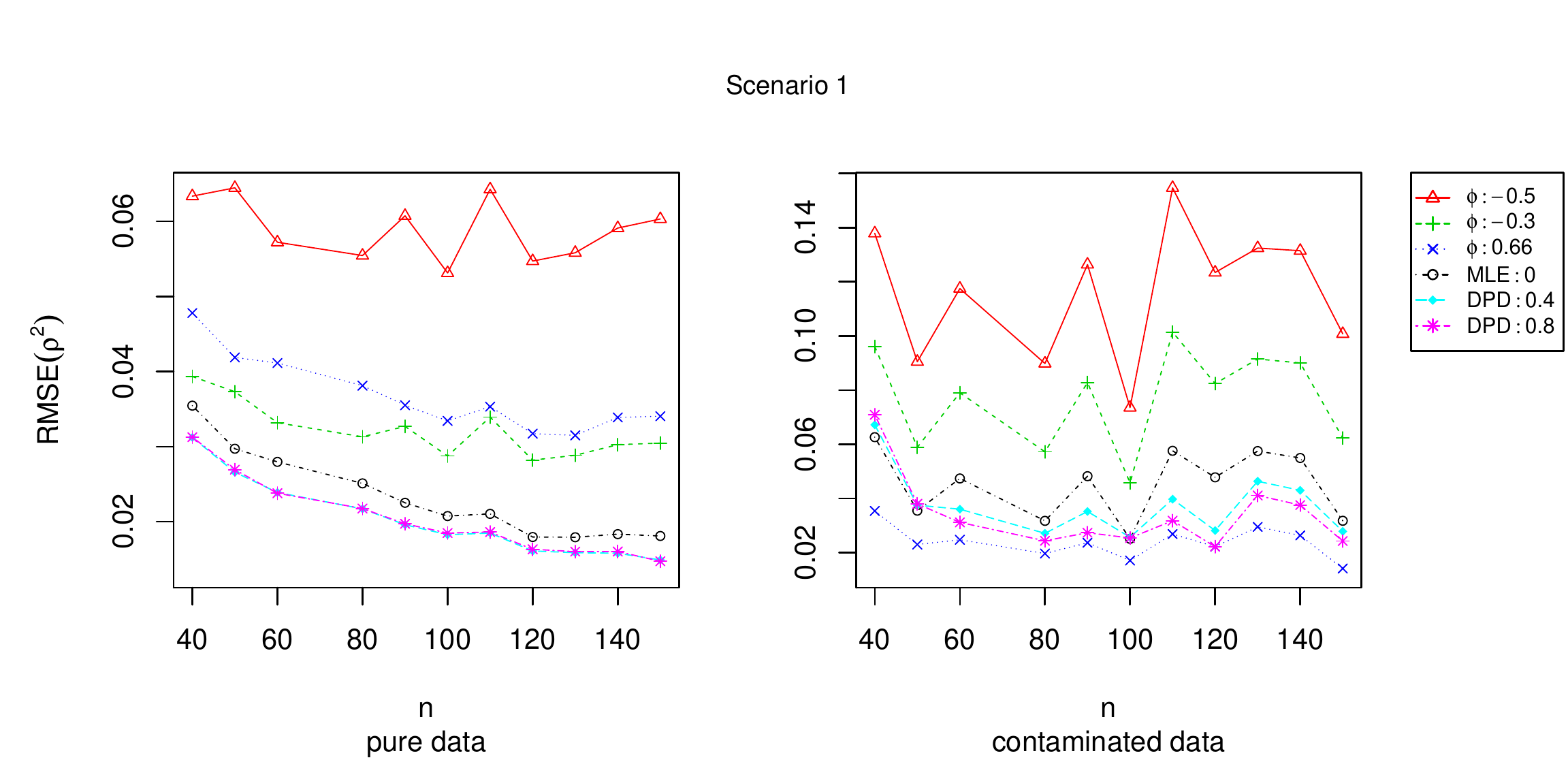}
\caption{Comparison among pseudo minimum phi−divergence and minimum quasi weighted DPD estimators}
\label{fig:S_rho_phi}
\end{figure}

\subsection{Comparison of minimum quasi weighted DPD estimators with pseudo minimum phi-divergence estimators}

In the previous Section we have illustrated how the quasi minimum quasi weighted DPD estimators present a
much more robust behavior than the maximum quasi weighted likelihood estimator for the estimation of $%
\boldsymbol{\beta}$ when a moderate/large sample size was considered. In
Castilla et al. (2018), another family of estimators, the pseudo minimum
phi-divergence estimators, was defined in the PLR model with
complex survey. In particular, pseudo minimum Cressie-Read divergence estimators
for positive tuning parameter $\lambda$ were studied. Results of the
simulation study suggested these estimators as an interesting alternative to
maximum quasi weighted likelihood estimator in terms of efficiency when a small sample size and a large intra
cluster correlation was considered. Nevertheless, the robustness of these
estimators was not studied.

In this section we want to make a general empirical comparison between minimum quasi weighted DPD estimator and pseudo minimum phi-divergence estimators (with positive and negative tuning parameters in the Cressie-Read subfamily) of $\boldsymbol{\beta}$, in terms of efficiency and robustness. Although they were not studied in the cited paper of Castilla et al. (2018), phi-divergences with negative tuning parameter and, in particular, the Hellinger distance ($\lambda=-0.5$), are known to have good robustness properties in some models. See for example Beran (1977)  Lindsay (1994) and Toma (2007). This seems to happen in the context of PLR model with complex survey too, when a moderate/large sample size is considered. Nevertheless, pseudo minimum phi-divergence estimators with positive tuning parameter present an important lack of robustness. This behaviour can be summarized in Figure \ref{fig:S_bar_phi}.

Let us now consider Scenario 3 of the previous section. A comparison between quasi minimum quasi weighted DPD estimators with $\lambda\in\{0.4,0.8\}$, classical maximum quasi weighted likelihood estimator, and pseudo minimum phi-divergence estimators with $\lambda\in\{-0.5,-0.3,0.66\}$ is made. Hellinger distance is shown to be, by far, the best choice when a contaminated scheme with low intra-cluster correlation is considered. With medium/high intra-correlation the behaviour turns to be the opposite. Minimum quasi weighted DPD estimators present a more stable behaviour, both in pure and contaminated schemes, with independence to the correlation parameter.

The same divergences are considered now in Scenario 1 for the estimation of the parameter $\rho^2$. The estimation of $\rho^2$ with pseudo minimum phi-divergence estimators is made by the method of Binder (Castilla et al., 2018) while the estimation of $\rho^2$ with minimum quasi weighted DPD estimators is made by the method of the estimating equations. Pseudo  minimum phi-divergence estimators with negative tuning parameter are not competitive neither in the pure nor in the contaminated schemes. Pseudo  minimum phi-divergence estimator with tuning parameter $\lambda=2/3$ is a good alternative to the minimum quasi weighted DPD estimators in contaminated data. This estimator showed also a good behaviour in the paper of Castilla et al. (2018).

\begin{figure}[h!!!!!!!!!!!!!]
\center
\includegraphics[scale=0.62]{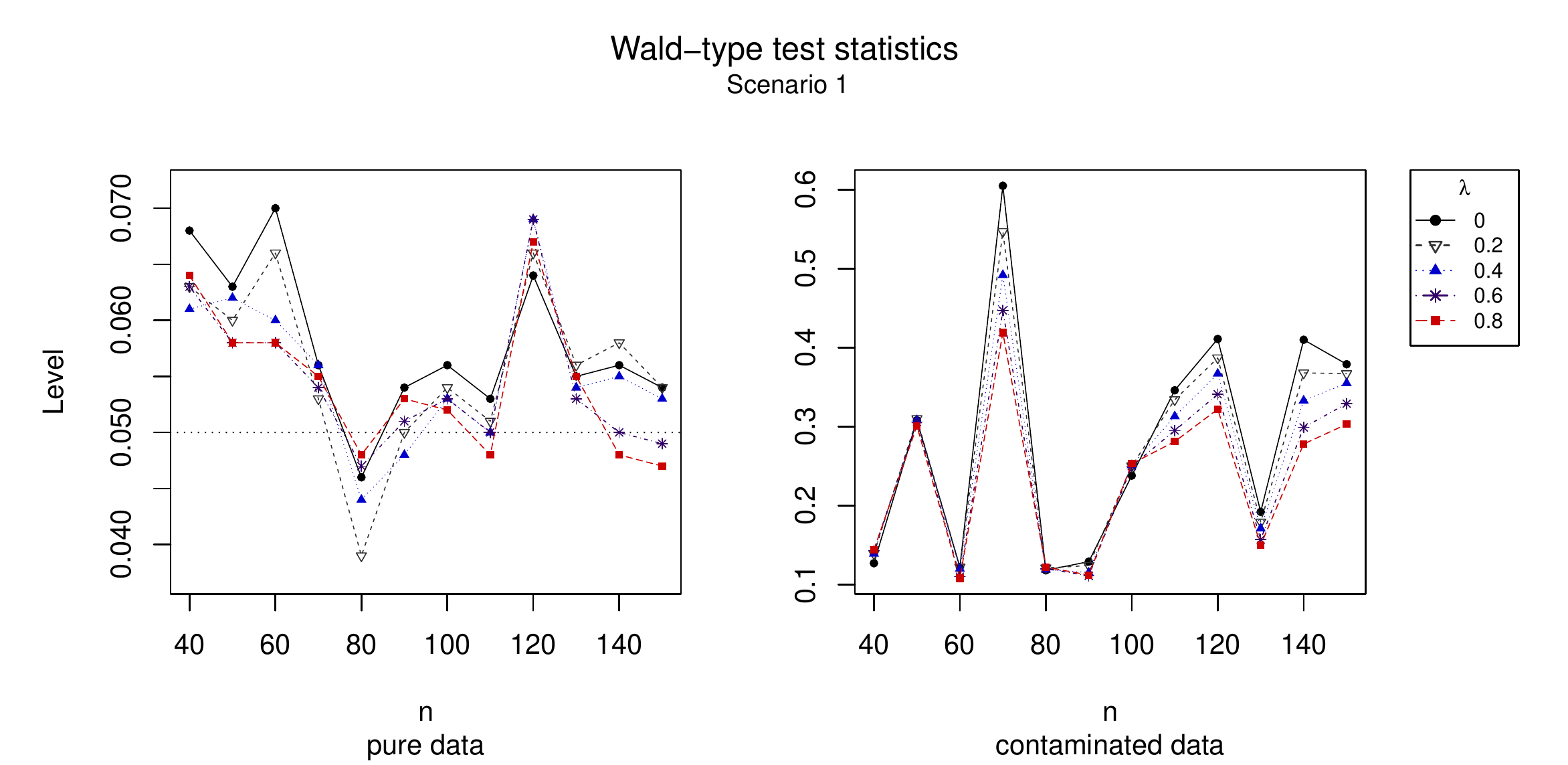} %
\includegraphics[scale=0.62]{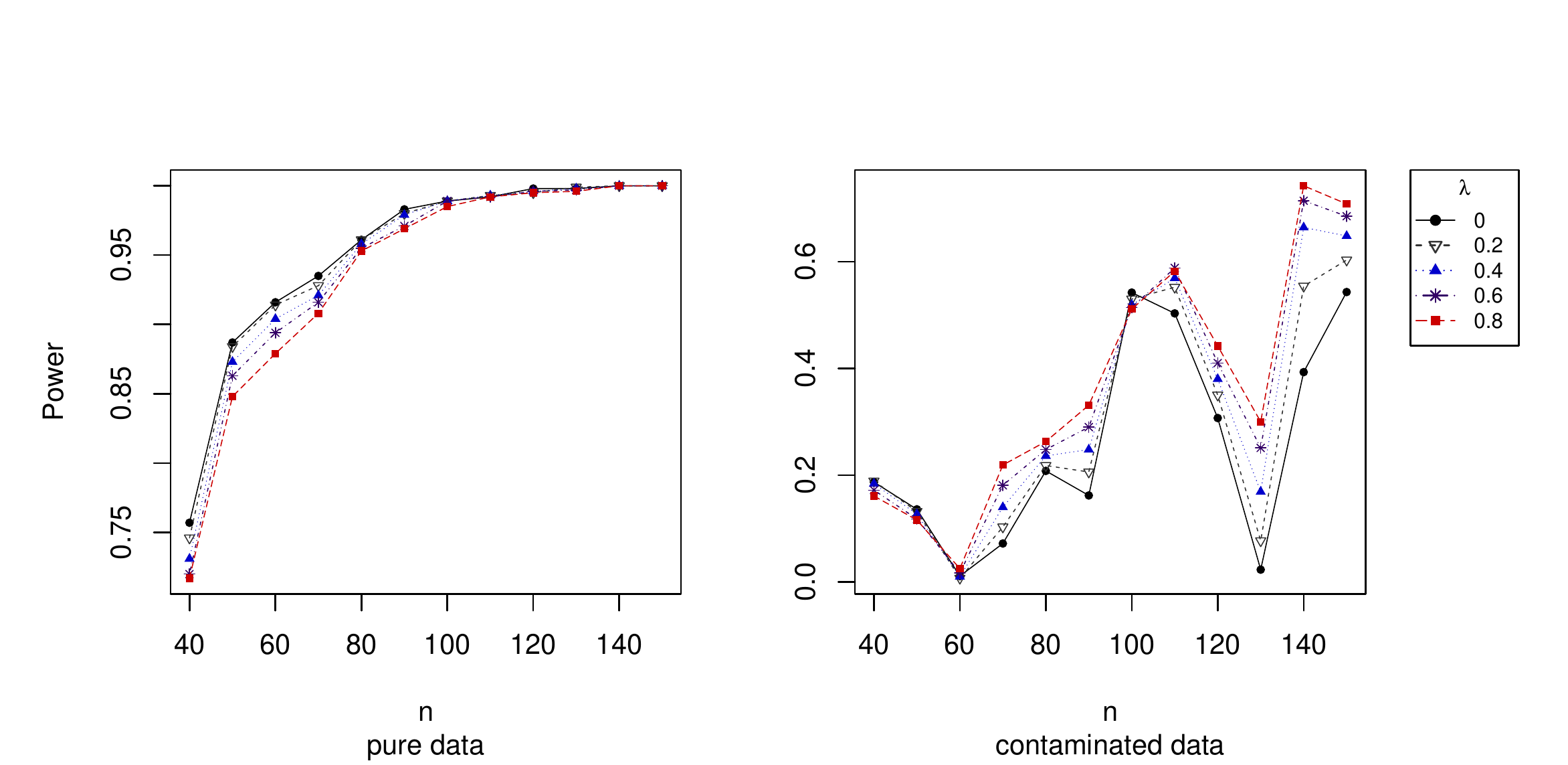}
\caption{Estimated Levels (top) and Powers (bottom) under pure (left) and contaminated (right) data. Scenario 1.}
\label{fig:TESTS}
\end{figure}
%\newpage
%%%%%%%%%%%%%%%%%%%%%%%%%%%%%%%%%
%%%%%%%%%%%%%%%%%%%%%%%%%%%55
%%%%%%%%%%%%%%%%%%%%%%%%%%%55

\subsection{Performance of the Wald-type tests}

With the same model as in Section \ref{sec:SimA} we now empirically study the robustness of the minimum quasi weighted DPD estimator based Wald-type tests for the PLR model. As in Castilla et al. (2018) we first study the level under the true null hypothesis $H_0:\beta_{02}=0.6$. For studying the power robustness,  the true data generating parameter value is considered as $\beta_{02}=1.08$.

Under Scenario 1, and both for pure and contaminated data, we compute observed levels and powers (measured as the proportion of test statistics exceeding the corresponding chi-square critical value), as can be seen in Figure \ref{fig:TESTS}. Under pure data, all levels present a very similar behaviour, while power attains their best value for classical maximum quasi weighted likelihood estimator. Under contamination, minimum quasi weighted DPD estimator based Wald-type tests for $\lambda>0$ present a more robust behaviour than maximum quasi weighted likelihood estimator, in concordance with previous results.

%\clearpage

\section{Concluding Remarks \label{secCon}}
%PLR model is a useful and popular method. Although most of classical literature deals with the cases of simple random sampling scheme,  we usually come across data which have been collected through a complex survey scheme. In this scheme, maximum quasi weighted  likelihood estimator becomes the base of most of the existing literature, despite it is known to present an important lack of robustness. 

We have presented minimum quasi weighted DPD estimators, as a robust alternative to classical approaches, in the modeling of categorical responses with associated covariates through polytomous logistic regression models. A Wald-type family of tests based on these estimators is also presented, for the problem of testing linear hypotheses on regression coefficients. The robustness of both estimators and tests are theoretically justified in terms of the influence function, which is shown to be bounded for the new procedures. An extensive simulation study is also provided, to empirically illustrate their robustness  along with a comparison with the pseudo minimum phi-divergence estimators of Castilla et al. (2018). The results clearly show how the proposed minimum quasi weighted DPD estimators seem to be the best choice for dealing with the robustness issue for a moderate sample size. 

%\section{Choice of the tuning parameter????}

%\clearpage

\appendix 
\section{Appendix  \label{App}}

\subsection*{Derivation of (\protect\ref{Un}) from (\protect\ref{2.4})}
\begin{proof}
From (\ref{2.4}) it holds ${\sum\limits_{h=1}^{H}}{\sum\limits_{i=1}^{n_{h}}}%
\boldsymbol{u}_{hi}(\boldsymbol{\beta })=\boldsymbol{0}_{d(k+1)}$, with%
\begin{equation*}
\boldsymbol{u}_{hi}(\boldsymbol{\beta })=w_{hi}\mathcal{\boldsymbol{\Delta }}%
^{\ast }(\boldsymbol{\pi }_{hi}\left( \boldsymbol{\beta }\right) )\mathrm{%
diag}^{-1}(\boldsymbol{\pi }_{hi}\left( \boldsymbol{\beta }\right) )\left[ 
\widehat{\boldsymbol{y}}_{hi}-m_{hi}\boldsymbol{\pi }_{hi}\left( \boldsymbol{%
\beta }\right) \right] \otimes \boldsymbol{x}_{hi}.
\end{equation*}%
The first term of $\boldsymbol{u}_{hi}(\boldsymbol{\beta })$ is%
$
w_{hi}\boldsymbol{\Delta }^{\ast }(\boldsymbol{\pi }_{hi}\left( \boldsymbol{%
\beta }\right) )\mathrm{diag}^{-1}\{\boldsymbol{\pi }_{hi}\left( \boldsymbol{%
\beta }\right) \}\{\widehat{\boldsymbol{y}}_{hi}-m_{hi}\boldsymbol{\pi }%
_{hi}\left( \boldsymbol{\beta }\right) \},
$
where%
\begin{align*}
w_{hi}\boldsymbol{\Delta }^{\ast }(\boldsymbol{\pi }_{hi}\left( \boldsymbol{%
\beta }\right) )\mathrm{diag}^{-1}\{\boldsymbol{\pi }_{hi}\left( \boldsymbol{%
\beta }\right) \}& =\left( \boldsymbol{I}_{d},\boldsymbol{0}_{d}\right) %
\left[ \mathrm{diag}(\boldsymbol{\pi }_{hi}\left( \boldsymbol{\beta }\right)
)-\boldsymbol{\pi }_{hi}\left( \boldsymbol{\beta }\right) \boldsymbol{\pi }%
_{hi}^{T}\left( \boldsymbol{\beta }\right) \right] \mathrm{diag}^{-1}\{%
\boldsymbol{\pi }_{hi}\left( \boldsymbol{\beta }\right) \} \\
& =\left( \boldsymbol{I}_{d},\boldsymbol{0}_{d}\right) (\boldsymbol{I}_{d+1}-%
\boldsymbol{\pi }_{hi}\left( \boldsymbol{\beta }\right) \boldsymbol{1}^{T}),
\end{align*}%
but $\boldsymbol{1}^{T}\{\widehat{\boldsymbol{y}}_{hi}-m_{hi}\boldsymbol{\pi 
}_{hi}\left( \boldsymbol{\beta }\right) \}=0$ and hence%
\begin{equation*}
\boldsymbol{u}_{hi}(\boldsymbol{\beta })=w_{hi}\left[ \widehat{\boldsymbol{y}%
}_{hi}-m_{hi}\boldsymbol{\pi }_{hi}\left( \boldsymbol{\beta }\right) \right]
\otimes \boldsymbol{x}_{hi}.
\end{equation*}
\end{proof}

\subsection*{Proof of Theorem \ref{th:u} }
\begin{proof}
The minimum quasi weighted DPD estimator of $\boldsymbol{\beta }$, is defined as 
\begin{align*}
\widehat{\boldsymbol{\beta }}_{\lambda ,Q}& =\underset{\boldsymbol{\beta }%
\in \mathbb{R}^{d(k+1)}}{\arg \min }\sum_{h=1}^{H}\sum_{i=1}^{n_{h}}w_{hi}%
\left\{ \left( m_{hi}\boldsymbol{\pi }_{hi}(\boldsymbol{\beta })-\frac{%
1+\lambda }{\lambda }\widehat{\boldsymbol{y}}_{hi}\right) ^{T}\boldsymbol{%
\pi }_{hi}^{\lambda }(\boldsymbol{\beta })\right\} \\
& =\underset{\boldsymbol{\beta }\in \mathbb{R}^{d(k+1)}}{\arg \min }%
\sum_{h=1}^{H}\sum_{i=1}^{n_{h}}w_{hi}\left[ \sum_{l=1}^{d+1}\left\{ \left(
\pi _{hil}(\boldsymbol{\beta })-\frac{1+\lambda }{\lambda }\frac{\widehat{y}%
_{hil}}{m_{hi}}\right) \pi _{hil}^{\lambda }(\boldsymbol{\beta })\right\} %
\right]
\end{align*}%
which can also be obtained by solving the system of equations $\boldsymbol{u}%
_{\lambda }\left( \boldsymbol{\beta }\right) =\boldsymbol{0}_{d(k+1)}$ where 
\begin{align*}
-\frac{1}{\lambda +1}\frac{\partial }{\partial \boldsymbol{\beta }}%
\sum_{h=1}^{H}\sum_{i=1}^{n_{h}}w_{hi}\left\{ \left( m_{hi}\boldsymbol{\pi }%
_{hi}(\boldsymbol{\beta })-\frac{1+\lambda }{\lambda }\widehat{\boldsymbol{y}%
}_{hi}\right) ^{T}\boldsymbol{\pi }_{hi}^{\lambda }(\boldsymbol{\beta }%
)\right\} & =\boldsymbol{0}_{d(k+1)}, \\
\sum_{h=1}^{H}\sum_{i=1}^{n_{h}}w_{hi}m_{hi}\sum_{l=1}^{d+1}\left\{ \frac{%
\widehat{y}_{hil}}{m_{hi}}-\pi _{hil}(\boldsymbol{\beta })\right\} \pi
_{hil}^{\lambda -1}(\boldsymbol{\beta })\frac{\partial }{\partial 
\boldsymbol{\beta }}\pi _{hil}(\boldsymbol{\beta })& =\boldsymbol{0}%
_{d(k+1)}, \\
\sum_{h=1}^{H}\sum_{i=1}^{n_{h}}w_{hi}m_{hi}\left[ \frac{\partial }{\partial 
\boldsymbol{\beta }}\boldsymbol{\pi }_{hi}^{T}\left( \boldsymbol{\beta }%
\right) \mathrm{diag}^{\lambda -1}\{\boldsymbol{\pi }_{hi}\left( \boldsymbol{%
\beta }\right) \}\{\frac{\widehat{\boldsymbol{y}}_{hi}}{m_{hi}}-\boldsymbol{%
\pi }_{hi}\left( \boldsymbol{\beta }\right) \}\right] & =\boldsymbol{0}%
_{d(k+1)}.
\end{align*}%
Now, taking into account that 
\begin{equation*}
\frac{\partial \pi _{hil}(\boldsymbol{\beta })}{\partial \beta _{uv}}%
=x_{iu}\pi _{hil}(\boldsymbol{\beta })\left\{ \delta _{jv}-\pi _{hiv}(%
\boldsymbol{\beta })\right\} ,\qquad u=1,...,k,v=1,...,d,
\end{equation*}%
we get 
\begin{align}
\frac{\partial \boldsymbol{\pi }_{hi}^{T}(\boldsymbol{\beta })}{\partial 
\boldsymbol{\beta }}& =\left( \boldsymbol{I}_{d},\boldsymbol{0}_{d}\right) 
\boldsymbol{\Delta }(\boldsymbol{\pi }_{hi}\left( \boldsymbol{\beta }\right)
)\otimes \boldsymbol{x}_{i}   =\boldsymbol{\Delta }^{\ast }(\boldsymbol{\pi }_{hi}\left( \boldsymbol{%
\beta }\right) )\otimes \boldsymbol{x}_{i},  \label{der}
\end{align}%
and hence the system of equations becomes 
\begin{equation*}
\sum_{h=1}^{H}\sum_{i=1}^{n_{h}}\left[ w_{hi}\boldsymbol{\Delta }^{\ast }(%
\boldsymbol{\pi }_{hi}\left( \boldsymbol{\beta }\right) )\mathrm{diag}%
^{\lambda -1}\{\boldsymbol{\pi }_{hi}\left( \boldsymbol{\beta }\right) \}\{%
\widehat{\boldsymbol{y}}_{hi}-m_{hi}\boldsymbol{\pi }_{hi}\left( \boldsymbol{%
\beta }\right) \}\right] \otimes \boldsymbol{x}_{hi}=\boldsymbol{0}_{d(k+1)}.
\end{equation*}
\end{proof}

\subsection*{Proof of Theorem \ref{Th1}}
\begin{proof}
By following formulas (3.3) and (3.4) of Ghosh and Basu (2013), it holds%
\begin{align*}
\mathbf{\Omega}_{\lambda }\left( \boldsymbol{\beta }\right) & =\mathrm{Var}\left[ 
\boldsymbol{U}_{\lambda }(\boldsymbol{\beta },\boldsymbol{X})\right] , \quad
\mathbf{\Psi}_{\lambda }\left( \boldsymbol{\beta }\right)  =\mathrm{E}\left[ -%
\frac{\partial }{\partial \boldsymbol{\beta }}\boldsymbol{U}_{\lambda }^{T}(%
\boldsymbol{\beta },\boldsymbol{X})\right] ,
\end{align*}%
which can be expressed as a limit of%
\begin{align*}
\mathbf{\Omega}_{n,\lambda }\left( \boldsymbol{\beta }\right) & =\widehat{\mathrm{%
E}}[\mathrm{Var}[\boldsymbol{U}_{\lambda }(\boldsymbol{\beta },\boldsymbol{X}%
)|\boldsymbol{X}]], \quad
\mathbf{\Psi}_{n,\lambda }\left( \boldsymbol{\beta }\right)  =\widehat{\mathrm{%
E}}\left[ \mathrm{E}\left[ -\tfrac{\partial }{\partial \boldsymbol{\beta }}%
\boldsymbol{U}_{\lambda }^{T}(\boldsymbol{\beta },\boldsymbol{X})|%
\boldsymbol{X}\right] \right] ,
\end{align*}%
where the summands given in (\ref{hs}) and (\ref{gs}) are%
\begin{align*}
\mathbf{\Omega}_{hi,\lambda }\left( \boldsymbol{\beta },\boldsymbol{\Sigma }%
_{hi}\right) & =\mathrm{Var}\left[ \boldsymbol{U}_{\lambda }(\boldsymbol{%
\beta },\boldsymbol{x}_{hi})\right] , \quad
\mathbf{\Psi}_{hi,\lambda }\left( \boldsymbol{\beta }\right)  =\mathrm{E}\left[
-\frac{\partial }{\partial \boldsymbol{\beta }}\boldsymbol{U}_{\lambda }^{T}(%
\boldsymbol{\beta },\boldsymbol{x}_{hi})\right] .
\end{align*}%
With respect to $\mathbf{\Omega}_{\lambda }\left( \boldsymbol{\beta }\right) =%
\mathrm{Var}\left[ \boldsymbol{U}_{\lambda }(\boldsymbol{\beta },\boldsymbol{%
X})\right] =\mathrm{E}\left[ \mathrm{Var}\left[ \boldsymbol{U}_{\lambda }(%
\boldsymbol{\beta },\boldsymbol{X})|\boldsymbol{X}\right] \right] +\mathrm{%
Var}\left[ \mathrm{E}\left[ \boldsymbol{U}_{\lambda }(\boldsymbol{\beta },%
\boldsymbol{X})|\boldsymbol{X}\right] \right] $, the first summand estimated
from the sample is%
\begin{align*}
\widehat{\mathrm{E}}[\mathrm{Var}[\boldsymbol{U}_{\lambda }(\boldsymbol{%
\beta },\boldsymbol{X})|\boldsymbol{X}]]& =\sum_{h=1}^{H}\sum_{i=1}^{n_{h}}%
\mathrm{Var}\left[ \boldsymbol{U}_{\lambda }(\boldsymbol{\beta },\boldsymbol{%
x}_{hi})\right] \widehat{\Pr }(\boldsymbol{X=x}_{hi})  =\frac{1}{n}\sum_{h=1}^{H}\sum_{i=1}^{n_{h}}\mathrm{Var}\left[ \boldsymbol{%
U}_{\lambda }(\boldsymbol{\beta },\boldsymbol{x}_{hi})\right] ,
\end{align*}%
and the second one%
\begin{align*}
\widehat{\mathrm{Var}}\left[ \mathrm{E}\left[ \boldsymbol{U}_{\lambda }(%
\boldsymbol{\beta },\boldsymbol{X})|\boldsymbol{X}\right] \right] & =\mathrm{%
Var}\left[ \sum_{h=1}^{H}\sum_{i=1}^{n_{h}}E\left[ \boldsymbol{U}_{\lambda }(%
\boldsymbol{\beta },\boldsymbol{x}_{hi})\right] \widehat{\Pr }(\boldsymbol{%
X=x}_{hi})\right] =\boldsymbol{0}_{d(k+1)},
\end{align*}%
for $\boldsymbol{\beta }=\boldsymbol{\beta }_{0}$, since $\mathrm{E}\left[ 
\boldsymbol{U}_{\lambda }(\boldsymbol{\beta }_{0},\boldsymbol{x}_{hi})\right]
=\boldsymbol{0}_{d(k+1)}$ from $\mathrm{E}[%
\widehat{\boldsymbol{Y}}_{hi}-m_{hi}\boldsymbol{\pi }_{hi}\left( \boldsymbol{%
\beta }_{0}\right) ]=\boldsymbol{0}_{d(k+1)}$. On the other hand,%
\begin{align*}
\mathrm{Var}\left[ \boldsymbol{U}_{\lambda }(\boldsymbol{\beta },\boldsymbol{%
x}_{hi})\right] & =\mathrm{Var}\left[ \left[ w_{hi}\boldsymbol{\Delta }%
^{\ast }(\boldsymbol{\pi }_{hi}\left( \boldsymbol{\beta }\right) )\mathrm{%
diag}^{\lambda -1}\{\boldsymbol{\pi }_{hi}\left( \boldsymbol{\beta }\right)
\}\{\widehat{\boldsymbol{Y}}_{hi}-m_{hi}\boldsymbol{\pi }_{hi}\left( 
\boldsymbol{\beta }\right) \}\right] \otimes \boldsymbol{x}_{hi}\right] \\
& =\left( w_{hi}^{2}\boldsymbol{\Delta }^{\ast }(\boldsymbol{\pi }%
_{hi}\left( \boldsymbol{\beta }\right) )\mathrm{diag}^{\lambda -1}\{%
\boldsymbol{\pi }_{hi}\left( \boldsymbol{\beta }\right) \}\mathrm{Var}[%
\widehat{\boldsymbol{Y}}_{hi}]\mathrm{diag}^{\lambda -1}\{\boldsymbol{\pi }%
_{hi}\left( \boldsymbol{\beta }\right) \}\boldsymbol{\Delta }^{\ast T}(%
\boldsymbol{\pi }_{hi}\left( \boldsymbol{\beta }\right) )\right) \otimes 
\boldsymbol{x}_{hi}\boldsymbol{x}_{hi}^{T}.
\end{align*}%
\newline
With respect to $\mathbf{\Psi}_{\lambda }\left( \boldsymbol{\beta }\right) =-%
\mathrm{E}\left[ \frac{\partial }{\partial \boldsymbol{\beta }}\boldsymbol{U}%
_{\lambda }^{T}(\boldsymbol{\beta },\boldsymbol{X})\right] $, estimated from
the sample is given by%
\begin{align*}
& \widehat{\mathrm{E}}\left[ \mathrm{E}\left[ -\tfrac{\partial }{\partial 
\boldsymbol{\beta }}\boldsymbol{U}_{\lambda }^{T}(\boldsymbol{\beta },%
\boldsymbol{X})|\boldsymbol{X}\right] \right] \\
& =-\sum_{h=1}^{H}\sum_{i=1}^{n_{h}}\mathrm{E}\left[ \frac{\partial }{%
\partial \boldsymbol{\beta }}\boldsymbol{U}_{\lambda }(\boldsymbol{\beta },%
\boldsymbol{x}_{hi})\right] \widehat{\Pr }(\boldsymbol{X=x}_{hi}) \\
& =-\frac{1}{n}\sum_{h=1}^{H}\sum_{i=1}^{n_{h}}\left[ w_{hi}\frac{\partial }{%
\partial \boldsymbol{\beta }}\left[ \boldsymbol{\Delta }^{\ast }(\boldsymbol{%
\pi }_{hi}\left( \boldsymbol{\beta }\right) )\mathrm{diag}^{\lambda -1}\{%
\boldsymbol{\pi }_{hi}\left( \boldsymbol{\beta }\right) \}\right] \mathrm{E}[%
\widehat{\boldsymbol{Y}}_{hi}-m_{hi}\boldsymbol{\pi }_{hi}\left( \boldsymbol{%
\beta }\right) ]\right] ^{T}\otimes \boldsymbol{x}_{hi}^{T} \\
& +\frac{1}{n}\sum_{h=1}^{H}\sum_{i=1}^{n_{h}}\left[ w_{hi}m_{hi}\boldsymbol{%
\Delta }^{\ast }(\boldsymbol{\pi }_{hi}\left( \boldsymbol{\beta }\right) )%
\mathrm{diag}^{\lambda -1}\{\boldsymbol{\pi }_{hi}\left( \boldsymbol{\beta }%
\right) \}\frac{\partial }{\partial \boldsymbol{\beta }}\boldsymbol{\pi }%
_{hi}\left( \boldsymbol{\beta }\right) \right] ^{T}\otimes \boldsymbol{x}%
_{hi}^{T} \\
& =\frac{1}{n}\sum_{h=1}^{H}\sum_{i=1}^{n_{h}}\left[ w_{hi}m_{hi}\boldsymbol{%
\Delta }^{\ast }(\boldsymbol{\pi }_{hi}\left( \boldsymbol{\beta }\right) )%
\mathrm{diag}^{\lambda -1}\{\boldsymbol{\pi }_{hi}\left( \boldsymbol{\beta }%
\right) \}\frac{\partial }{\partial \boldsymbol{\beta }}\boldsymbol{\pi }%
_{hi}\left( \boldsymbol{\beta }\right) \right] ^{T}\otimes \boldsymbol{x}%
_{hi}^{T} \\
& =\frac{1}{n}\sum_{h=1}^{H}\sum_{i=1}^{n_{h}}\left[ w_{hi}m_{hi}\boldsymbol{%
\Delta }^{\ast }(\boldsymbol{\pi }_{hi}\left( \boldsymbol{\beta }\right) )%
\mathrm{diag}^{\lambda -1}\{\boldsymbol{\pi }_{hi}\left( \boldsymbol{\beta }%
\right) \}\boldsymbol{\Delta }^{\ast T}(\boldsymbol{\pi }_{hi}\left( 
\boldsymbol{\beta }\right) )\right] \otimes \boldsymbol{x}_{hi}\boldsymbol{x}%
_{hi}^{T},
\end{align*}%
where the third equality is again true for $\boldsymbol{\beta }=\boldsymbol{%
\beta }_{0}$. The derivations are omitted for $\lambda =0$, in which case
similar ideas as $\lambda >0$ might be followed.
\end{proof}

\subsection*{Proof of Theorem \ref{th:c}}

\begin{proof}
We have 
\begin{align*}
Po_{W_{n}\left( \widehat{\boldsymbol{\beta }}_{\lambda ,Q}\right) }\left( 
\boldsymbol{\beta }^{\ast }\right) & =\Pr \left( W_{n}\left( \widehat{%
\boldsymbol{\beta }}_{\lambda ,Q}\right) >\chi _{r,\alpha }^{2}\right) =\Pr
\left( n\left( l_{\widehat{\boldsymbol{\beta }}_{\lambda ,Q}}(\widehat{%
\boldsymbol{\beta }}_{\lambda ,Q})-l_{\boldsymbol{\beta }^{\ast }}(%
\boldsymbol{\beta }^{\ast })\right) >\chi _{r,\alpha }^{2}-nl_{\boldsymbol{%
\beta }^{\ast }}(\boldsymbol{\beta }^{\ast })\right) \\
& =\Pr \left( \sqrt{n}\left( l_{\widehat{\boldsymbol{\beta }}_{\lambda ,Q}}(%
\widehat{\boldsymbol{\beta }}_{\lambda ,Q})-l_{\boldsymbol{\beta }^{\ast }}(%
\boldsymbol{\beta }^{\ast })\right) >\frac{\chi _{r,\alpha }^{2}}{\sqrt{n}}-%
\sqrt{n}l_{\boldsymbol{\beta }^{\ast }}(\boldsymbol{\beta }^{\ast })\right) .
\end{align*}%
Now we are going to get the asymptotic distribution of the random variable $%
\sqrt{n}\left( l_{\widehat{\boldsymbol{\beta }}_{\lambda ,Q}}(\widehat{%
\boldsymbol{\beta }}_{\lambda ,Q})-l_{\boldsymbol{\beta }^{\ast }}(%
\boldsymbol{\beta }^{\ast })\right) $.

It is clear that $l_{\widehat{\boldsymbol{\beta }}_{\lambda ,Q}}(\widehat{%
\boldsymbol{\beta }}_{\lambda ,Q})$ and $l_{\widehat{\boldsymbol{\beta }}%
_{\lambda ,Q}}(\boldsymbol{\beta }^{\ast })$ have the same asymptotic
distribution because $\widehat{\boldsymbol{\beta }}_{\lambda ,Q}\underset{%
n\rightarrow \infty }{\overset{P}{\longrightarrow }}\boldsymbol{\beta }%
^{\ast }.$ A first Taylor expansion of $l_{\widehat{\boldsymbol{\beta }}%
_{\lambda ,Q}}(\boldsymbol{\beta }^{\ast })$ at $\widehat{\boldsymbol{\beta }%
}_{\lambda ,Q}$ around $\boldsymbol{\beta }^{\ast }$ gives%
\begin{equation*}
l_{\widehat{\boldsymbol{\beta }}_{\lambda ,Q}}(\boldsymbol{\beta }^{\ast
})-l_{\boldsymbol{\beta }^{\ast }}(\boldsymbol{\beta }^{\ast })=\left. \frac{%
\partial l_{\boldsymbol{\beta }}(\boldsymbol{\beta }^{\ast })}{\partial 
\boldsymbol{\beta }^{T}}\right\vert _{\boldsymbol{\beta }=\boldsymbol{\beta }%
^{\ast }}(\widehat{\boldsymbol{\beta }}_{\lambda ,Q}-\boldsymbol{\beta }%
^{\ast })+o_{p}\left( \left\Vert \widehat{\boldsymbol{\beta }}_{\lambda ,Q}-%
\boldsymbol{\beta }^{\ast }\right\Vert \right) .
\end{equation*}%
Therefore,%
\begin{equation*}
\sqrt{n}\left( l_{\widehat{\boldsymbol{\beta }}_{\lambda ,Q}}(\widehat{%
\boldsymbol{\beta }}_{\lambda ,Q})-l_{\boldsymbol{\beta }^{\ast }}(%
\boldsymbol{\beta }^{\ast })\right) \underset{n\rightarrow \infty }{\overset{%
L}{\longrightarrow }}N\left( 0,\sigma ^{2}\left( \boldsymbol{\beta }^{\ast
}\right) \right)
\end{equation*}%
being 
\begin{equation*}
\sigma ^{2}\left( \boldsymbol{\beta }^{\ast }\right) =\left. \frac{\partial
l_{\boldsymbol{\beta }}(\boldsymbol{\beta }^{\ast })}{\partial \boldsymbol{%
\beta }^{T}}\right\vert _{\boldsymbol{\beta }=\boldsymbol{\beta }^{\ast }}%
\widehat{\mathbf{Q}}_{n,\lambda }\left( \boldsymbol{\beta }^{\ast }\right)
\left. \frac{\partial l_{\boldsymbol{\beta }}(\boldsymbol{\beta }^{\ast })}{%
\partial \boldsymbol{\beta }}\right\vert _{\boldsymbol{\beta }=\boldsymbol{%
\beta }^{\ast }}.
\end{equation*}%
%??? Now the result follows.+
\end{proof}

\subsection*{Proof of Corollary \ref{cor1}}
\begin{proof}
It is based on the weak consistency of $\widehat{\boldsymbol{\beta }}%
_{\lambda ,Q}$ as well as in the continuity with respect to $\boldsymbol{%
\beta }$\ of the elements of different matrices. Based on Theorem \ref{Th1},
it holds $\lim_{n\rightarrow \infty }\mathrm{E}[\widehat{\boldsymbol{\beta }}%
_{\lambda ,Q}]=\boldsymbol{0}_{d(k+1)}$ and $\lim_{n\rightarrow \infty }%
\mathrm{Var}[\widehat{\boldsymbol{\beta }}_{\lambda ,Q}]=\boldsymbol{O}%
_{d(k+1)\times d(k+1)}$ and hence $\widehat{\boldsymbol{\beta }}_{\lambda ,Q}%
\underset{n\rightarrow \infty }{\overset{P}{\longrightarrow }}\boldsymbol{%
\beta }_{0}$.\newline
\end{proof}

\subsection*{Proof of Theorem \ref{Th1b}}
\begin{proof}
Let 
\begin{align*}
\widehat{\mathbf{\Omega}}_{n,\lambda }(\widehat{\boldsymbol{\beta }}_{\lambda
,Q})& =\widehat{\mathrm{E}}[\mathrm{Var}[\boldsymbol{U}_{\lambda }(\widehat{%
\boldsymbol{\beta }}_{\lambda ,Q},\boldsymbol{X})|\boldsymbol{X}]] \\
& =\widehat{\mathrm{E}}[\mathrm{E}[\boldsymbol{U}_{\lambda }(\widehat{%
\boldsymbol{\beta }}_{\lambda ,Q},\boldsymbol{X})\boldsymbol{U}_{\lambda
}^{T}(\widehat{\boldsymbol{\beta }}_{\lambda ,Q},\boldsymbol{X})|\boldsymbol{%
X}]-\mathrm{E}[\boldsymbol{U}_{\lambda }(\widehat{\boldsymbol{\beta }}%
_{\lambda ,Q},\boldsymbol{X})|\boldsymbol{X}]\mathrm{E}[\boldsymbol{U}%
_{\lambda }^{T}(\widehat{\boldsymbol{\beta }}_{\lambda ,Q},\boldsymbol{X})|%
\boldsymbol{X}]]
\end{align*}%
be another possible consistent estimator of $\mathbf{\Omega}_{\lambda }\left( 
\boldsymbol{\beta }\right) $, alternative to $\mathbf{\Omega}_{n,\lambda }\left( 
\boldsymbol{\beta }\right) $. Since 
\begin{equation*}
\widehat{\mathrm{E}}\left[ \mathrm{E}[\boldsymbol{U}_{\lambda }(\widehat{%
\boldsymbol{\beta }}_{\lambda ,Q},\boldsymbol{X})|\boldsymbol{X}]\mathrm{E}[%
\boldsymbol{U}_{\lambda }^{T}(\widehat{\boldsymbol{\beta }}_{\lambda ,Q},%
\boldsymbol{X})|\boldsymbol{X}]]\right] =\sum_{h=1}^{H}\sum_{i=1}^{n_{h}}%
\mathrm{E}[\boldsymbol{U}_{\lambda }(\widehat{\boldsymbol{\beta }}_{\lambda
,Q},\boldsymbol{x}_{hi})]\mathrm{E}[\boldsymbol{U}_{\lambda }(\widehat{%
\boldsymbol{\beta }}_{\lambda ,Q},\boldsymbol{x}_{hi})]\widehat{\Pr }(%
\boldsymbol{X=x}_{hi}),
\end{equation*}%
and $\mathrm{E}[\boldsymbol{U}_{\lambda }(\widehat{\boldsymbol{\beta }}%
_{\lambda ,Q},\boldsymbol{X})|\boldsymbol{x}_{hi}]=0$, for being $\mathrm{E}[%
\widehat{\boldsymbol{Y}}_{hi}-m_{hi}\boldsymbol{\pi }_{hi}(\widehat{%
\boldsymbol{\beta }}_{\lambda ,Q})]=\boldsymbol{0}_{k+1}$, for all $(h,i)\in
\{1,...,H\}\times \{1,...,m_{hi}\}$ according to the PLR model with complex
design, it holds%
\begin{align*}
\widehat{\mathbf{\Omega}}_{n,\lambda }(\boldsymbol{\beta })& =\widehat{\mathrm{E}}%
[\mathrm{E}[\boldsymbol{U}_{\lambda }(\boldsymbol{\beta },\boldsymbol{X})%
\boldsymbol{U}_{\lambda }^{T}(\boldsymbol{\beta },\boldsymbol{X})|%
\boldsymbol{X}]  =\frac{1}{n}\sum_{h=1}^{H}\sum_{i=1}^{n_{h}}\boldsymbol{U}_{\lambda }(%
\boldsymbol{\beta },\boldsymbol{x}_{hi})\boldsymbol{U}_{\lambda }^{T}(%
\boldsymbol{\beta },\boldsymbol{x}_{hi})
\end{align*}%
is another possible consistent estimator of $\mathbf{\Omega}_{\lambda }\left( 
\boldsymbol{\beta }\right) $, alternative to $\mathbf{\Omega}_{n,\lambda }\left( 
\boldsymbol{\beta }\right) $.
\end{proof}

\subsection*{Proof of Corollary \ref{cor2} }
\begin{proof}
Section a) is straightforward considering the expressions of matrix $\mathbf{%
G}_{n,\lambda }(\boldsymbol{\beta })$ and the related ones. For section b)
we consider the vector%
\begin{equation*}
\boldsymbol{Z}_{hi}^{\ast }\left( \boldsymbol{\beta }\right) =\sqrt{m}\Delta
^{-1/2}\left( \boldsymbol{\pi }_{hi}^{\ast }\left( \boldsymbol{\beta }%
\right) \right) (\tfrac{\widehat{\boldsymbol{Y}}_{hi}^{\ast }}{m}-%
\boldsymbol{\pi }_{hi}^{\ast }\left( \beta \right) ).
\end{equation*}%
Taking into account that 
$
\mathrm{Var}[\widehat{\boldsymbol{Y}}_{hi}^{\ast }]=m\nu \boldsymbol{\Delta }
\left( \boldsymbol{\pi }_{hi}^{\ast }\left( \beta \right) \right) ,
$
we have 
\begin{equation*}
\mathrm{E}\left[ \boldsymbol{Z}_{hi}^{\ast }\left( \boldsymbol{\beta }%
\right) \right] =\boldsymbol{0}\text{ and }\mathrm{Var}\left[ \boldsymbol{Z}%
_{hi}^{\ast }\left( \boldsymbol{\beta }\right) \right] =\nu \boldsymbol{I}%
_{d},
\end{equation*}%
for $h=1,...,H$ and $i=1,...,n_{h}$. An unbiased estimator of $\mathrm{Var}%
\left[ \boldsymbol{Z}_{hi}^{\ast }\left( \boldsymbol{\beta }\right) \right] $
is 
\begin{equation*}
\widehat{\mathrm{Var}}\left[ \boldsymbol{Z}_{hi}^{\ast }(\boldsymbol{\beta })%
\right] =\frac{1}{n}\sum\limits_{h=1}^{H}\sum\limits_{i=1}^{n_{h}}%
\boldsymbol{Z}_{hi}^{\ast }(\widehat{\boldsymbol{\beta }}_{\lambda ,Q})%
\boldsymbol{Z}_{hi}^{\ast T}(\widehat{\boldsymbol{\beta }}_{\lambda ,Q}),
\end{equation*}%
i.e. $\mathrm{E}\left[ \widehat{\mathrm{Var}}\left[ \boldsymbol{Z}%
_{hi}^{\ast }\left( \boldsymbol{\beta }\right) \right] \right] =\mathrm{Var}%
\left[ \boldsymbol{Z}_{hi}^{\ast }\left( \boldsymbol{\beta }\right) \right] $%
\ for which the trace is%
\begin{align*}
\mathrm{E}\left[ \mathrm{trace}\left( \widehat{\mathrm{Var}}\left[ 
\boldsymbol{Z}_{hi}^{\ast }(\boldsymbol{\beta })\right] \right) \right] & =%
\mathrm{trace}\left( \mathrm{Var}\left[ \boldsymbol{Z}_{hi}^{\ast }\left( 
\boldsymbol{\beta }\right) \right] \right) , \\
\mathrm{E}\left[ \frac{1}{n}\sum\limits_{h=1}^{H}\sum\limits_{i=1}^{n_{h}}%
\boldsymbol{Z}_{hi}^{\ast }(\boldsymbol{\beta })\boldsymbol{Z}_{hi}^{\ast T}(%
\boldsymbol{\beta })\right] & =\nu d, \\
\mathrm{E}\left[ \frac{1}{nd}\sum\limits_{h=1}^{H}\sum\limits_{i=1}^{n_{h}}%
\boldsymbol{Z}_{hi}^{\ast }(\boldsymbol{\beta })\boldsymbol{Z}_{hi}^{\ast T}(%
\boldsymbol{\beta })\right] & =\nu .
\end{align*}%
Since $\widehat{\boldsymbol{\beta }}_{\lambda ,Q}$ is consistent of $%
\boldsymbol{\beta }$, it holds that $\frac{1}{nd}\sum\limits_{h=1}^{H}\sum%
\limits_{i=1}^{n_{h}}\boldsymbol{Z}_{hi}^{\ast }(\widehat{\boldsymbol{\beta }%
}_{\lambda ,Q})\boldsymbol{Z}_{hi}^{\ast T}(\widehat{\boldsymbol{\beta }}%
_{\lambda ,Q})$ is consistent of $\nu $, but%
\begin{equation*}
\frac{1}{nd}\sum\limits_{h=1}^{H}\sum\limits_{i=1}^{n_{h}}\boldsymbol{Z}%
_{hi}^{\ast }(\widehat{\boldsymbol{\beta }}_{\lambda ,Q})\boldsymbol{Z}%
_{hi}^{\ast T}(\widehat{\boldsymbol{\beta }}_{\lambda ,Q})=\frac{1}{nd}%
\sum\limits_{h=1}^{H}\sum\limits_{i=1}^{n_{h}}(\widehat{\boldsymbol{Y}}%
_{hi}^{\ast }-m\boldsymbol{\pi }_{hi}^{\ast }(\widehat{\boldsymbol{\beta }}%
_{\lambda ,Q}))^{T}\frac{1}{m}\Delta ^{-1}\left( \boldsymbol{\pi }%
_{hi}^{\ast }\left( \boldsymbol{\beta }\right) \right) (\widehat{\boldsymbol{%
Y}}_{hi}^{\ast }-m\boldsymbol{\pi }_{hi}^{\ast }(\widehat{\boldsymbol{\beta }%
}_{\lambda ,Q})),
\end{equation*}%
is in principle different in shape in comparison with (\ref{vM}) with%
\begin{equation*}
\sum_{h=1}^{H}\sum_{i=1}^{n_{h}}\sum_{j=1}^{d+1}\frac{(\widehat{Y}%
_{hij}-m\pi _{hij}(\widehat{\boldsymbol{\beta }}_{\lambda ,Q}))^{2}}{m\pi
_{hij}(\widehat{\boldsymbol{\beta }}_{\lambda ,Q})}=\sum\limits_{h=1}^{H}%
\sum\limits_{i=1}^{n_{h}}(\widehat{\boldsymbol{Y}}_{hi}-m\boldsymbol{\pi }%
_{hi}(\widehat{\boldsymbol{\beta }}_{\lambda ,Q}))^{T}\frac{1}{m}\Delta
^{-}\left( \boldsymbol{\pi }_{hi}^{\ast }\left( \boldsymbol{\beta }\right)
\right) (\widehat{\boldsymbol{Y}}_{hi}-m\boldsymbol{\pi }_{hi}(\widehat{%
\boldsymbol{\beta }}_{\lambda ,Q})),
\end{equation*}%
where $\Delta ^{-}\left( \boldsymbol{\pi }_{hi}^{\ast }\left( \boldsymbol{%
\beta }\right) \right) =\mathrm{diag}^{-1}\left( \boldsymbol{\pi }%
_{hi}\left( \boldsymbol{\beta }\right) \right) $. To assure that both
expressions are equivalent we count with the result related to invariance of
quadratic forms with generalized variances, Lemma 1a, of Moore (1977), from
which is concluded that%
\begin{align*}
& (\widehat{\boldsymbol{Y}}_{hi}-m\boldsymbol{\pi }_{hi}(\widehat{%
\boldsymbol{\beta }}_{\lambda ,Q}))^{T}\frac{1}{m}\Delta ^{-}\left( 
\boldsymbol{\pi }_{hi}^{\ast }\left( \boldsymbol{\beta }\right) \right) (%
\widehat{\boldsymbol{Y}}_{hi}-m\boldsymbol{\pi }_{hi}(\widehat{\boldsymbol{%
\beta }}_{\lambda ,Q})) \\
& =(\widehat{\boldsymbol{Y}}_{hi}^{\ast }-m\boldsymbol{\pi }_{hi}^{\ast }(%
\widehat{\boldsymbol{\beta }}_{\lambda ,Q}))^{T}\frac{1}{m}\Delta
^{-1}\left( \boldsymbol{\pi }_{hi}^{\ast }\left( \boldsymbol{\beta }\right)
\right) (\widehat{\boldsymbol{Y}}_{hi}^{\ast }-m\boldsymbol{\pi }_{hi}^{\ast
}(\widehat{\boldsymbol{\beta }}_{\lambda ,Q})).
\end{align*}
\end{proof}

\subsection*{Proof of Theorem \ref{th2}}
\begin{proof}
The details are omitted for being very similar to Theorem \ref{Th1}
conditioning within each stratum, i.e.%
\begin{align*}
\mathbf{\Omega}_{i,\lambda }^{(h)}\left( \boldsymbol{\beta },\boldsymbol{\Sigma }%
_{hi}\right) & =\widehat{\mathrm{E}}[\mathrm{Var}[\boldsymbol{U}_{\lambda }(%
\boldsymbol{\beta },\boldsymbol{X}_{h})|\boldsymbol{X}_{h}]], \quad
\mathbf{\Omega}_{\lambda }^{(h)}\left( \boldsymbol{\beta }\right)  =\mathrm{Var}%
\left[ \boldsymbol{U}_{\lambda }(\boldsymbol{\beta },\boldsymbol{X}_{h})%
\right] , \\
\mathbf{\Psi}_{i,\lambda }^{(h)}\left( \boldsymbol{\beta }\right) & =-\widehat{%
\mathrm{E}}\left[ \mathrm{E}\left[ \tfrac{\partial }{\partial \boldsymbol{%
\beta }}\boldsymbol{U}_{\lambda }^{T}(\boldsymbol{\beta },\boldsymbol{X}%
_{h})|\boldsymbol{X}_{h}\right] \right] , \quad
\mathbf{\Psi}_{\lambda }^{(h)}\left( \boldsymbol{\beta }\right)  =-\mathrm{E}%
\left[ \frac{\partial }{\partial \boldsymbol{\beta }}\boldsymbol{U}_{\lambda
}^{T}(\boldsymbol{\beta },\boldsymbol{X}_{h})\right] .
\end{align*}
\end{proof}

\subsection*{Proof of Theorem \ref{th:wald}}
\begin{proof}
We have 
\begin{align*}
\boldsymbol{M}^{T}\widehat{\boldsymbol{\beta }}_{\lambda ,Q}-\boldsymbol{l}& 
\boldsymbol{=M}^{T}\text{ }\boldsymbol{\mathbf{\beta }}_{n}-l\boldsymbol{+M}%
^{T}\left( \widehat{\boldsymbol{\beta }}_{\lambda ,Q}-\text{ }\boldsymbol{%
\mathbf{\beta }}_{n}\right) \\
& =\boldsymbol{M}^{T}\text{ }\boldsymbol{\mathbf{\beta }}_{0}+\boldsymbol{%
M^{T}}n^{-1/2}\boldsymbol{\boldsymbol{d}}-\boldsymbol{l+L^{T}\left( \widehat{%
\boldsymbol{\beta }}_{\lambda ,Q}-\text{ }\boldsymbol{\mathbf{\beta }}%
_{n}\right) } \\
& \boldsymbol{=M}\boldsymbol{^{T}}n^{-1/2}\boldsymbol{\boldsymbol{d+M}%
^{T}\left( \widehat{\boldsymbol{\beta }}_{\lambda ,Q}-\text{ }\boldsymbol{%
\mathbf{\beta }}_{n}\right) .}
\end{align*}%
Therefore,%
\begin{equation*}
\boldsymbol{M}^{T}\text{ }\widehat{\boldsymbol{\beta }}_{\lambda ,Q}-%
\boldsymbol{l=M^{T}}n^{-1/2}\boldsymbol{\boldsymbol{d+M}^{T}\left( \widehat{%
\boldsymbol{\beta }}_{\lambda ,Q}-\text{ }\boldsymbol{\mathbf{\beta }}%
_{n}\right) .}
\end{equation*}%
We know, under $H_{1,n}$ that %??? 
\begin{equation*}
\sqrt{n}\left( \widehat{\boldsymbol{\beta }}_{\lambda ,Q}-\boldsymbol{%
\boldsymbol{\mathbf{\beta }}_{n}}\right) \underset{n\longrightarrow \infty }{%
\overset{\mathcal{L}}{\longrightarrow }}\mathcal{N}\left( \mathbf{0},\text{ }%
\mathbf{Q}_{\lambda }\boldsymbol{\left( \boldsymbol{\beta }_{0}\right) }%
\right)
\end{equation*}%
and 
\begin{equation*}
\sqrt{n}\left( \boldsymbol{M}^{T}\text{ }\widehat{\boldsymbol{\beta }}%
_{\lambda ,Q}-\boldsymbol{c}\right) \underset{n\longrightarrow \infty }{%
\overset{\mathcal{L}}{\longrightarrow }}\mathcal{N}\left( M\boldsymbol{^{T}%
\boldsymbol{d}},\text{ }M\boldsymbol{^{T}\mathbf{\Psi}_{\lambda }\left( 
\boldsymbol{\beta }_{0}\right) M}\right) .
\end{equation*}%
%It is well known the following result: "If $\boldsymbol{Z\in }\mathcal{N}%
%\left( \boldsymbol{\mu ,\Sigma }\right) ,$ $\boldsymbol{\Sigma }$ is a
%symmetric projection of rank $k$ and $\boldsymbol{\Sigma \mu =\mu ,}$ then $%
%\boldsymbol{Z}^{T}\boldsymbol{Z}$ is a chi-square distribution with $k$
%degrees of freedom and noncentrality parameter $\boldsymbol{\mu }^{T}%
%\boldsymbol{\mu }^{"}\boldsymbol{.}$
\noindent
The Wald-type test statistics can be written as the quadratic form 
$
W_{n}\left( \widehat{\boldsymbol{\beta }}_{\lambda ,Q}\right) =\boldsymbol{Z}
^{T}\boldsymbol{Z}
$
with 
\begin{equation*}
\boldsymbol{Z}=\sqrt{n}\left[ M\boldsymbol{^{T}\mathbf{\Psi}_{\lambda }\left( 
\boldsymbol{\beta }_{0}\right) M}\right] ^{-1/2}\left( \boldsymbol{M}^{T}%
\text{ }\widehat{\boldsymbol{\beta }}_{\lambda ,Q}-\boldsymbol{l}\right)
\end{equation*}%
and 
\begin{equation*}
\boldsymbol{Z}\underset{n\longrightarrow \infty }{\overset{\mathcal{L}}{%
\longrightarrow }}\mathcal{N}\left( \left[ M\boldsymbol{^{T}\mathbf{\Psi}%
_{\lambda }\left( \boldsymbol{\beta }_{0}\right) M}\right] ^{-1/2}%
\boldsymbol{M^{T}d,I}_{r\times r}\right) ,
\end{equation*}%
where $\boldsymbol{I}$ is the identity $r\times r$ matrix. The application
of the previous gives i). The noncentrality parameter is 
$
\boldsymbol{d}^{T}\boldsymbol{M}\left[ M\boldsymbol{^{T}\mathbf{\Psi}_{\lambda
}\left( \boldsymbol{\beta }_{0}\right) M}\right] ^{-1}\boldsymbol{M^{T}d.}
$
Result ii) follows using relation (\ref{G}).
\end{proof}

\end{document}